\documentclass[usenatbib]{mn2e}
\usepackage{natbib}
\usepackage{hyperref}
\usepackage{times}
\usepackage{epsfig}
\usepackage{graphicx}
\usepackage[export]{adjustbox}
\usepackage{blindtext}
\usepackage{enumitem}
\usepackage{pdflscape}
\usepackage{array}
\usepackage{ragged2e}
\def\mh1{$M_{\rm H_{I}}$}

\def\etal{{\it et al.}\thinspace}

\title[]{The nuclear properties and extended morphologies of powerful radio galaxies: the roles of host galaxy and environment }
\author [H. Miraghaei, et al]{H. Miraghaei$^{1}$, P. N. Best$^2$ \\
$^{1}$School of Astronomy, Institute for Research in Fundamental Sciences, PO Box 19395-5531, Tehran, Iran\\
$^{2}$Institute for Astronomy (IfA), University of Edinburgh, Royal Observatory, Blackford Hill, EH9 3HJ Edinburgh, U.K.\\
}
\begin{document}
\date{}
\pagerange{\pageref{firstpage}--\pageref{lastpage}} \pubyear{}

\maketitle

\label{firstpage}
\begin{abstract}

Powerful radio galaxies exist as either compact or extended sources,
with the extended sources traditionally classified by their radio
morphologies as Fanaroff--Riley (FR) type I and II sources. FRI/II and
compact radio galaxies have also been classified by their optical
spectra into two different types: high excitation (HERG; quasar-mode)
and low excitation (LERG; jet-mode). We present a catalogue of visual
morphologies for a complete sample of $>$1000 1.4-GHz-selected
extended radio sources from the Sloan Digital Sky Survey. We study the
environment and host galaxy properties of FRI/II and compact sources,
classified into HERG/LERG types, in order to separate and distinguish
the factors that drive the radio morphological variations from those
responsible for the spectral properties. Comparing FRI LERGs with FRII
LERGs at fixed stellar mass and radio luminosity, we show that FRIs
typically reside in richer environments and are hosted by smaller
galaxies with higher mass surface density; this is consistent with
extrinsic effects of jet disruption driving the FR dichotomy. Using
matched samples of HERGs and LERGs, we show that HERG host galaxies
are more frequently star-forming, with more evidence for disk-like
structure than LERGs, in accordance with currently-favoured models of
fundamentally different fuelling mechanisms. Comparing FRI/II LERGs
with compact LERGs, we find the primary difference is that compact
objects typically harbour less massive black holes. This suggests that
lower-mass black holes may be less efficient at launching stable radio
jets, or do so for shorter times. Finally, we investigate rarer
sub-classes: wide-angle tail, head-tail, FR-hybrid and double-double
sources.

\end{abstract}

\section{Introduction}
\label{sec:Intro}

Powerful radio galaxies display a very wide range of properties, both in
their extended radio morphologies and in their optical
spectra. Historically, the luminous radio galaxy population has been
sub-divided in two different manners. Fanaroff \& Riley (1974) classified
sources according to their radio morphologies as type I (FRI), in which
the peak of radio emission is located near the core (edge--darkened), and
type II (FRII), in which the peak of surface brightness is at the edge of
the radio lobes far from the center of emission (edge--brightened). An
alternative classification scheme is based on the relative intensity of
high and low excitation lines in the optical spectrum (cf. Hine \& Longair
1979; Laing et al. 1994), comprising high-excitation radio galaxies
(HERGs) and low-excitation radio galaxies (LERGs). The HERG and LERG
populations are believed to represent intrinsically different types of
objects (Best \& Heckman 2012). HERGs show high accretion rate (giving a
total luminosity $>$ 0.01 L$_{\rm Edd}$, where L$_{\rm Edd}$ is the
Eddington luminosity) powered via accretion of cold gas, which may be
brought in through mergers or interactions, or through secular processes
such as non-axisymmetric perturbations or star forming winds. In contrast,
LERGs show low accretion rates ($<$ 0.01 L$_{\rm Edd}$) and are believed to be
powered primarily via accretion of hot intergalactic gas (Heckman \& Best
2014; Yuan \& Narayan 2014). The host galaxies of the HERG and LERG
populations are also different, with HERGs typically being hosted by
galaxies of lower stellar mass, bluer colours, lower concentration (more
disk-like) and lower black hole mass (Best \& Heckman 2012). The redshift
evolution is also different for the two samples: HERGs show rapid cosmic
evolution, while LERGs have little or no redshift evolution (Best
et~al.\ 2014; Pracy et~al.\ 2016).

Two main descriptions have been proposed for the origin of the FR
dichotomy. Early studies highlighted the different emission line
properties of FRIs and FRIIs (Zirbel \& Baum 1995), and proposed that FRIs
and FRIIs might be intrinsically different classes of objects, according
to their central black hole parameters or jet content (Baum, Zirbel \&
O'Dea 1995; Meliani, Keppens \& Sauty 2010). However, these emission line
differences may be driven by the LERG/HERG classification, since in the
samples studied there was a large overlap between the FRI and LERG
populations, and FRIIs with HERGs. An alternative hypothesis is that the
FR dichotomy is extrinsic, driven by the role of the host galaxy and
surrounding environment. In this scenario, FRI and FRII radio galaxies are
considered as fundamentally the same class of objects, with FRIs hosting
less powerful jets which get disrupted by interactions in a dense
surrounding environment (Kaiser \& Best 2007). In support of this model,
several studies report an increased prevalence of FRIs in denser
environment compared to FRIIs (Hill \& Lilly 1991; Gendre et~al.\ 2010, 2013).
 In addition, the discovery of sources with hybrid morphologies,
which are FRI on one side and FRII on the other side, strongly supports
extrinsic models (Gopal-Krishna \& Wiita 2000; Gawronski et~al.\ 2006;
Ceglowski, Gawronski \& Kunert-Bajraszewska 2013). However the origin of
the FR dichotomy might be related to a combination of environment and
central engine properties (Wold, Lacy \& Armus 2007). The FR dichotomy has
also been investigated through hydrodynamical and magneto-hydrodynamical
simulations that have argued the jet disruptions can emerge from
Kelvin-Helmholtz instabilities (Perucho et~al.\ 2010), jet-stellar wind
interactions (Wykes et~al.\ 2015) or magnetic instabilities (Porth \&
Komissarov 2015; Tchekhovskoy \& bromberg 2015); these all favour the
extrinsic scenario.
 
Studies of the evolution of the space density of FR radio galaxies over
redshift (Clewley \& Jarvis 2004; Sadler et~al.\ 2007; Rigby, Best \&
Snellen 2008; Gendre et~al.\ 2010), the host galaxy properties (Heckman et
al. 1994; Baum et~al.\ 1992, 1995; Govoni et~al.\ 2000; Scarpa \& Urry 2001)
and the black hole accretion mechanism (Gendre et~al.\ 2013) have all
helped us to understand the nature of FR dichotomy; however, it is still
unclear what combination of intrinsic and extrinsic scenarios gives the
most realistic description. A primary reason for this is because the
extended morphologies of radio galaxies show a strong dependence on radio
luminosity, and so does their HERG/LERG nature, and thus disentangling the
two effects is challenging.

An example of this is the study of the cosmic evolution of the different
radio source classes. It has been long-established that less powerful
radio sources show less cosmic evolution than more powerful samples
(e.g. Dunlop \& Peacock 1993 and references therein; Rigby et al. 2011). It
has therefore been concluded that FRI sources with low radio powers show
no redshift evolution (Clewley \& Jarvis 2004), while higher power FRI or
FRII have rapid redshift evolution (Rigby, Best \& Snellen 2008; Gendre,
Best \& Wall 2010). However, none of these previous studies took HERG/LERG
classifications into account. More recent studies of the redshift
evolution of HERG/LERG objects have shown that HERGs evolve very strongly
and LERGs show little cosmic evolution, indicating that the luminosity
dependence of the cosmic evolution might be driven by the changing
relative contributions of HERG/LERG populations with luminosity (Best \&
Heckman 2012; Best et al. 2014).

The motivation for the present study is to separate FRI/II dependencies
from HERG/LERG dependencies, using four samples of FRI HERGs, FRI LERGs,
FRII HERGs and FRII LERGs to investigate FRI/FRII and HERG/LERG properties
independently. The host galaxy properties, together with environmental and galaxy
interaction parameters will be investigated for all sub-samples.

 Additionally, while FR radio galaxies have been identified by their
extended morphologies, there is a class of compact radio sources
which have no extended components in their radio structure. 
Baldi, Capetti \& Giovannini (2015) suggested a new type of
radio galaxies called FR0, corresponding to compact sources in this study,
which are more core dominated and display less extended radio emission
compared to FRI/II radio galaxies. These compact
radio galaxies dominate the population at lower radio luminosities
(e.g. Best et~al.\ 2005a).
Studying compact radio sources can help us to
determine whether these objects go through the same evolutionary path as
the extended sources, and may shed light on the origin of different radio
morphologies observed for radio galaxies. Previous studies have claimed
that these compact sources may be: i) the same as the extended sources,
but with radio jets and lobes viewed at a small angle to their axis
(Blandford \& K${\ddot{o}}$nigl 1979; Fanti et~al.\ 1990); ii) young radio
sources at the early stage of their evolution, which will later become FRI
or FRII (Fanti et~al.\ 1995); iii) short-lived radio galaxies, whose jets
get disrupted due to the low jet bulk speed (hence unable to sustain extended radio
jets), perhaps caused by a lower black hole spin (Baldi, Capetti \& Giovannini
2015) or interaction with the dense gas (O'Dea \& Baum 1997; Alexander
2000); iv) a fundamentally different class of objects which do not have
potential of developing extended radio jets.  In this study, we
investigate the environment and host galaxy properties of compact radio
sources, comparing them with the extended sources in order to examine the
above scenarios.

The layout of our paper is as follows. The radio source samples and
classifications are presented in Section 2. Results considering the
overall differences of FRIs and FRIIs (irrespective of HERG/LERG
classification) are shown in Section 3, and compared to the
literature. Our main result, considering FRI/FRII, HERG/LERG and
compact/extended comparisons using matched samples that remove other
dependencies, are presented in Section 4. In Section 5 we discuss radio
galaxies with special and complex morphologies, for which our
classification produces significant samples. We summarise and draw
conclusions in Section \ref{sec:summery}.  Throughout the paper we assumed
a $\Lambda CDM$ cosmology with the following parameters: $\Omega_m=0.3$,
$\Omega_\Lambda=0.7$ and $H_0=100 h$ km s$^{-1}$ Mpc$^{-1}$ where $h$ =
0.70.\\

\section {Sample and classification }
\label{sec:Sample}

\begin{table*}

\begin{tabular}{rrrrrrrrr}
\hline
Plate & Julian & Fibre & RA      & Dec.        & z &  Log[L$_{rad,t}$]& Size         &  FR     \\
ID    & Date   & ID    &  J2000  &  J2000      &   &                  &  in radio       &  class   \\
      &        &       & (h)     & ($^{\circ}$)&   &  (W Hz$^{-1}$)   & (arcsec)     &          \\    
\hline 

267 & 51608 & 34 & 9.9446584 & -0.02334 & 0.1392 & 24.92 & 107.66 & 200 \\
267 & 51608 & 205 & 9.8472382 & -0.88775 & 0.2715 & 25.08 & 30.00 & 200 \\
267 & 51608 & 260 & 9.8285522 & -0.84008 & 0.0810 & 24.42 & 43.06 & 300 \\
269 & 51910 & 257 & 10.0313580 & -0.87805 & 0.1364 & 24.86 & 124.24 & 100 \\
271 & 51883 & 93 & 10.3095760 & -0.83961 & 0.3410 & 25.47 & 27.38 & 200 \\
273 & 51957 & 633 & 10.6016020 & 0.10189 & 0.0968 & 25.06 & 145.39 & 100 \\
274 & 51913 & 218 & 10.6525950 & -0.78773 & 0.0952 & 23.98 & 115.91 & 210 \\
275 & 51910 & 617 & 10.8205810 & 0.99589 & 0.1065 & 24.44 & 71.47 & 210 \\
276 & 51909 & 314 & 10.8225380 & -0.66806 & 0.0387 & 23.94 & 87.16 & 100 \\
276 & 51909 & 440 & 10.8330150 & 0.32231 & 0.0390 & 23.34 & 83.86 & 100 \\
279 & 51608 & 34 & 11.3555320 & -0.22246 & 0.1010 & 25.24 & 85.64 & 102 \\
284 & 51662 & 114 & 11.9204590 & -0.52610 & 0.1322 & 24.83 & 41.03 & 102 \\
285 & 51663 & 190 & 11.9948950 & -0.53164 & 0.1782 & 24.64 & 123.28 & 210 \\
286 & 51999 & 267 & 12.0303620 & -0.50945 & 0.3282 & 25.21 & 14.41 & 200 \\
287 & 52023 & 266 & 12.1842470 & -0.33479 & 0.3192 & 25.48 & 12.41 & 200 \\
287 & 52023 & 573 & 12.2428980 & 0.79107 & 0.2510 & 25.09 & 109.39 & 200 \\
288 & 52000 & 490 & 12.3237460 & 0.68112 & 0.4062 & 25.56 & 15.78 & 200 \\
288 & 52000 & 502 & 12.3424470 & 0.07151 & 0.1585 & 24.74 & 135.33 & 100 \\
290 & 51941 & 291 & 12.5461130 & -0.92355 & 0.2050 & 24.87 & 133.36 & 200 \\
291 & 51660 & 42 & 12.7461690 & -1.01928 & 0.1468 & 24.39 & 12.19 & 200 \\
...&...&...&...&...&...&...&...&... \\
...&...&...&...&...&...&...&...&... \\
\hline  
\end{tabular}
\caption{Properties of the 1329 extended radio galaxies with z$>$0.03. The first 20
  sources are listed here: the full table is available
  electronically. Columns 1 to 3 are the SDSS identifications of the target
  sources. The next three columns are the coordinates and redshift of the
  sample objects. Column 7 is total radio luminosity. Column 8 is the size
  of the radio source in arcsec. Column 9 indicates the morphological
  classification of the radio source. This is expressed in three digits.
  The first (left-most) digit indicates the FR class: (1) represents FRI,
  (2) is for FRII, (3) for hybrid and (4) unclassifiable. The second
  (middle) digit indicates whether the FR classification is consider
  certain (0) or less secure (1). The third (right-most) digit highlights
  any special nature of the sources: (0) stands for normal, (1) for a
  double-double source, (2) for a wide-angle tailed source, (3) for
  diffuse, and (4) for head-tail radio galaxies. An example of each class
  is presented in Figure~$\ref{14}$, as detailed in Table~$\ref{table4}$.
\label{table1}
}
\end{table*}

\begin{table*}

\begin{tabular}{rrrrrrrrl}
\hline
Object&Plate & Julian & Fibre & RA      & Dec.        & z &   FR    & Note \\
      &ID    & Date   & ID    &  J2000  &  J2000      &   &  class &   \\
      &      &        &       & (h)     & ($^{\circ}$)&   &        &  \\    
\hline 
A1&450&51908&38& 9.2855509 &  55.15227 &   0.1820&100& certain FRI\\
A2&2422&54096&67&8.4179606  & 12.73467  &  0.3216 &200&certain FRII\\
A3&759&52254&12&8.2815938  & 39.18779 &   0.4654 &300&FR hybrid\\
A4&904&52381&307&10.1744710 &  53.05367  &  0.3411&400&unclassifiable\\
B1&596&52370&221& 11.1525290 & 63.47019  &  0.4263 &110&uncertain FRI\\
B2&1202&52672&463& 9.6429687  & 45.33995  &  0.4501&210&uncertain FRII\\
B3&1724&53859&275& 15.6371080  &  7.95388  &  0.3566&310&uncertain FR hybrid\\
B4&1603&53119&165& 11.0489000  & 11.25012 &   0.4747 &400&unclassifiable\\
C1&2750&54242&325& 14.8007230  & 14.78097  &  0.2089 &102&wide-angle-tailed FRI\\
C2&796&52401&492& 15.7547800 &  50.79831  &  0.4309&201&double-double FRII\\
C3&1833&54561&586& 15.3127010 &   6.23225 &   0.1021&104&head-tail FRI\\
C4&814&52370&117& 16.3080980  & 44.57584  &  0.1966&103&diffuse FRI\\
\hline  
\end{tabular}
\caption{List of objects presented in Fig.~$\ref{14}$. Column 1 represents the
  object label according to their rows and column in Fig.~$\ref{14}$. Columns 2-4
  are the SDSS identifications of the target sources. The next three
  columns are the coordinates and redshift of the objects. Column 8 is the
  FR class of the objects as described in Table~$\ref{table1}$. Column 9
  is a note describing the type of the radio galaxies.
\label{table4}
}
\end{table*}

\begin{figure*}
\center
\epsfig{file=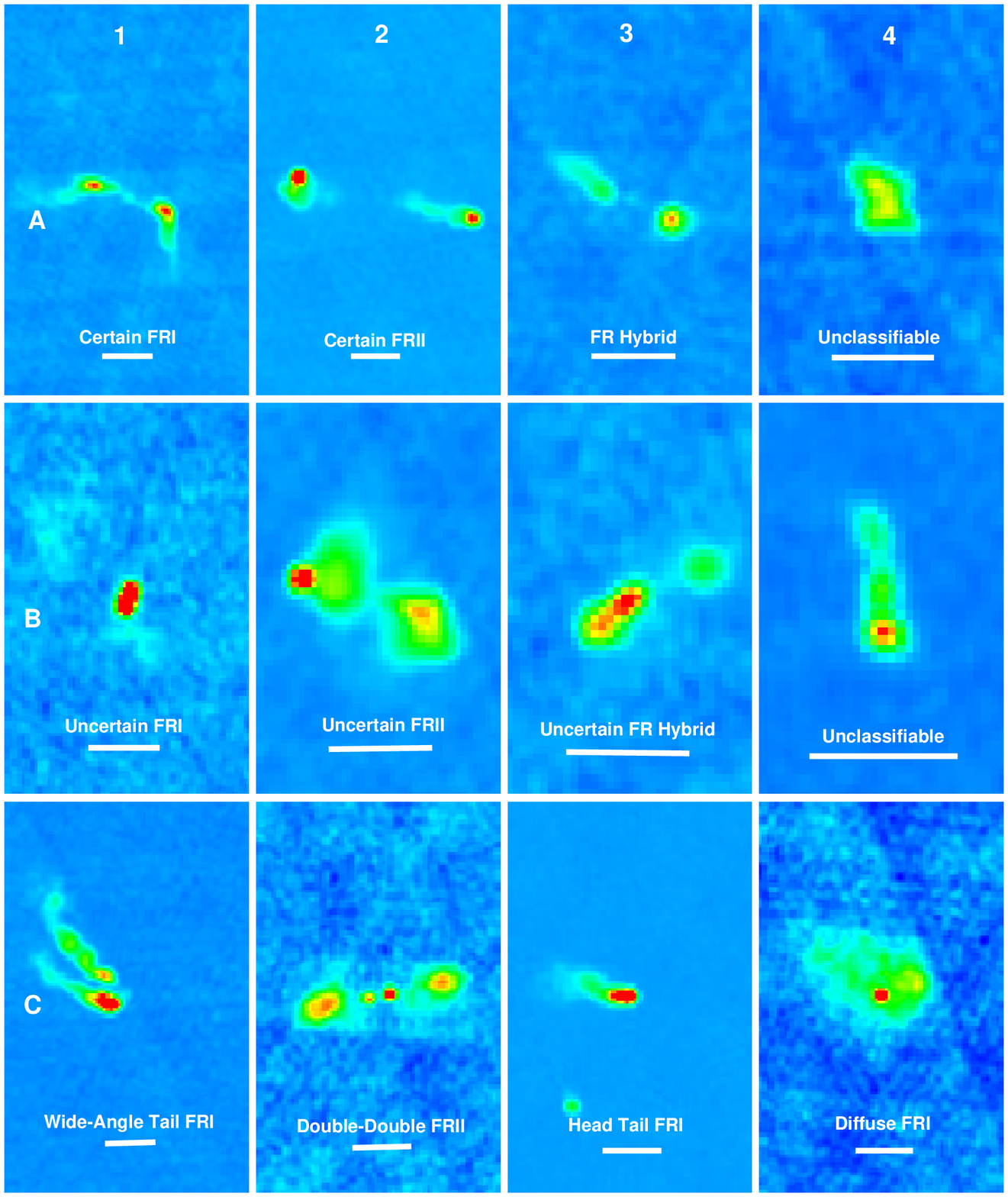,width=6.5in,height=7.8in}
\caption{Examples of different classes of extended radio sources. The list
  of objects presented is given in Table~$\ref{table4}$, according to their row and
  column labels. The white bars are 30 arcsec length scale. For each source, 
  the optical host galaxy position is precisely at the centre of the panel.}
\label{14}
\end{figure*}

 In this section, we describe the selection criteria of FRI, 
FRII and compact sources, the method that we applied for FRI/FRII 
and HERG/LERG classifications and the host galaxy and environment properties of the sample that we use
in this study.

\subsection {Sample selection; global constraints}
 The radio source sample and the parent galaxy sample are taken from Best
\& Heckman (2012), who have cross-matched the seventh data release (DR7;
Abazajian et~al.\ 2009) of the Sloan Digital Sky Survey (SDSS; York
et~al.\ 2000) with the National Radio Astronomy Observatory (NRAO) Very
Large Array (VLA) Sky Survey (NVSS; Condon et~al.\ 1998) and the Faint
Images of the Radio Sky at Twenty centimetres (FIRST) survey (Becker,
White \& Helfand 1995), following the techniques of Best et~al. (2005a). We
applied a lower redshift cut of $z>0.03$ due to the
large angular size of the nearby sources (and potential errors in
catalogued higher-level SDSS parameters) and considered objects only within 
the SDSS `main galaxy' or `luminous red galaxy' samples. 
A 40 mJy flux density cut was also applied so that there would be sufficient
signal-to-noise in any extended structures to allow morphological
classification. Radio sources classified as having an AGN host
(rather than having radio emission dominated by star formation; see Best
\& Heckman 2012) were selected. 

To investigate a sample of extended sources, and classify their morphologies,
 the sub-sample of sources with multiple components in either the FIRST or 
NVSS imaging (see Best et~al. 2005a) were considered.  Sources contained
 within a single FIRST component were not considered because it would be
nearly impossible to judge their morphology. 

\subsection {Morphological classification}

FRI/FRII Classification was
primarily based upon the original definition of the two classes (Fanaroff
\& Riley 1974), namely whether the distance between the peak of the
emission on the opposite sides of the radio source was larger (FRII) or
smaller (FRI) than half of the total size of the radio source. 
The extended radio sample were visually examined in order to
morphologically classify them (cf. Best 2009). However,
the relatively poor angular resolution of FIRST and the low sensitivity to
extended low surface brightness structures limits the ability to determine
both source sizes and peak locations, particularly for smaller sources. A
degree of human interpretation was therefore required. An additional flag
was therefore introduced to the classification, to note whether it was
secure, or less certain. In a few cases a visual
examination showed that the source had been incorrectly flagged as
extended, and these were removed from the extended sample. In total, there were
1329 genuinely morphologically classified extended sources.

\begin{table}
\center
\begin{tabular}{lrrr}
\hline
 & FRI & FRII & Compact      \\
\hline
HERG &5 & 8 & 5    \\
LERG &92 & 32 &  103   \\
Total &97 & 40 &  108   \\
\hline  
\end{tabular}
\caption{Numbers of sources in the sample of galaxies (z $<$ 0.1) with different classifications used in this study.  
\label{table5}
}
\end{table}

\begin{landscape}
\begin{table}
\begin{tabular}{>{\RaggedLeft}p{0.5cm}>{\RaggedLeft}p{0.5cm}>{\RaggedLeft}p{0.5cm}>{\RaggedLeft}p{1.2cm}>{\RaggedLeft}p{1.2cm}>{\RaggedLeft}p{1.2cm}>{\RaggedLeft}p{0.5cm}>{\RaggedLeft}p{0.7cm}>{\RaggedLeft}p{0.5cm}>{\RaggedLeft}p{0.5cm}>{\RaggedLeft}p{0.5cm}>{\RaggedLeft}p{1cm}>{\RaggedLeft}p{0.7cm}>{\RaggedLeft}p{0.8cm}>{\RaggedLeft}p{0.8cm}>{\RaggedLeft}p{0.8cm}>{\RaggedLeft}p{0.8cm}>{\RaggedLeft}p{0.8cm}>{\RaggedLeft}p{0.6cm}>{\RaggedLeft}p{0.6cm}>{\RaggedLeft}p{0.6cm}}
\hline
Plate & Julian & Fibre& RA      & Dec.        & z & M$_{\star}$ & R$_{50}$   &  g-r & D$_{4000}$ &  C &$\mu_{50}$&M$_{BH}$ & $\eta$ & Q & n & PCA1 & PCA2 &  L$_{\rm [OIII]}$&HERG/& compact/ \\
ID    & Date   & ID    &  J2000  &  J2000        &  &  &   &   &  &   && &  &  &  &  &  & &LERG& FR1/FR2 \\  
    &    &     &   (h)     & ($^{\circ}$)      &  &  & (kpc)  &   &  &   && &  &  &  &  &  & && \\    
\hline 
1016 & 52759 & 293 & 11.751812 & 53.64800 & 0.06901 & 11.46 & 7.04 & 1.01 & 2.00 & 3.22 & 8.96636 & 8.69 & 0.323 & ${\textendash}$0.433 & 0.477 & 1.753 & 0.133 & 6.11 & 0 & 2\\
1017 & 52706 & 284 & 11.836466 & 53.72242 & 0.06031 & 11.25 & 7.93 & 1.07 & 2.07 & 2.68 & 8.65305 & 8.38 & 0.510 & 0.874 & 0.602 & 2.709 & 0.724 & 5.97 & 0 & 2\\
1044 & 52468 & 504 & 14.136240 & 52.68005 & 0.08287 & 11.63 & 12.15 & 1.06 & 1.94 & 2.58 & 8.66213 & 8.44 & 0.286 & ${\textendash}$0.288 & 0.301 & 1.804 & 0.255 & 6.20 & 0 & 0\\
1044 & 52468 & 602 & 14.197066 & 52.81670 & 0.07649 & 11.43 & 6.33 & 0.94 & 1.70 & 3.42 & 9.02853 & 8.61 & 0.466 & ${\textendash}$0.805 & 0.903 & 1.673 & ${\textendash}$0.226 & 6.37 & 0 & 0\\
1046 & 52460 & 612 & 14.589354 & 50.85637 & 0.09969 & 11.22 & 6.65 & 1.08 & 1.64 & 3.09 & 8.77514 & 7.98 & ${\textendash}$1.512 & ${\textendash}$0.763 & 0.301 & ${\textendash}$0.231 & 1.727 & 6.41 & 0 & 0\\
1169 & 52753 & 172 & 15.998331 & 44.70899 & 0.04173 & 11.04 & 3.54 & 1.07 & 1.89 & 3.26 & 9.14332 & 7.96 & ${\textendash}$0.447 & ${\textendash}$1.296 & 0.477 & 0.493 & 0.372 & 6.23 & 0 & 0\\
1170 & 52756 & 473 & 16.134591 & 43.16347 & 0.08490 & 11.49 & 7.61 & 1.04 & 2.03 & 2.97 & 8.92854 & 8.41 & 0.418 & 1.100 & 0.477 & 2.753 & 0.948 & 6.15 & 0 & 1\\
1176 & 52791 & 637 & 16.983614 & 32.49423 & 0.06274 & 11.22 & 4.42 & 1.04 & 2.00 & 3.24 & 9.13068 & 8.53 & 0.715 & ${\textendash}$1.119 & 0.845 & 1.729 & ${\textendash}$0.655 & 6.33 & 0 & 1\\
1184 & 52641 & 415 & 8.313819 & 4.10876 & 0.09462 & 11.21 & 5.02 & 1.01 & 1.85 & 3.18 &  9.01026 & 8.40 & 0.265 & ${\textendash}$2.108 & 0.477 & 0.706 & ${\textendash}$0.801 & 6.17 & 0 & 0\\
1186 & 52646 & 613 & 8.615519 & 5.54502 & 0.09927 & 11.30 & 8.74 & 0.99 & 1.64 & 2.42 &  8.61879 & 7.66 & ${\textendash}$1.358 & ${\textendash}$1.156 & 0.301 & ${\textendash}$0.313 & 1.344 & 7.23 & 1 & 2\\
1192 & 52649 & 448 & 9.049780 & 6.32909 & 0.07708 & 11.31 & 6.55 & 1.01 & 2.03 & 3.07 & 8.87890  & 8.37 & ${\textendash}$0.621 & ${\textendash}$1.512 & 0.301 & 0.195 & 0.415 & 5.99 & 0 & 2\\
1198 & 52669 & 611 & 9.021818 & 40.11377 & 0.09617 & 11.12 & 5.71 & 1.01 & 1.91 & 3.58 & 8.80738 & 8.27 & ${\textendash}$0.176 & ${\textendash}$1.066 & 0.699 & 0.893 & 0.245 & 6.15 & 0 & 1\\
1220 & 52723 & 37 & 11.104728 & 8.97591 & 0.08502 & 11.35 & 7.83 & 1.04 & 2.04 & 3.15 & 8.76413 & 8.51 & ${\textendash}$1.339 & ${\textendash}$0.684 & 0.477 & ${\textendash}$0.015 & 1.605 & 6.21 & 0 & 1\\
1222 & 52763 & 423 & 11.288224 & 9.95881 & 0.07891 & 11.07 & 3.95 & 1.02 & 1.92 & 3.22 & 9.07829 & 8.51 & ${\textendash}$0.794 & ${\textendash}$0.945 & 0.477 & 0.362 & 0.919 & 6.19 & 0 & 0\\
1237 & 52762 & 433 & 10.201777 & 8.69253 & 0.09844 & 11.32 & 8.25 & 1.04 & 2.02 & 3.24 & 8.68821 & 8.41 & 0.621 & ${\textendash}$0.564 & 0.602 & 1.967 & ${\textendash}$0.235 & 6.38 & 0 & 1\\
1238 & 52761 & 296 & 10.320063 & 8.07998 & 0.08654 & 11.32 & 8.55 & 1.04 & 2.06 & 3.21 & 8.65719 & 8.19 & ${\textendash}$1.120 & ${\textendash}$1.674 & 0.477 & ${\textendash}$0.387 & 0.805 & 6.15 & 0 & 1\\
1238 & 52761 & 550 & 10.389618 & 8.86699 & 0.06260 & 11.22 & 3.72 & 1.03 & 2.05 & 3.56 & 9.27952 & 8.60 & ${\textendash}$1.143 & ${\textendash}$0.692 & 0.477 & 0.172 & 1.408 & 6.09 & 0 & 1\\
1265 & 52705 & 158 & 8.093057 & 24.16399 & 0.05968 & 10.99 & 3.27 & 0.96 & 1.68 & 3.26 & 9.16118 & 8.17 & ${\textendash}$0.065 & ${\textendash}$3.100 & 0.699 & ${\textendash}$0.202 & ${\textendash}$1.067 & 7.79 & 1 & 2\\
1269 & 52937 & 228 & 8.667324 & 29.81740 & 0.06484 & 11.24 & 9.28 & 0.95 & 1.42 & 2.51 & 8.50579 & 8.26 & ${\textendash}$1.381 & 0.538 & 0.000 & 0.667 & 2.368 & 7.78 & 1 & 2\\
1272 & 52989 & 114 & 9.126363 & 32.95634 & 0.04906 & 11.04 & 6.96 & 1.12 & 1.49 & 2.41 & 8.55564 & 7.74 & ${\textendash}$0.486 & ${\textendash}$1.767 & 0.000 & 0.176 & 0.133 & 5.71 & 0 & 0\\
...&...&...&...&...&...&...&...&...&...&...&...&...&...&...&...&...&...&...&...&... \\
...&...&...&...&...&...&...&...&...&...&...&...&...&...&...&...&...&...&...&...&...\\
\hline  
\end{tabular}
\caption{Properties of FR radio galaxies and compact radio sources used in
  our analysis. The first 20 sources are listed here: the full table is
  available electronically. Columns 1 to 3 are the SDSS identifications of the
  target sources. The next three columns are the coordinates and redshift of
  the sample objects. Columns 7-13 are host galaxies properties: log
  [stellar mass/ solar mass] (M$_{\star})$, galaxy size in kpc (R$_{50}$),
  colour (g-r), 4000$\AA$ break strength (D$_{4000}$),
  C=R$_{90}$/R$_{50}$, log [half-light surface mass density] ($\mu_{50}$) and log
  [black hole mass/ solar mass] (M$_{BH}$) respectively. Columns 14-18 are
  environment properties of the sources: density ($\eta$), tidal (Q),
  richness (log [n]), PCA1 and PCA2 respectively. Column 19 is log
  [L$_{\rm [OIII]}$/ L$_{\rm sun}$] as we described in Section 2.5. Column 20 is
  the HERG (1) and LERG (0) class of sources. Column 21 is the compact (0)
  and extended (1 for FRI and 2 for FRII) labels for target sources.
\label{table2}
}
\end{table}
\end{landscape}

Some sources presented morphologies which did not fit obviously into an
FRI or FRII morphology. 35 sources were classified as hybrid sources
(Gopal-Krishna \& Wiita 2000), which display an FRI-like morphology on one
side of the nucleus and an FRII-like morphology on the opposite side. A
further 40 sources were deemed to be unclassifiable. Additionally some
sources presented interesting morphologies which were given additional
flags. These are 5 double-double sources (Schoenmakers et~al.\ 2000), 9
head-tail sources (Rudnick \& Owen 1976) and 53 wide-angle tailed sources
(cf. Owen \& Rudnick 1976). Examples of each of these sources are shown in
Figure~$\ref{14}$. The source classifications for the full sample of
sources are given in Table~$\ref{table1}$.

In addition to classifying the morphology, the visual analysis confirmed
the NVSS and FIRST components of which the sources were comprised,
allowing total fluxes (and hence luminosities) and radio source sizes to
be determined. These properties are also provided in
Table~$\ref{table1}$. Total radio fluxes were obtained by summing across
the NVSS component fluxes; these should be reliable for sources with sizes
upto $\approx 500$ arcsec (the largest angular size observable in snapshot
observations with the VLA at 1.4\,GHz in D-array configuration), but may
be underestimated for sources larger than this size (of which there are
only 3).  Source sizes are determined from the maximum angular separation
of the centroids of the catalogued NVSS and/or FIRST components of the
source. These may marginally underestimate the sizes of poorly resolved
sources, but should provided a good approximation. Furthermore, the sizes
of FRI sources are likely to be underestimated, since the emission from
these gets progressively fainter with distance from the nucleus, and the
most extended emission is likely to be missed by the short observations of
NVSS and FIRST.

\begin{figure}
\center
\epsfig{file=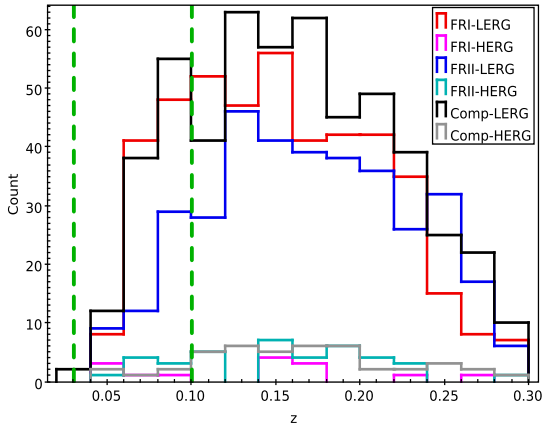,width=2.8in,height=2.2in}
\caption{The redshift distribution of FR radio galaxies and compact radio
  AGN, separated into HERG and LERG classifications. The vertical lines
  are upper and lower redshift limit cuts applied to the all samples for
  the analysis in this paper.}
\label{RedshiftDis}
\end{figure}

\begin{figure}
\center
\epsfig{file=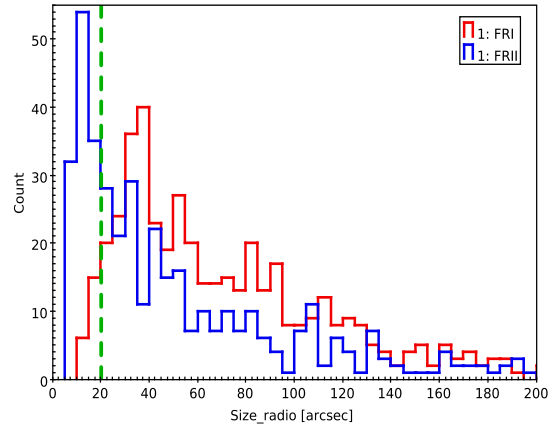,width=2.8in,height=2.2in}
\caption{The distribution of radio sizes of the FR radio galaxies. The
  vertical line is the lower limit cut we applied to FRI and FRII
  samples for the analysis in this paper.}
\label{RadioSize}
\end{figure}

\begin{figure}
\center
\epsfig{file=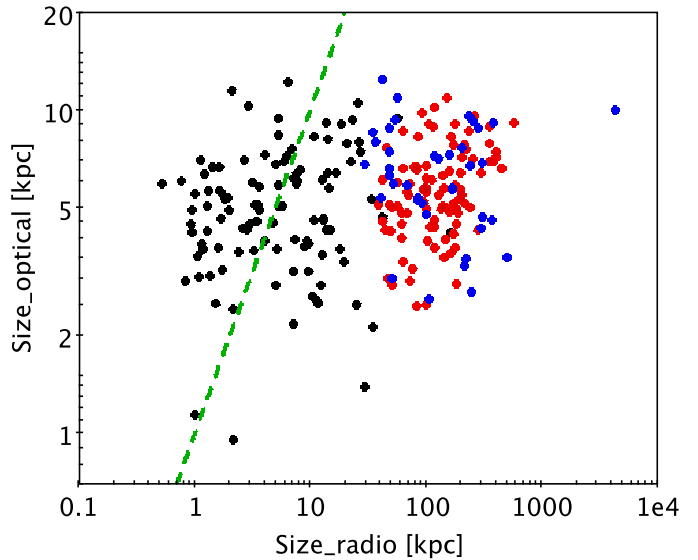,width=3.5in,height=2.9in}
\caption{The radio and optical sizes (in kpc) of the compact (black), FRI (red) and FRII (blue) sources. 
The diagonal line represents equality between the both scales.}
\label{Physize}
\end{figure}

\subsection {Selection of subsamples for analysis} 

For the goal of this paper, additional constraints have also been applied to the sample.
We applied an upper redshift cut of $z<0.1$ due to the
completeness limits in spectroscopic classification of
HERGs and LERGs (see Section 2.4), and in finding companion galaxies for environmental
studies (see Section 2.5). The redshift distribution of FRI and FRII radio sources are shown in
 Fig.~$\ref{RedshiftDis}$.  FRI sources have, on average, lower redshifts than FRIIs, as
expected in a flux-limited sample since they are typically found at lower
luminosities. 

The distribution of angular sizes of FRI/FRII populations are shown in 
Fig.~$\ref{RadioSize}$.  It is notable that there is a dearth of FRI sources with angular sizes below 20 arcsec.
This is believed to be due to the biases in classification of the sources arising from the 
low angular resolution and low surface brightness sensitivity of FIRST. Small-scale FRI sources, 
especially those whose extended emission is faint, may well be catalogued by FIRST as a 
single (albeit extended) component, and thus excluded from our analysis (which was restricted 
to multi-component sources; see Section 2.1).
To avoid these biases in our further analysis, we hereafter
restrict both FRI and FRII samples to the objects with angular sizes above
20$''$. The final sample size for each classification is presented in
Table~$\ref{table5}$.

For comparison with the extended sample, we also define a sample of compact radio sources.
 These correspond to those sources identified as single-component FIRST sources in the Best
\& Heckman (2012) sample, with the same additional constraints: i) objects only within 
the SDSS `main galaxy' or `luminous red galaxy' samples, 
ii) with the redshift cuts of 0.03 $<$ z $<$ 0.1, iii) and the flux density cut of S $>$ 40 mJy. The redshift distribution 
of compact sources are also shown in Fig.~$\ref{RedshiftDis}$.

The optical sizes of the host galaxies compared to the radio sizes (in kpc) of the compact 
and extended radio sources are displayed in Fig.~$\ref{Physize}$. 
The diagonal line represents equality between the two scales (though, note these are 
differently defined, as the radio size is the full size of the sources whereas the 
optical size is a half-light radius). Compact radio sources
are distributed around the equality line while extended sources are at the larger radio sizes.
Therefore, the FRI and FRII sources typically extend well beyond their host galaxies and
are large enough to be affected by the surrounding environment as well as conditions within 
their host galaxy which are both subjects of this study. 
 We will consider both the extended and compact samples in Section 4.

\subsection {HERGS/LERGS classification}

The method used to classify sources into the two class of HERG and LERG
has been extensively described by Best \& Heckman (2012) who considered
the ratios of four high excitation lines ([OIII], [NII], [SII] and [OI])
to the H$_{\alpha}$ and H$_{\beta}$ emission lines, and also the
equivalent width of the [OIII] emission line. They used the line-ratio
diagnostic diagrams from Kewley et~al.\ (2006) and also Cid Fernandes
et~al.\ (2011). We adopt the source classifications derived by Best \&
Heckman. Note that these were only complete out to $z=0.1$, which is one
reason why this was adopted as an upper redshift limit in our analysis.
The radio sample including HERG/LERG classifications is presented in
Table~$\ref{table2}$.

Using these classifications together with the morphological classifications from
Section~2.2, the sample has been divided into six sub-samples of FRI
HERGs, FRI LERGs, FRII HERGs, FRII LERGs, compact HERGs and compact LERGs with which we can study how
the environment and host galaxy properties relate separately to the
HERG/LERG and FRI/FRII/compact classifications. The sample size for each
classification is presented in Table~$\ref{table5}$, while Fig.~$\ref{RedshiftDis}$ shows
the redshift distributions of the six sub-samples. The LERG sources are
dominant in the sub-samples of FRI, FRII and compact, although the fraction of
HERGs is highest for the FRIIs. Since there are relatively
few sources classified as HERGs, when considering the FRI/II comparison in
the rest of the paper we focus only on the FR LERGs.

\subsection {Host galaxy and environment properties of the sample}

Host galaxy properties for the radio sources are extracted from the
value-added spectroscopic catalogues produced by the group from the Max
Planck Institute for Astrophysics, and Johns Hopkins University (cf. Brinchmann et al. 2004).
 In particular, the parameters used in this paper
are defined and estimated as follow:

\begin{description}[align=left]
\item  [stellar mass] (M$_{\star}$ or Mass) is derived from the extinction-corrected
 optical luminosity using the mass to light ratio (Kauffmann, Heckman \& White 2003).
\item [black hole mass] (M$_{BH}$) is estimated using the velocity
  dispersion (${\sigma}_{\star}$) of the galaxy and the relation between
  the velocity dispersion and the black hole mass given in Tremaine
  et~al.\ (2002): log(M$_{BH}$/M$_{\star}$)=8.13 + 4.02 log
  (${\sigma}_{\star}$/200km s$^{-1}$).
\item [absolute magnitude] (M$_{r}$) is the SDSS r-band absolute
  magnitude.
\item [size] which is R$_{50}$, defined by the radius containing 50 percent
  of the galaxy light in the r-band.
\item [half-light surface mass density] ($\mu_{50}$) which is calculated
  using the relation: $\mu_{50}$ = 0.5 M$_{\star}$/ ($\pi$
  R$_{50}$$^{2}$).
\item [concentration] (C) calculated from the relation:
  C=R$_{90}$/R$_{50}$, where R$_{90}$ is the radius containing 90 percent
  of the r-band galaxy light. Galaxies with high concentration index
  (C$>$2.6) are typically bulge-dominated systems whereas galaxies with C
  $<$ 2.6 are mostly disk-dominated systems (see Kauffmann et al. 2003).
\item [colour] (g--r) at rest-frame.
\item [4000$\AA$ break] (D$_{4000}$) which is strength of the 4000$\AA$
  break of the galaxy optical spectrum, and is small for young stellar
  populations and large for old, metal-rich galaxies, thus giving a guide
  to the age of the galaxy.
\item [OIII luminosity] which is calculated from the detected [OIII]~5007
  emission line provided this line is detected with a S/N ratio above
  2.5. In the case of no detection, we have used the corresponding upper
  limit luminosity of the 2.5 sigma flux density.
\item [total radio luminosity] (L$_{rad,t}$) is calculated from the total
  radio flux obtained by summing across the NVSS component fluxes.
\item [core radio luminosity] (L$_{rad,c}$) is calculated from the radio
  flux of the central FIRST component of the galaxy.
\end{description}

To obtain the environment and galaxy interaction parameters we
cross-matched the main catalogue with the environmental catalogue from
Sabater, Best \& Argudo-Fernandez (2013).
They defined and estimated three interacting parameters of density, tidal
force (hereafter tidal) and richness, as follows:

\begin{description}[align=left]
\item [density] ($\eta$) which is defined from the distance ($r_{10}$ in
  Mpc) to the 10th nearest neighbour, $\eta =log[ 10/(4\pi
    r_{10}^{3} /3)]$.
\item [tidal interaction] (Q) which is defined by the relative tidal
  forces exerted by companions (i) with respect to the internal binding
  forces of the target galaxy (t). Here, $R$ is the radius of the target
  galaxy, $d$ is the distance between the target and the companion, and
  $Lr$ is the corrected luminosity of the galaxy in r band. $$Q_{t}= log
  \left[\sum_{i} (Lr_{t}/Lr_{i}){(2R_{t}/d_{i,t})}^{3} \right] $$
\item [richness] (n) is the number of galaxies in the cluster or group
  to which the target galaxy belongs, as derived from the friends-of-friends
  catalogue of Tago et~al. (2010).
\end{description}

Sabater et~al. (2013) also carried out a principal component analysis, to combine
the density and tidal parameters in a way that removes the observed
correlation between these two parameters. They thus introduced two new
parameters:

\begin{description}[align=left]
\item [PCA1] which traces the overall interaction level and environmental
  density of a galaxy.
\item [PCA2] in which a higher value traces higher one-on-one interactions
  and a lower value traces galaxies that are relatively isolated for their overall environment.
\end{description}

The host galaxy and environment properties of the samples are listed in
Table~$\ref{table2}$.

\section {Overall properties of the samples}
\label{sec:overal}

\begin{figure*}
\center

\epsfig{file=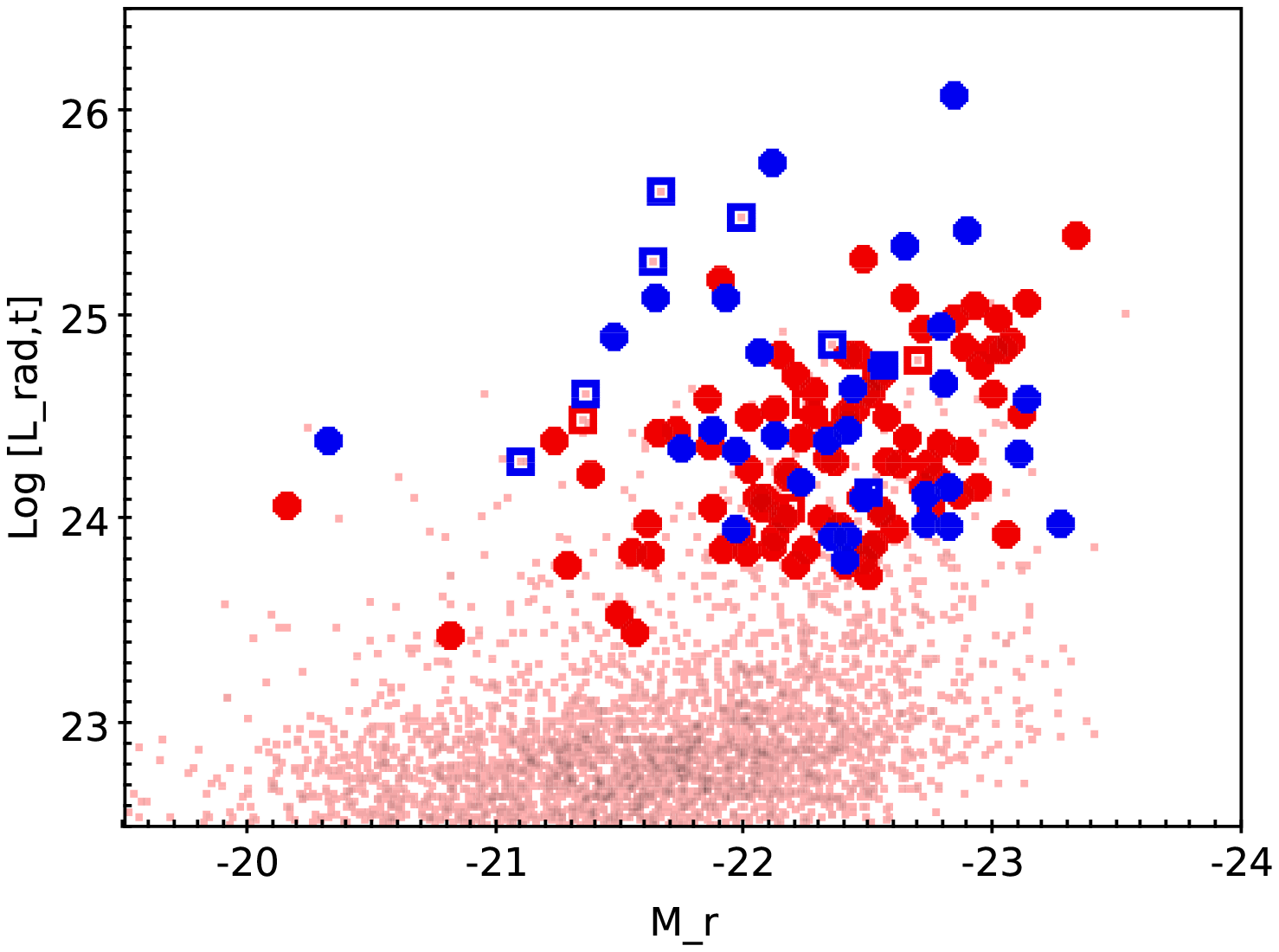,width=3.2in,height=2.6in}
\epsfig{file=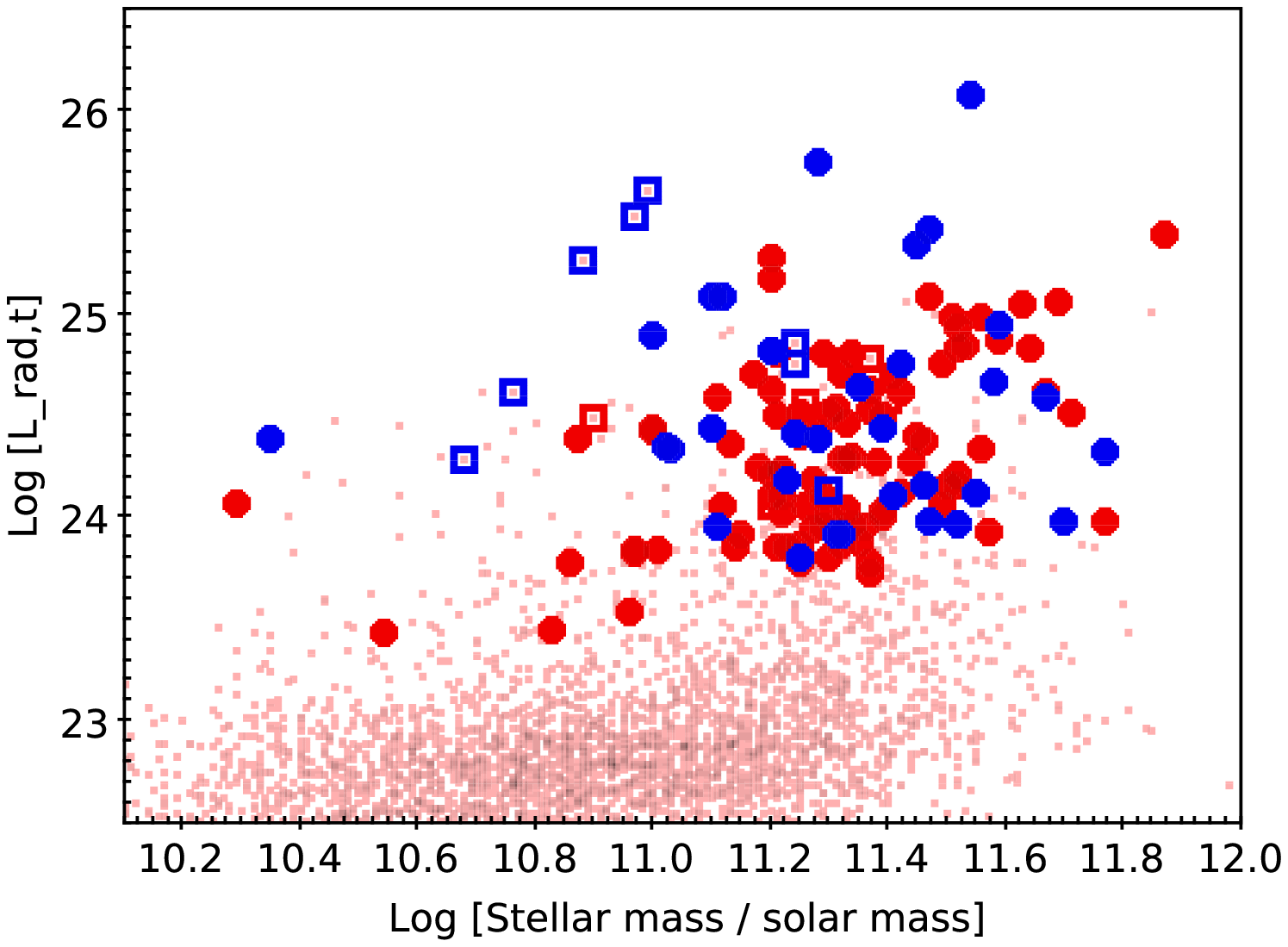,width=3.2in,height=2.6in}
\caption{Total radio luminosity of the FRI (red) and FRII (blue) radio
  galaxies versus r-band absolute magnitude (left panel) and stellar mass
  (right panel). The filled circles represent LERGs and open squares are
  HERGs. Pink circles are all galaxies (hereafter galaxies with 0.03$<$z$<$0.1).}
\label{RadioOpt}
\end{figure*}

\begin{figure*}
\center
\epsfig{file=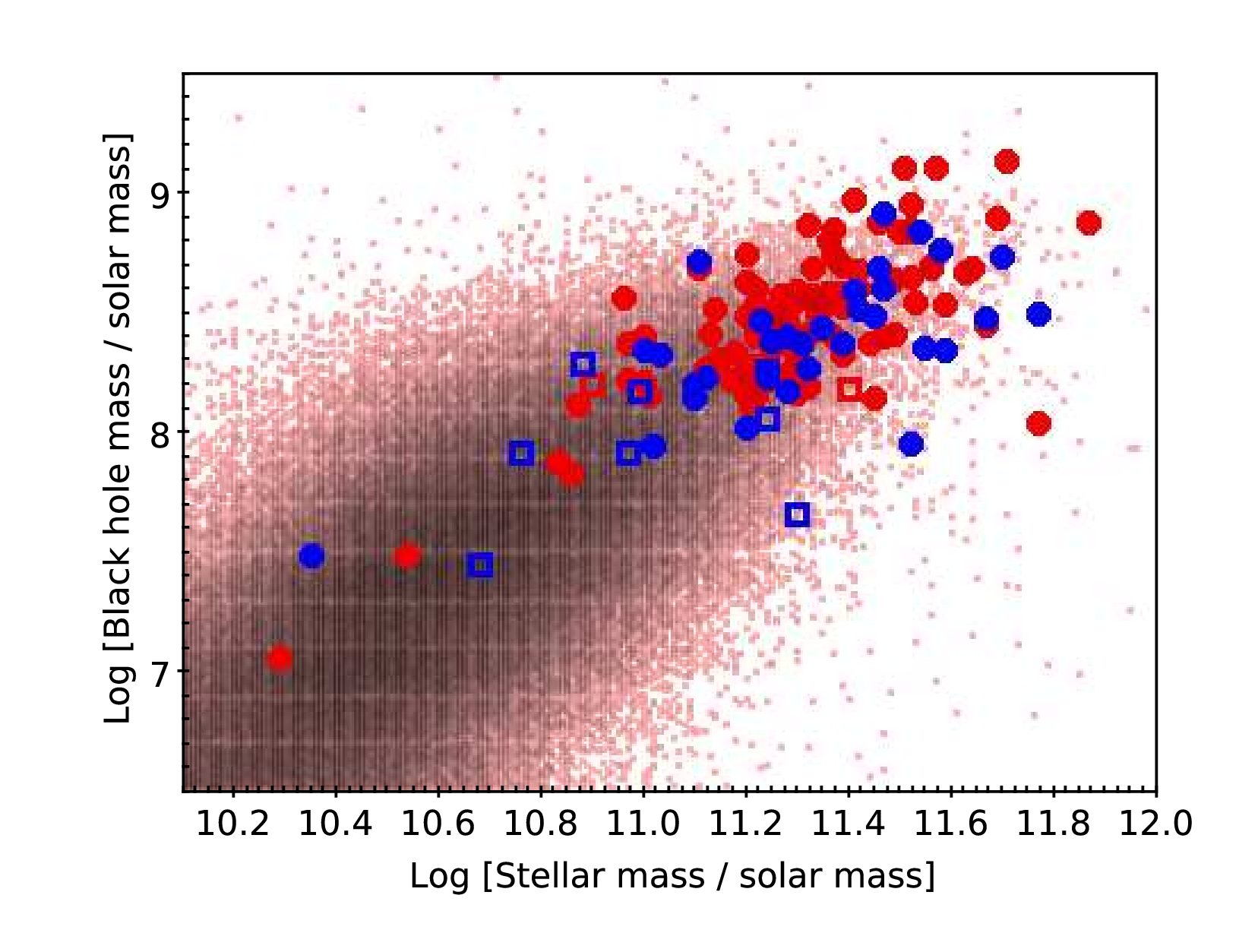,width=3.2in,height=2.6in}
\epsfig{file=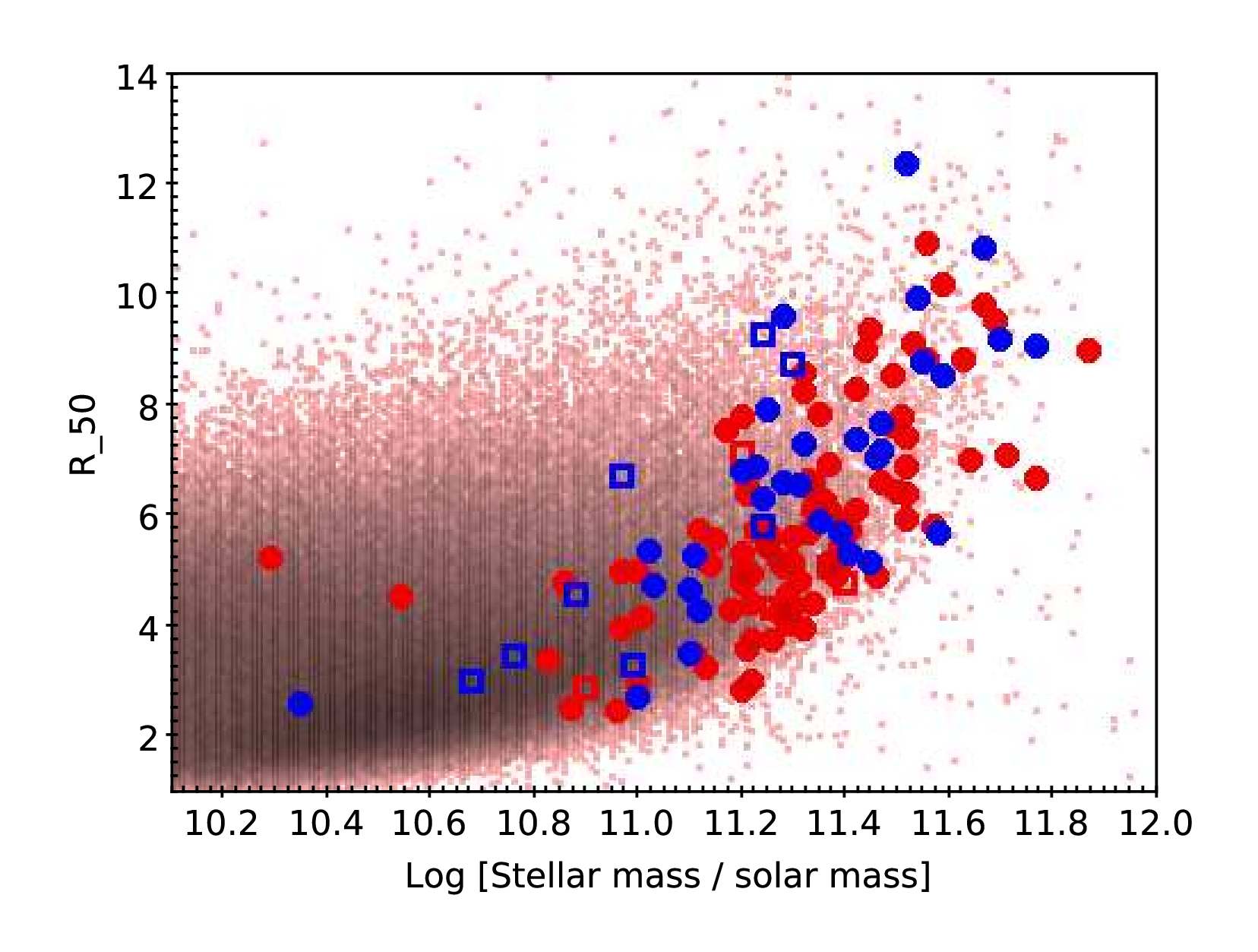,width=3.2in,height=2.6in}
\epsfig{file=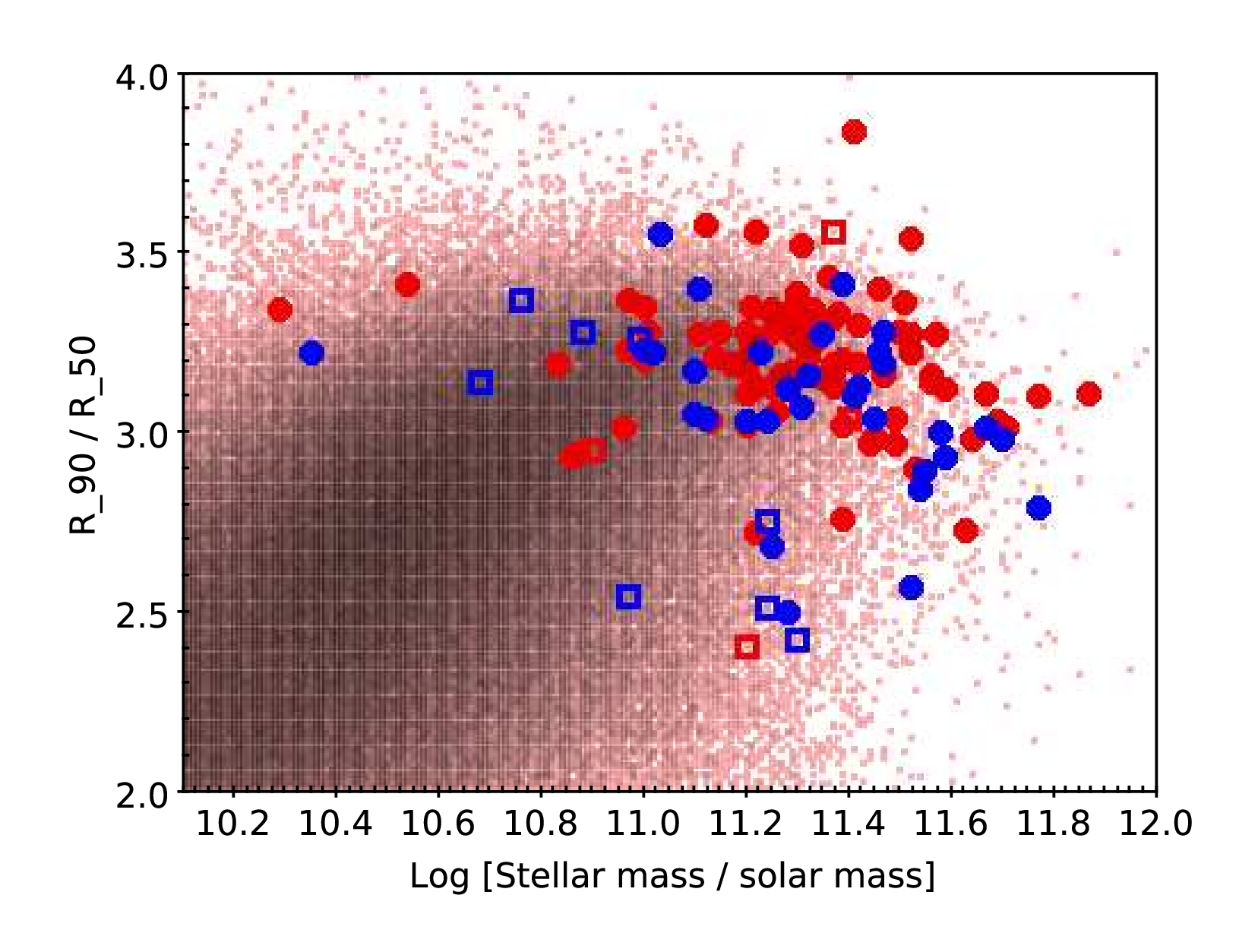,width=3.2in,height=2.6in}
\epsfig{file=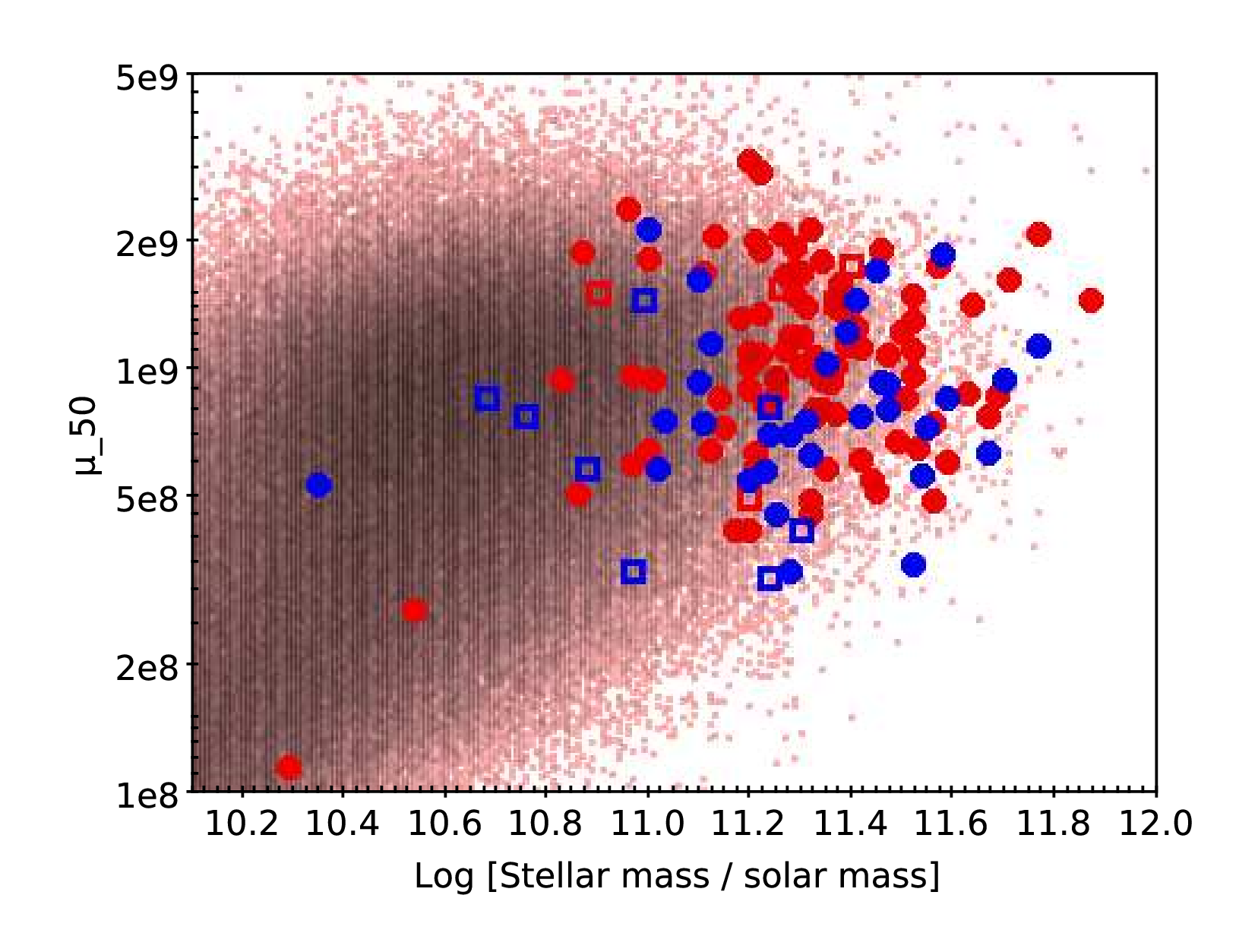,width=3.2in,height=2.6in}
\epsfig{file=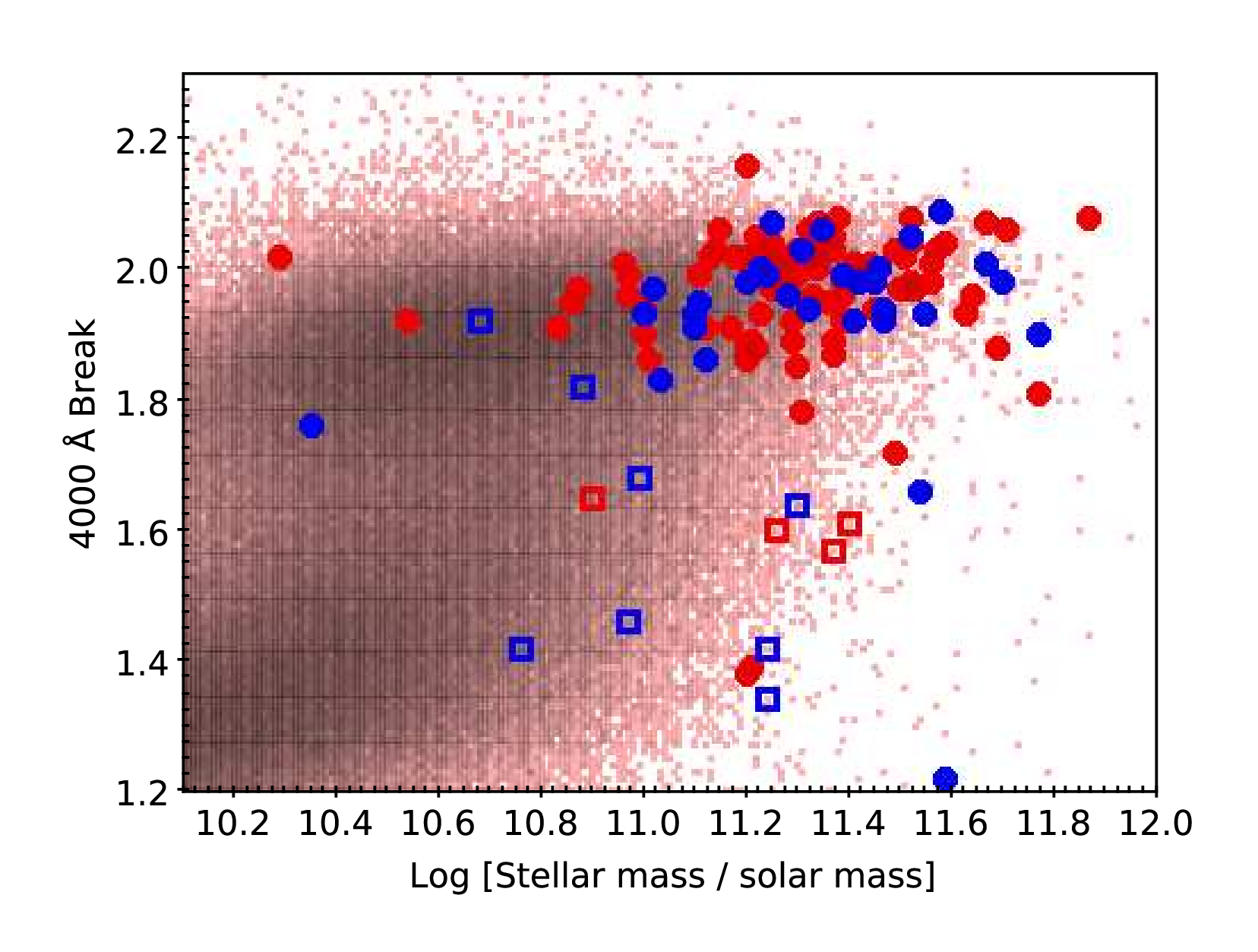,width=3.2in,height=2.6in}
\epsfig{file=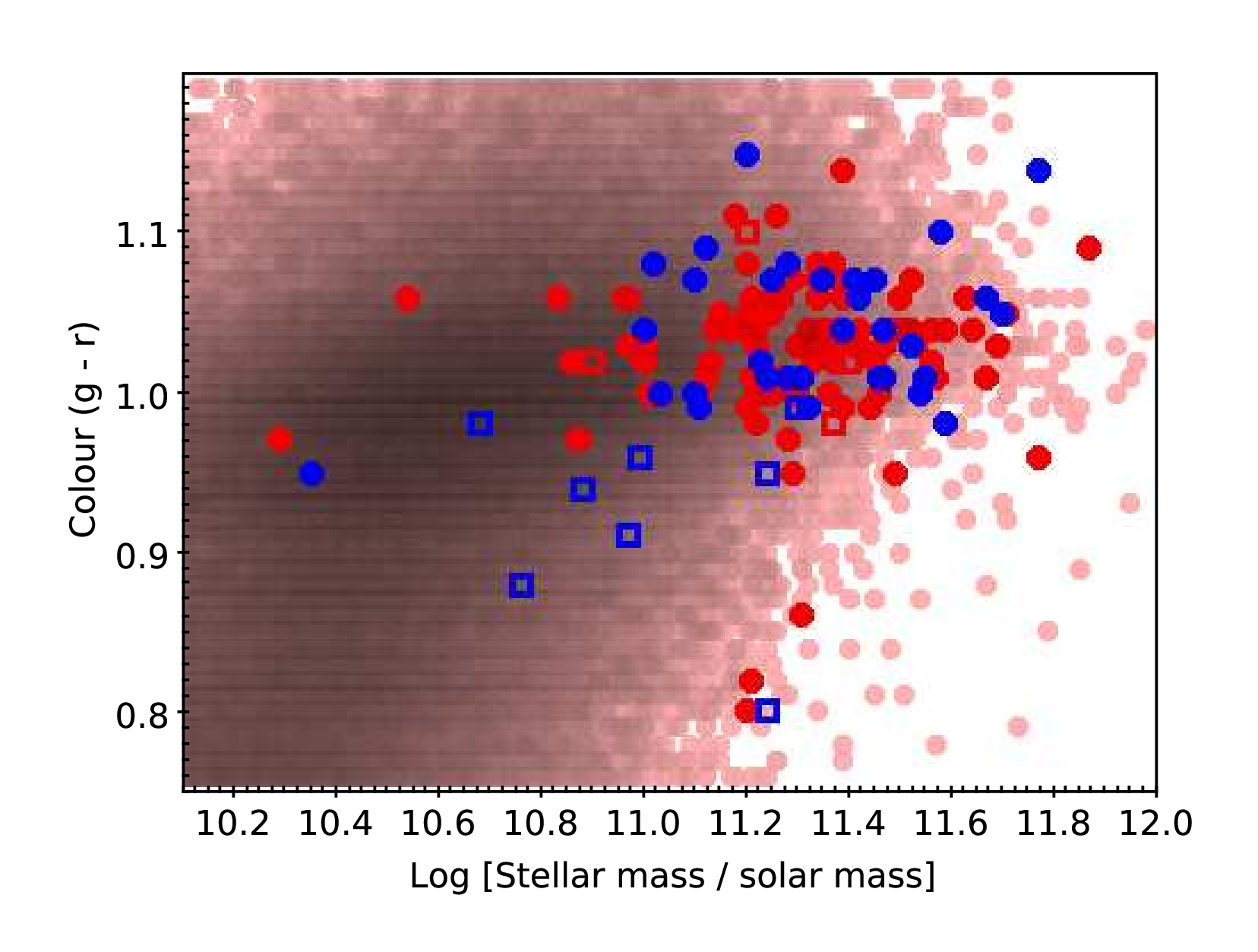,width=3.2in,height=2.6in}

\caption{The host galaxy properties of the FRI (red) and FRII (blue) radio
  galaxies versus the stellar mass. The six plots show the black hole
  mass, galaxy half-light radius (R$_{50}$), concentration index
  (R$_{90}$/R$_{50}$), the stellar mass surface density ($\mu_{50}$), the
  4000\AA\ break strength and the g--r colour. The filled circles
  represent LERGs and open squares are HERGs. Pink circles are all
  galaxies.}
\label{host}
\end{figure*}

\begin{figure*}
\center
\epsfig{file=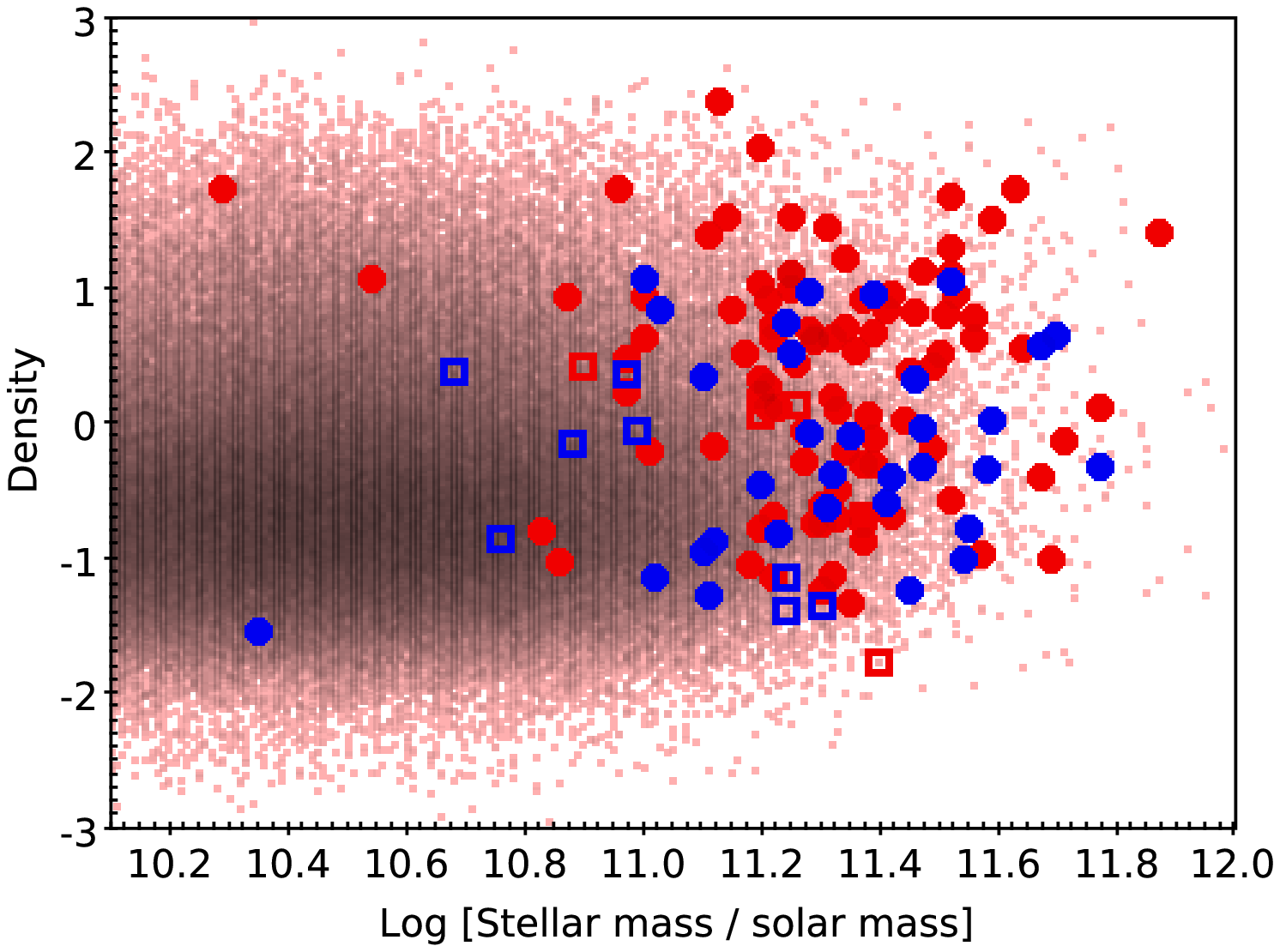,width=3.2in,height=2.6in}
\epsfig{file=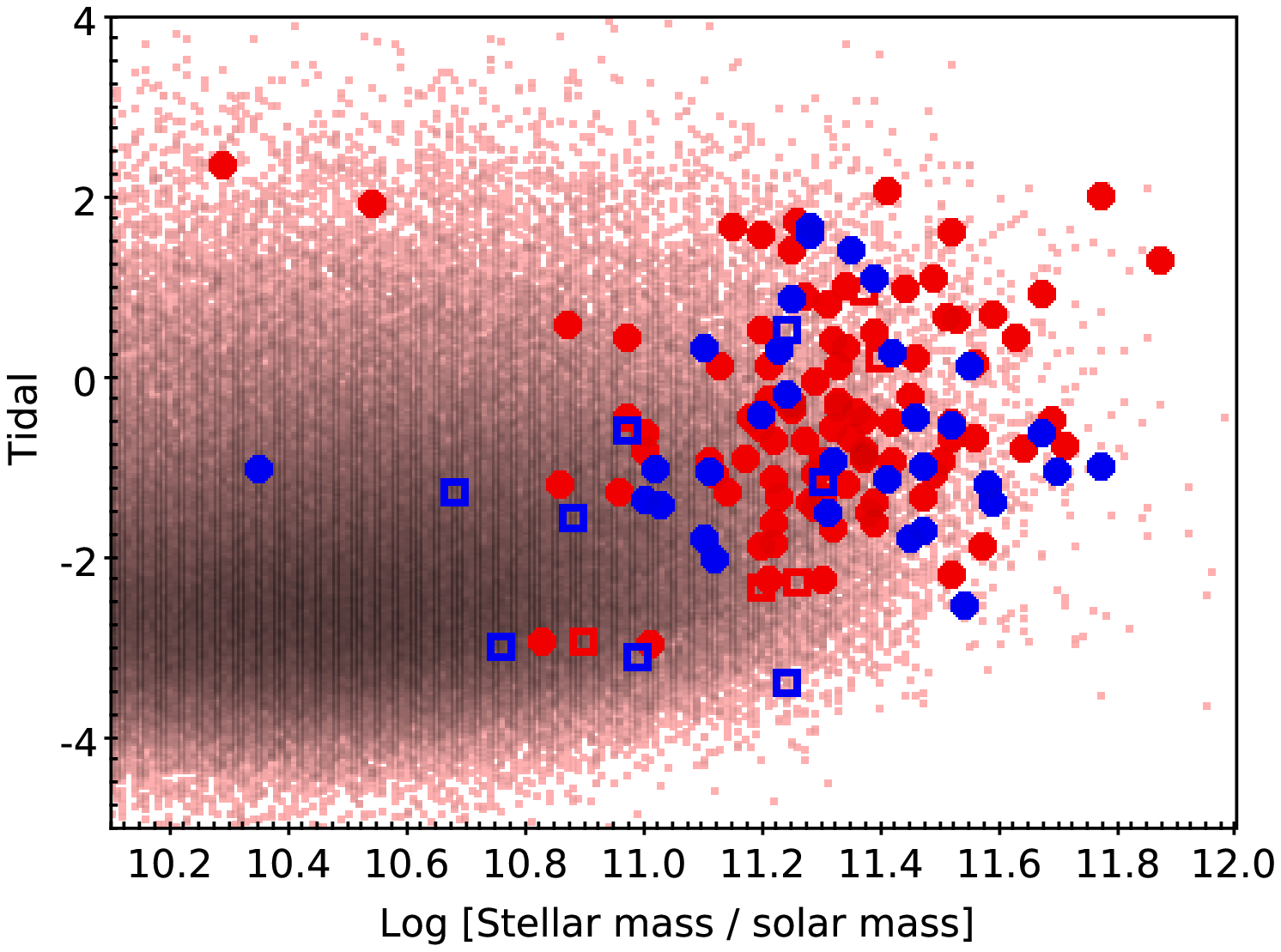,width=3.2in,height=2.6in}
\epsfig{file=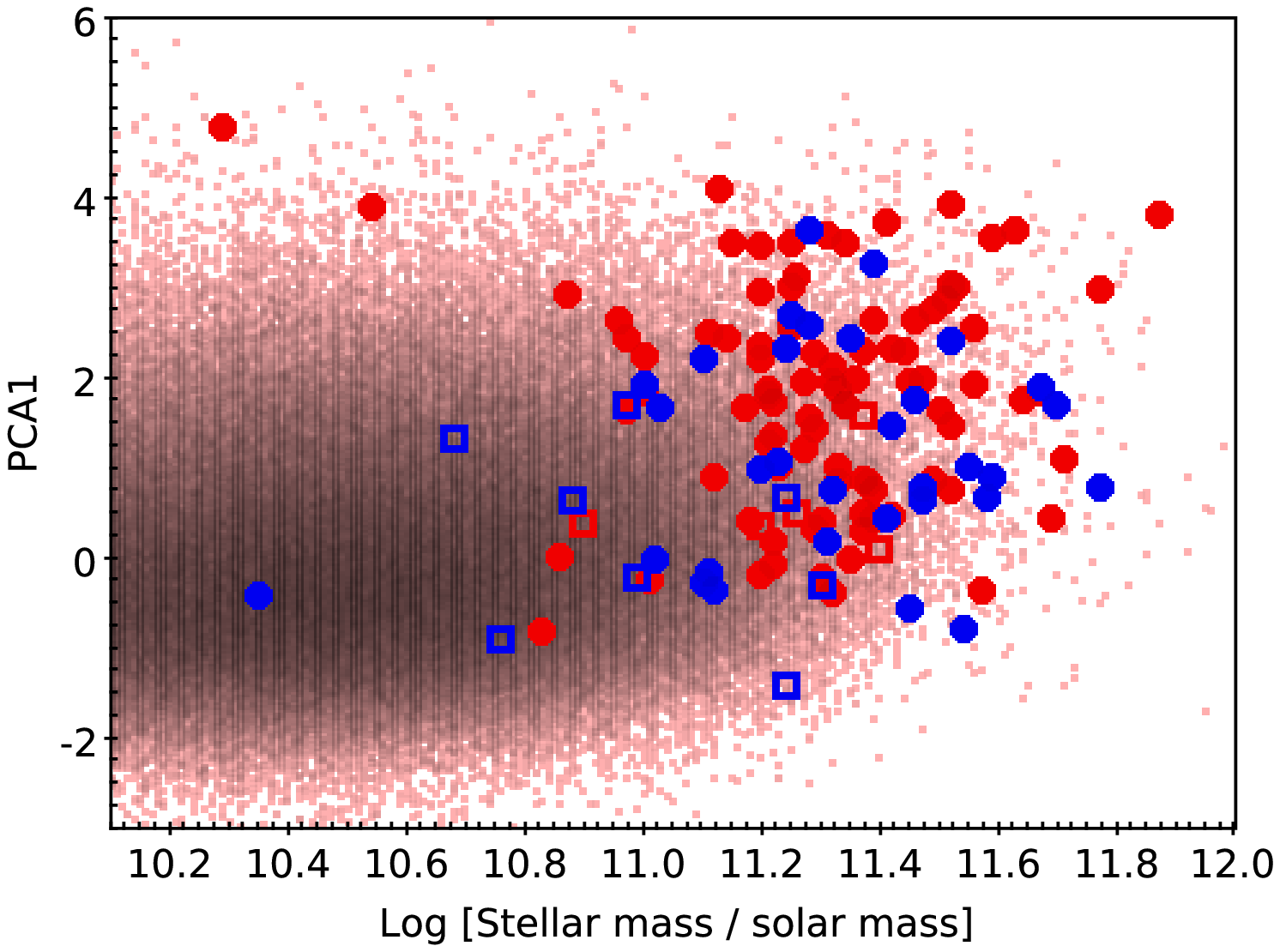,width=3.2in,height=2.6in}
\epsfig{file=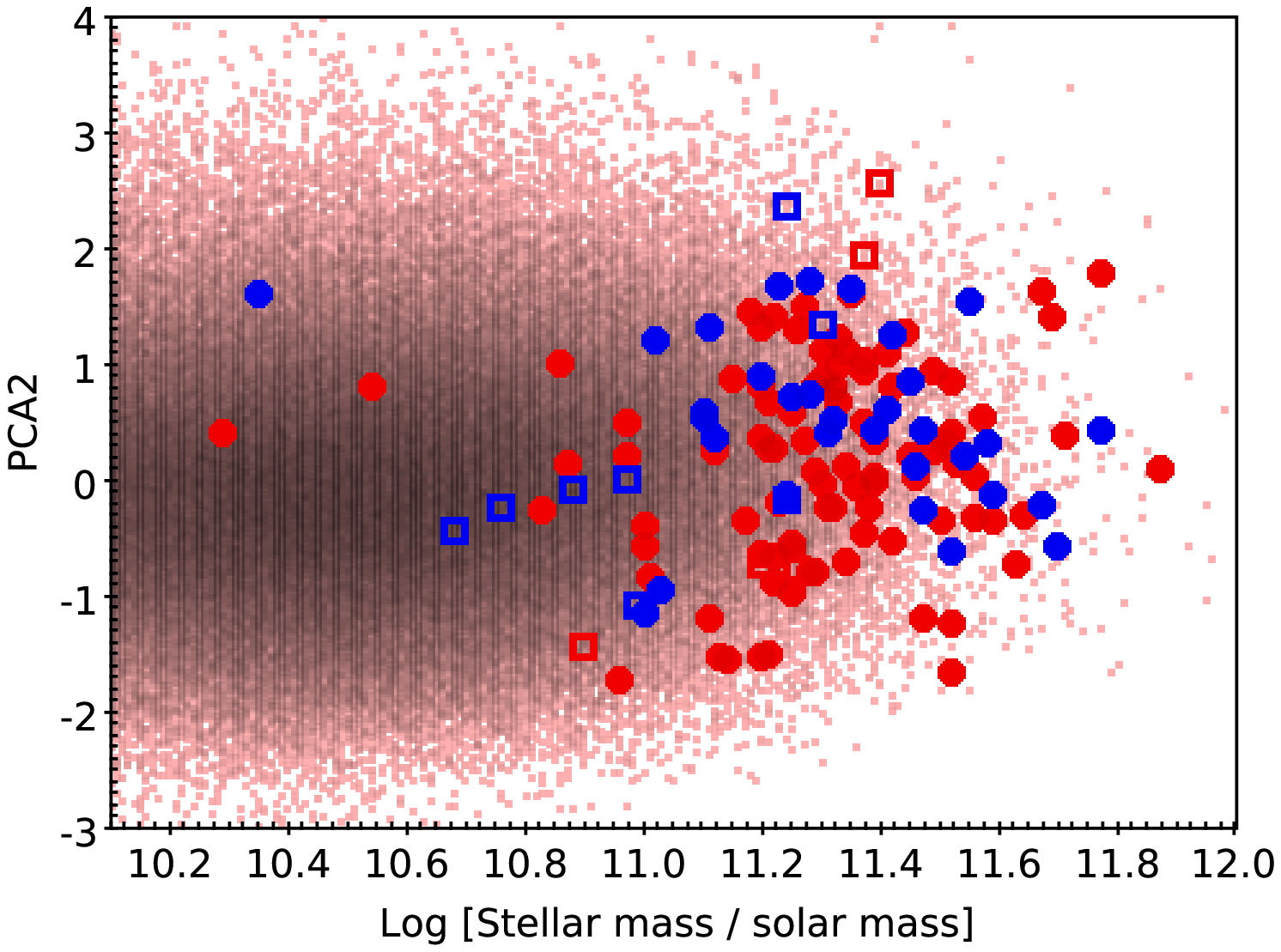,width=3.2in,height=2.6in}

\caption{The environmental properties of the FRI (red) and FRII (blue)
  radio galaxies. The upper panels show the local galaxy density and the
  tidal interaction parameter plotted against stellar mass and the lower panels show two principal component parameters
  derived by Sabater et al (2013), each plotted against stellar mass.
  The filled circles represent LERGs and open squares are
  HERGs. Pink circles are all galaxies.}
\label{envir}
\end{figure*}

\begin{figure*}
\center
\epsfig{file=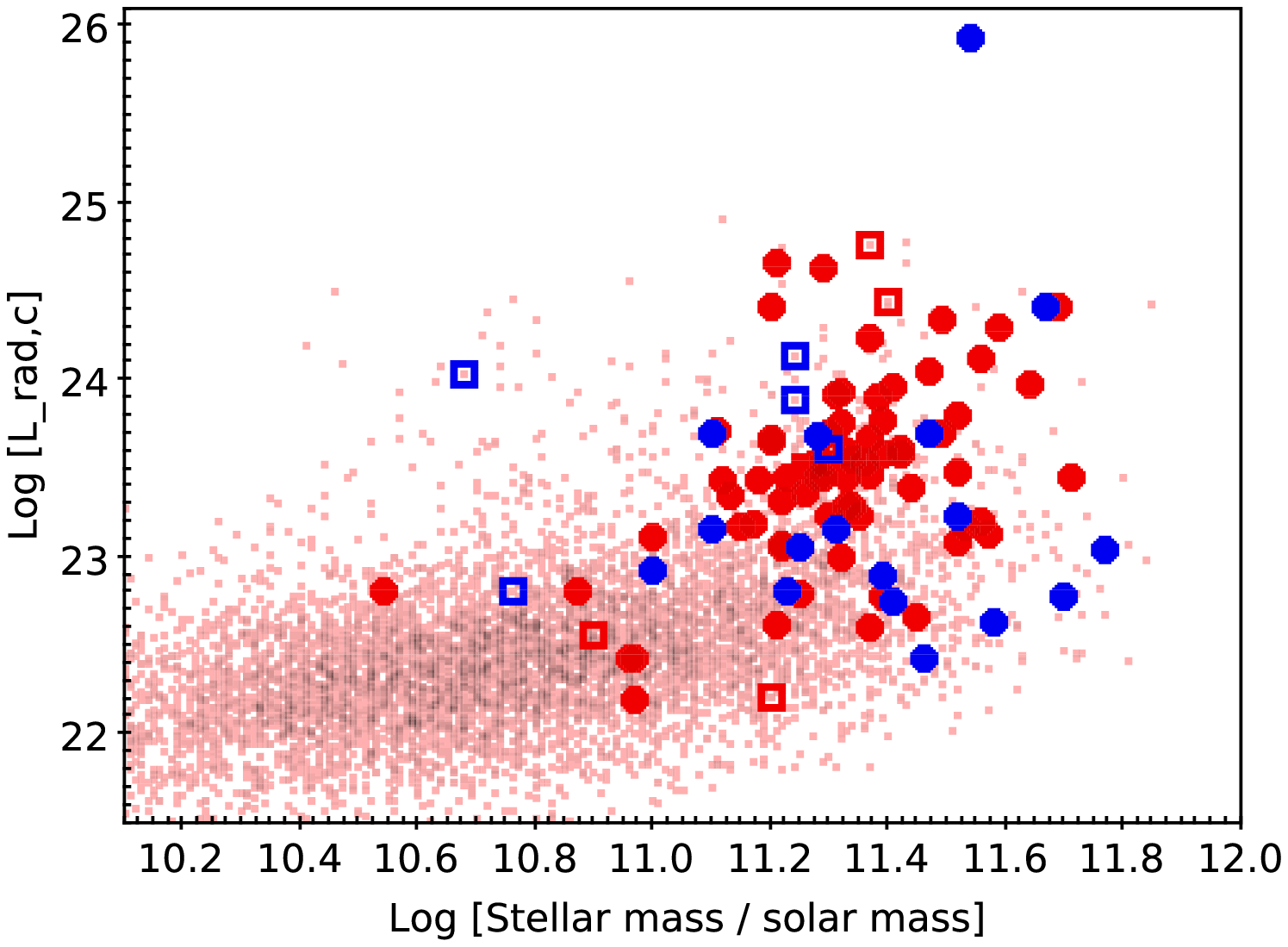,width=3.2in,height=2.6in}
\epsfig{file=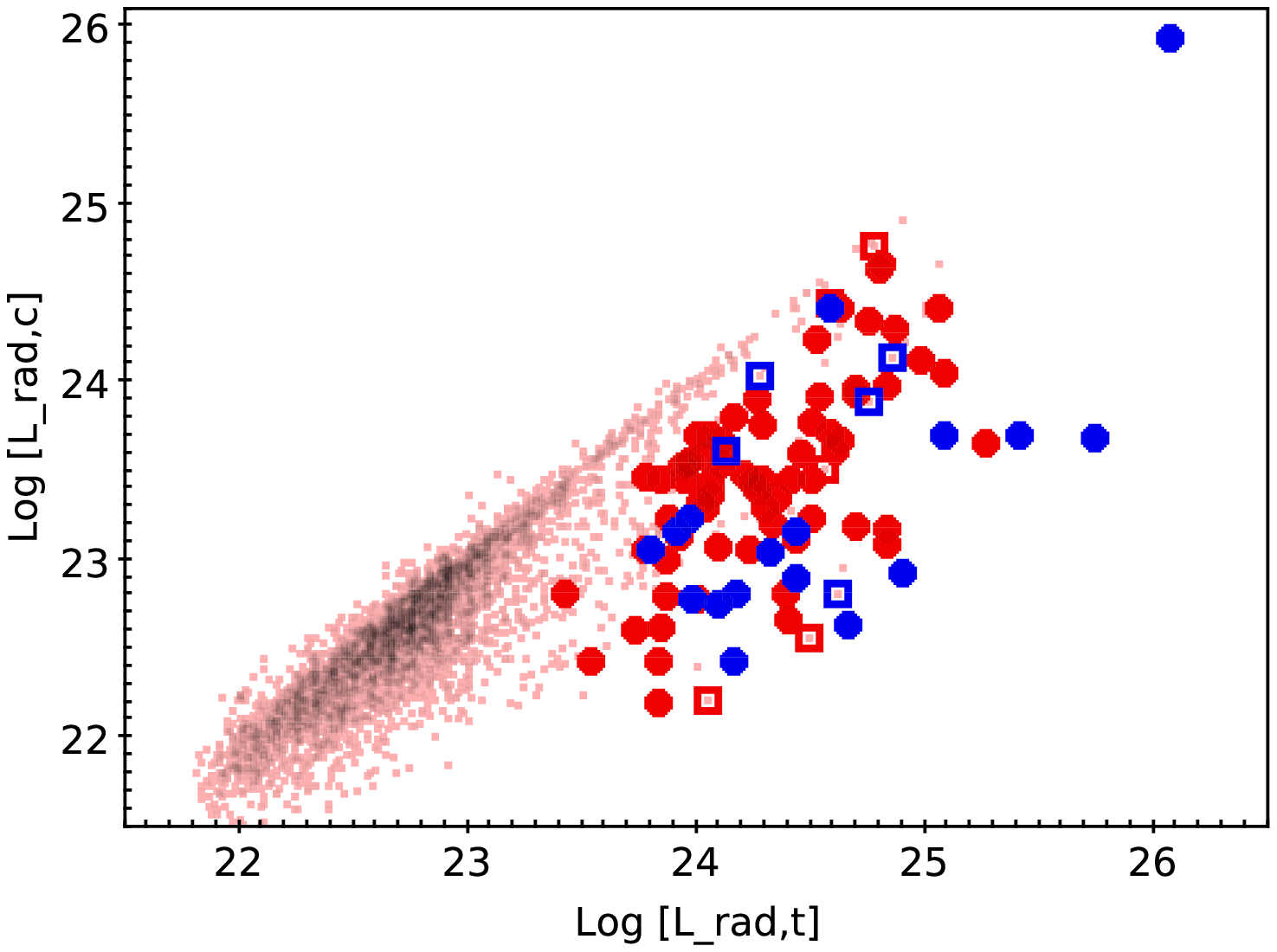,width=3.2in,height=2.6in}

\caption{The core radio luminosity of the FRI (red) and FRII (blue) radio
  galaxies versus stellar mass (left panel) and total radio luminosity
  (right panel). The filled circles represent LERGs and open squares are
  HERGs. Pink circles are all galaxies.}
\label{core}
\end{figure*}

The FRI/II and HERG/LERG dichotomies have been explored extensively in the
literature for both the host galaxy properties and the environment. In
this section, we look at the overall properties of both classifications
and compare our results with the previous studies. The sample selection
has been described in Section 2 and the results are presented in
Figs.~$\ref{RadioOpt}$--~$\ref{core}$ . The left panel of Fig.~$\ref{RadioOpt}$ shows how FR galaxies are distributed
in L$_{rad,t}$--M$_{r}$ plane. Although there is a tendency for the high
radio luminosity objects to be FRIIs, the sharp division line between the
two class of objects that has been previously reported by Ledlow \& Owen
(1996) is not seen; this result was also found by Best (2009). We observe
the same distribution in L$_{rad,t}$--M$_{\star}$ plane (Fig.~$\ref{RadioOpt}$, right
panel). Since there is a tight correlation between stellar mass and
optical absolute magnitude of the galaxies, we use stellar mass and total
radio luminosity as the main parameters in the rest of the paper.

Fig.~$\ref{host}$ displays the host galaxy properties of the radio galaxies, compared
to the underlying galaxy population. The radio galaxy hosts lie at the tip
of the M$_{\star}$--M$_{BH}$ distribution (Fig.~$\ref{host}$, top-left), as expected
since the most massive galaxies are more likely to host a radio-loud AGN
(Best et~al.\ 2005b). They also follow the stellar mass-black hole mass
correlation line. The top-right panel of Fig.~$\ref{host}$ shows the same behaviour
for the R$_{50}$ versus M$_{\star}$ relation: the radio galaxies reside
along the upper mass envelope of the galaxy population but with a
comparable size distribution to underlying galaxies of the same mass. The
concentration (R$_{90}$/R$_{50}$) and half-light surface mass density
($\mu_{50}$) versus stellar mass diagrams (Fig.~$\ref{host}$, middle-left and
middle-right) illustrate a clear tail of radio galaxies away from the main
distribution of the galaxy population. This is also seen in the 4000$\AA$
break distribution, and to a lesser extent the g--r colour, with a scatter
of sources towards bluer colours and younger stellar ages respectively
(Fig.~$\ref{host}$, bottom-right and bottom-left).

Fig.~$\ref{envir}$ compares the environmental parameters of the radio galaxies with
the full galaxy population. It can be seen that the radio galaxies are
typically distributed towards higher density (Fig.~$\ref{envir}$ , top-left), tidal
(Fig.~$\ref{envir}$  top-right) and PCA1 (Fig.~$\ref{envir}$  lower-left) parameters compared to
other galaxies of the same mass. On the other hand, there is no
significant offset between the radio galaxies and the underlying
population in the PCA2 (one-on-one interactions) parameter (Fig.~$\ref{envir}$ 
lower-right). 

Concentrating on the properties of the two FR samples, the FRIs and FRIIs (red {\it vs} blue points)
have broadly similar distributions in mass while FRIIs tend to have lower
black hole masses. In contrast, Wold et~al. (2007) argue that FRI and FRII
have the same black hole mass distribution, but they also argue that for
FRIs with low-excitation spectra the black hole masses correlates with
radio luminosity, so sample selection limits might explain this difference. The
important point is that the black-hole to stellar mass ratio appears lower
for FRIIs than FRIs. No remarkable differences are observed in the
distribution of sizes of the host galaxies (R$_{50}$).  FRIIs tend to have
lower concentration and lower $\mu_{50}$ than FRIs, and a larger
proportion of the FRIIs than FRIs lie within the tails towards lower
colour and lower 4000$\AA$ break. In a similar study, Raimann
et~al.\ (2005) showed that the stellar populations of FR I galaxies are,
on average, older than those of FRIIs. Comparing the environments of FRIs
with FRIIs, on average FRIIs clearly reside in lower density environments
than FRIs, and are affected by slightly lower tidal forces. They also have
lower PCA1 that confirms they are typically in lower density regions. Both
samples show similar PCA2 distributions, indicating that the small
differences in tidal forces might be a projection effect associated with
the denser environments of FRIs. These result are all consistent with the
previous studies that claim FRI radio galaxies are in denser environment
(e.g.\ Prestage and Peacock 1988, Hill and Lilly 1991, Gendre et al 2013).
Finally, FRIs have also brighter cores in radio, which is expected as this
is part of their definition (Fig.~$\ref{core}$, left panel).

As the plots show, many of these differences might also have emerged from
a study of the HERG/LERG dichotomy, as clear differences are also seen
between HERG and LERG objects at those parameters. For instance, HERGs
appear to have higher total radio luminosity, lower black hole mass, bluer
colours, and reside in lower density environments than LERGs (Figs~$\ref{RadioOpt}$-~$\ref{core}$).
The result for the black hole mass has been previously reported by Best
\& Heckman (2012) while both higher (Smith \& Heckman 1989) and lower
(Gendre et~al.\ 2013) galaxy interaction have been reported for the HERG
sources. It is noticeable that in some properties, the HERG/LERG
separation appears to be a stronger driving factor than FRI/II
differences: in particular, it is predominantly the HERG population (both
FRI and FRII) which have weaker 4000$\AA$ breaks and bluer colours than
typical galaxies of their stellar mass. Therefore, a lot of observed
differences between FRIs and FRIIs may be caused by the HERG/LERG nature of the FR sources, and this
issue has caused lots of misunderstanding and confusion in the study of FR
radio galaxies when HERG/LERG classification is not taken into account.

In order to obtain a clean picture of FRI/FRII differences and understand
their causes to explain the morphological dichotomy observed at radio
galaxies, we need to remove possible HERG/LERG biases. We also need to remove biases with
the host galaxy mass and radio luminosity, since Figs ~$\ref{RadioOpt}$-~$\ref{core}$ make clear
that many parameters correlate strongly with these properties. The method
we adopt for that in the next section is to construct populations of FRI
vs FRII, HERG vs LERG and Compact vs Extended sources, having the same
distribution of stellar mass, total radio luminosity or core radio
luminosity, redshift and excitation class. We only confine the matching
criteria to these parameters, in order to keep the sample of each type
large enough to achieve robust statistics.

\section{Matched samples}

For each of FRI/FRII, HERG/LERG and Compact/Extended sources, we construct
matched sample of objects in the L$_{\rm radio}$--M$_{\star}$ plane by randomly
selecting pairs within a certain tolerance (detailed below) in L$_{\rm radio}$
and M$_{\star}$ (two dimensional (2D) matching). Hence, we remove all of
the mass-dependent and luminosity-dependent effects seen in the set of
plots discussed in Section 3. We then consider the normalized cumulative
histogram of each physical parameter for each class of objects, and apply
the Kolmogorov-Smirnov (KS) test to assess whether the matched samples are
consistent with being drawn from the same parent population. The KS test
calculates the significance of the maximum difference (D) between two
distributions, and assigns a probability (P) according to the parameter D
and the number of objects in the samples. We consider differences with a
probability above 95 percent to be significant. Finally, we repeat the analysis
up to 1000 times (with different random selections for the source pairing)
and calculate the average D and then the significance from that.

We also constructed three dimensional (3D) matched samples by adding
redshift (z) to L$_{\rm radio}$ and M$_{\star}$, and repeat the previous steps
using a matched sample in L$_{radio}$--M$_{\star}$--z space. In this way,
we remove any effect of cosmic evolution (expected to be small) and more importantly 
any potential redshift biases in parameter
estimation, in addition to the mass and luminosity. This results in
smaller samples, due to the more restrictive matching requirements. The
3D-matching results are in good agreement with the 2D-matching results
(but with larger uncertainties), so we only report 2D results in this
section.

\begin{table*}
\begin{tabular}{l*{8}{r}}  
\hline
\multicolumn{1}{l}{ Sample} & \multicolumn{2}{c}{FRII-FRI} & \multicolumn{2}{c}{HERG-LERG}  & \multicolumn{4}{c}{Compact-Extended}\\ 
\multicolumn{1}{c}{ } & \multicolumn{2}{c}{} & \multicolumn{2}{c}{}  & \multicolumn{4}{c}{}\\ 
\multicolumn{1}{l}{Matched properties}  & \multicolumn{2}{c}{L$_{rad,t}$-M$_{\star}$} & \multicolumn{2}{c}{L$_{rad,t}$-M$_{\star}$} & \multicolumn{2}{c}{L$_{rad,t}$-M$_{\star}$} & \multicolumn{2}{c}{L$_{rad,c}$-M$_{\star}$}\\
\multicolumn{1}{l}{Sample size}  & \multicolumn{2}{c}{M=N=77} & \multicolumn{2}{c}{M=15, N=45} & \multicolumn{2}{c}{M=N=81} & \multicolumn{2}{c}{M=N=58}\\
\multicolumn{1}{l}{Significance thresholds}  & \multicolumn{2}{c}{D$_{95}$=0.22, D$_{99}$=0.26} & \multicolumn{2}{c}{D$_{95}$=0.40, D$_{99}$=0.48} & \multicolumn{2}{c}{D$_{95}$=0.21, D$_{99}$=0.26} & \multicolumn{2}{c}{D$_{95}$=0.25, D$_{99}$=0.30}\\
\hline 
L$_{rad,c}$   & 0.58 & $>$99$\%$ & -0.36 & - & -0.68 & $>$99$\%$&- & -\\
L$_{rad,t}$  & - & - & - & - & - & -&0.57 & $>$99$\%$ \\
R$_{50}$ & -0.35 & $>$99$\%$ & -0.10 & - & 0.17 & -&-0.05 & - \\
 g-r& 0.21 & - & 0.36 & - & 0.18 & -&-0.07 & - \\
4000$\AA$ break & 0.20 & - & 0.74 & $>$99$\%$ & 0.33 & $>$99$\%$&0.06 & - \\
 R$_{90}$/R$_{50}$& 0.32 & $>$99$\%$ & 0.45 & $>$95$\%$ & 0.17 & -&0.20 & - \\
$\mu_{50}$ & 0.38 & $>$99$\%$ & -0.30 & - & -0.20 & -&0.22 & - \\
M$_{BH}$ & 0.35 & $>$99$\%$ & 0.47 & $>$95$\%$ & 0.16 & -&0.35 & $>$99$\%$ \\
Density  & 0.36 & $>$99$\%$ & 0.43 & $>$95$\%$ & 0.12 & -&0.09 & -\\
Tidal  & 0.15 & - & 0.42 & $>$95$\%$ & 0.18 & -&0.14 & -\\
Richness  & 0.28 & $>$99$\%$ & 0.24 & - & 0.13 & -&0.14 & -\\
PCA1  & 0.30 & $>$99$\%$ & 0.53 & $>$99$\%$ & 0.13 & -&0.09 & -\\
PCA2 & -0.31 & $>$99$\%$ & -0.19 & - & 0.19 & -&0.07 & - \\
 L$_{[OIII]}$& 0.38 & $>$99$\%$ & -0.90 & $>$99$\%$& -0.30 & $>$99$\%$ &-0.13 & -  \\

\hline
\end{tabular}
\caption{The result of KS test for three sets of comparisons: i) FRI and
  FRII radio galaxies ii) HERGs and LERG sources iii) compact and extended
  source. They have been cross-matched in luminosity-mass plane where
  L$_{rad,t}$ represents total radio luminosity and L$_{rad,c}$ represents
  core radio luminosity. The first column in each set shows the KS
  differences D, and the second column shows confidence level for the
  estimated differences. We only indicate the significance above 95
  percent. Positive values show that the first mentioned sample in each
  set (Compact, FRII and HERG) has lower value for the declared
  characteristic and negative sign means that the first mentioned sample
  has higher value. For example, FRIIs have higher R$_{50}$ compared to
  FRIs with $>$ 99 percent confidence and HERGs have lower black hole mass
  with $>$ 95 percent confidence. The typical uncertainty of the D values (the
  standard deviation out of 1000 iterations) is 0.01-0.03 which we have
  considered to report the probabilities. M and N are the sizes
  of the first and the second mentioned sample in each set which 
  we have used to calculate significance thresholds (D). D$_{95}$ and D$_{99}$ are the level
  of D needed for 95$\%$ and 99$\%$ significance respectively.}
\label{table3}
\end{table*}

\begin{figure*}
\center
\epsfig{file=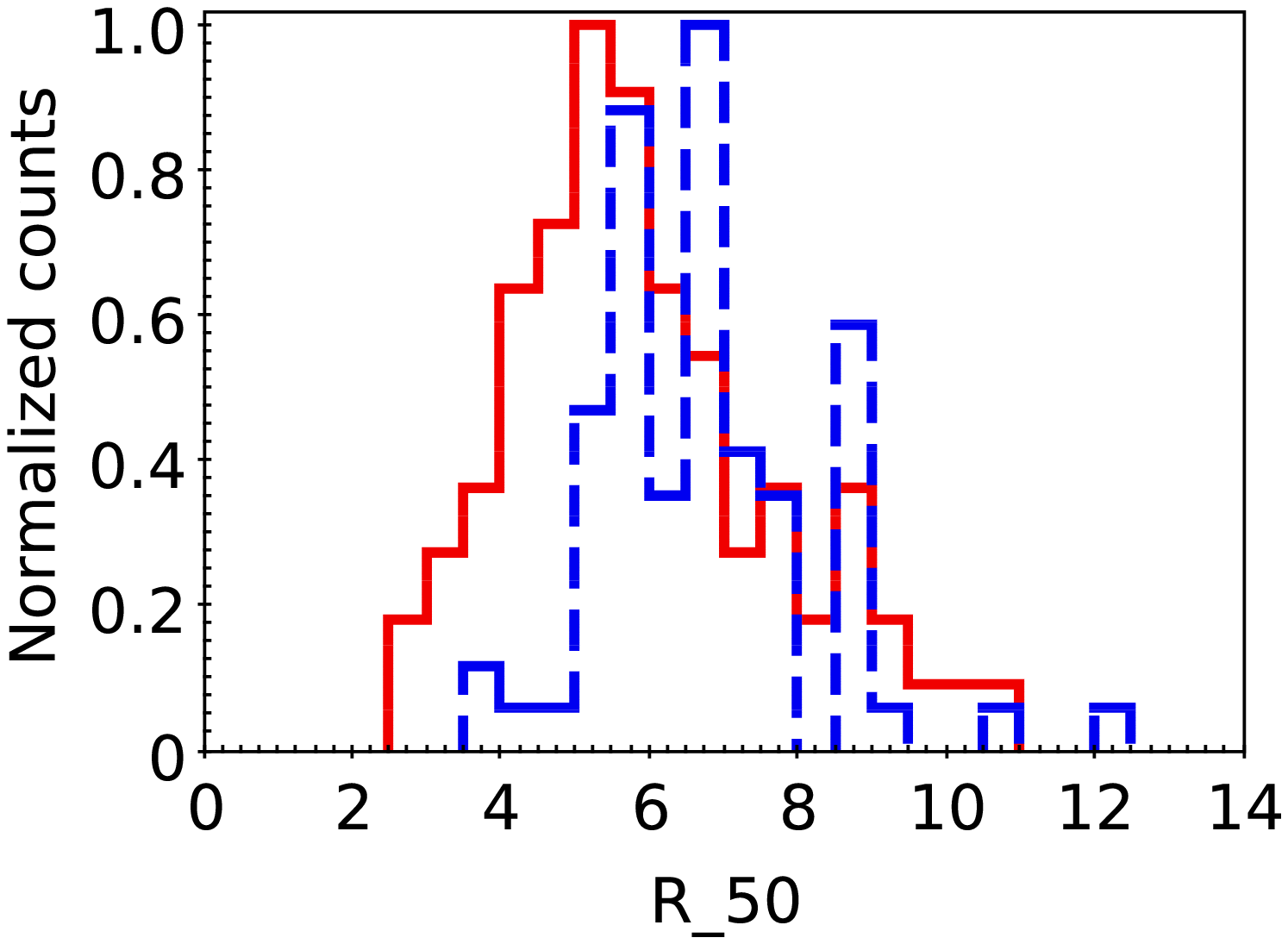,width=1.6in,height=1.3in}
\epsfig{file=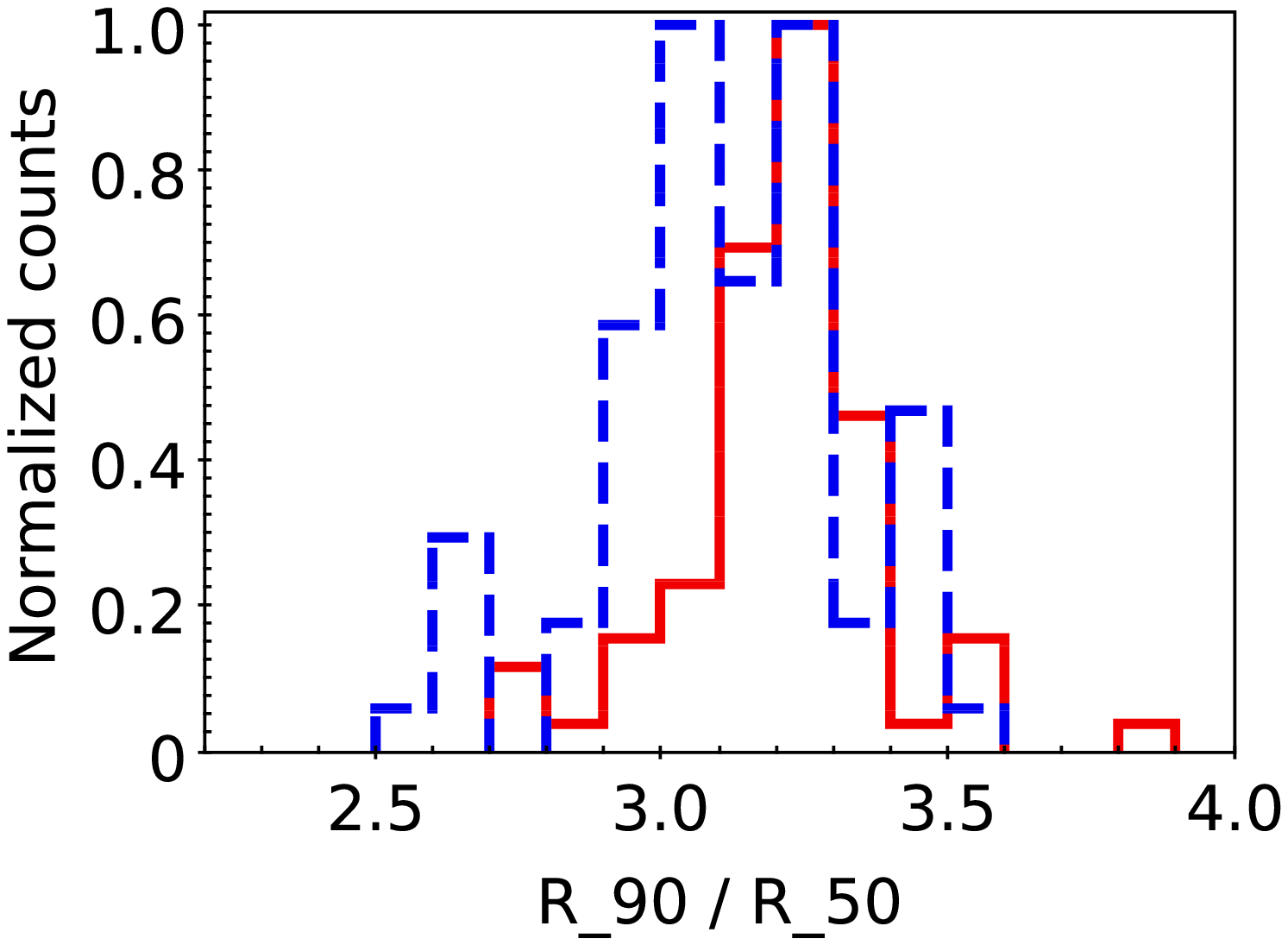,width=1.6in,height=1.3in}
\epsfig{file=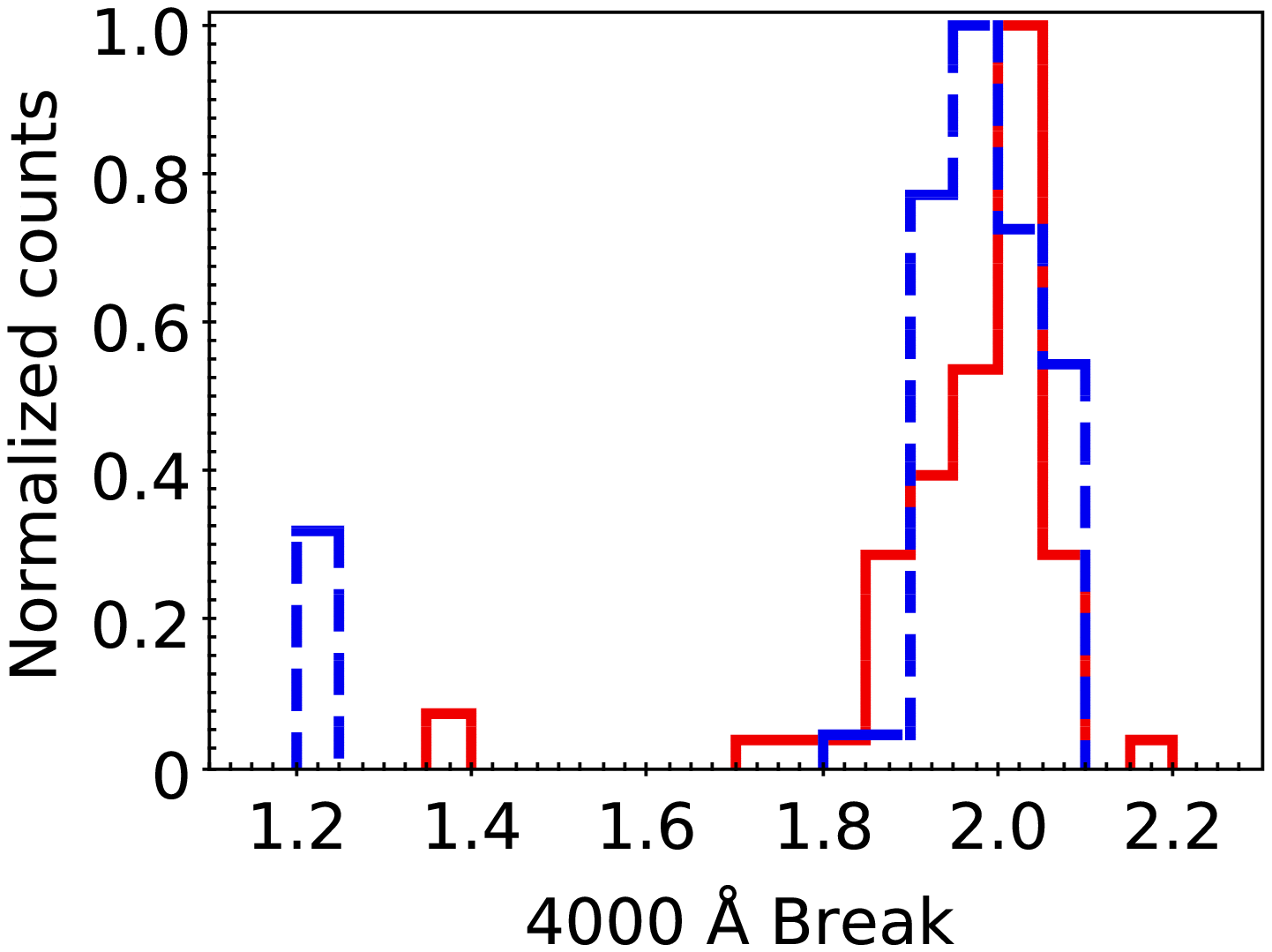,width=1.6in,height=1.3in}
\epsfig{file=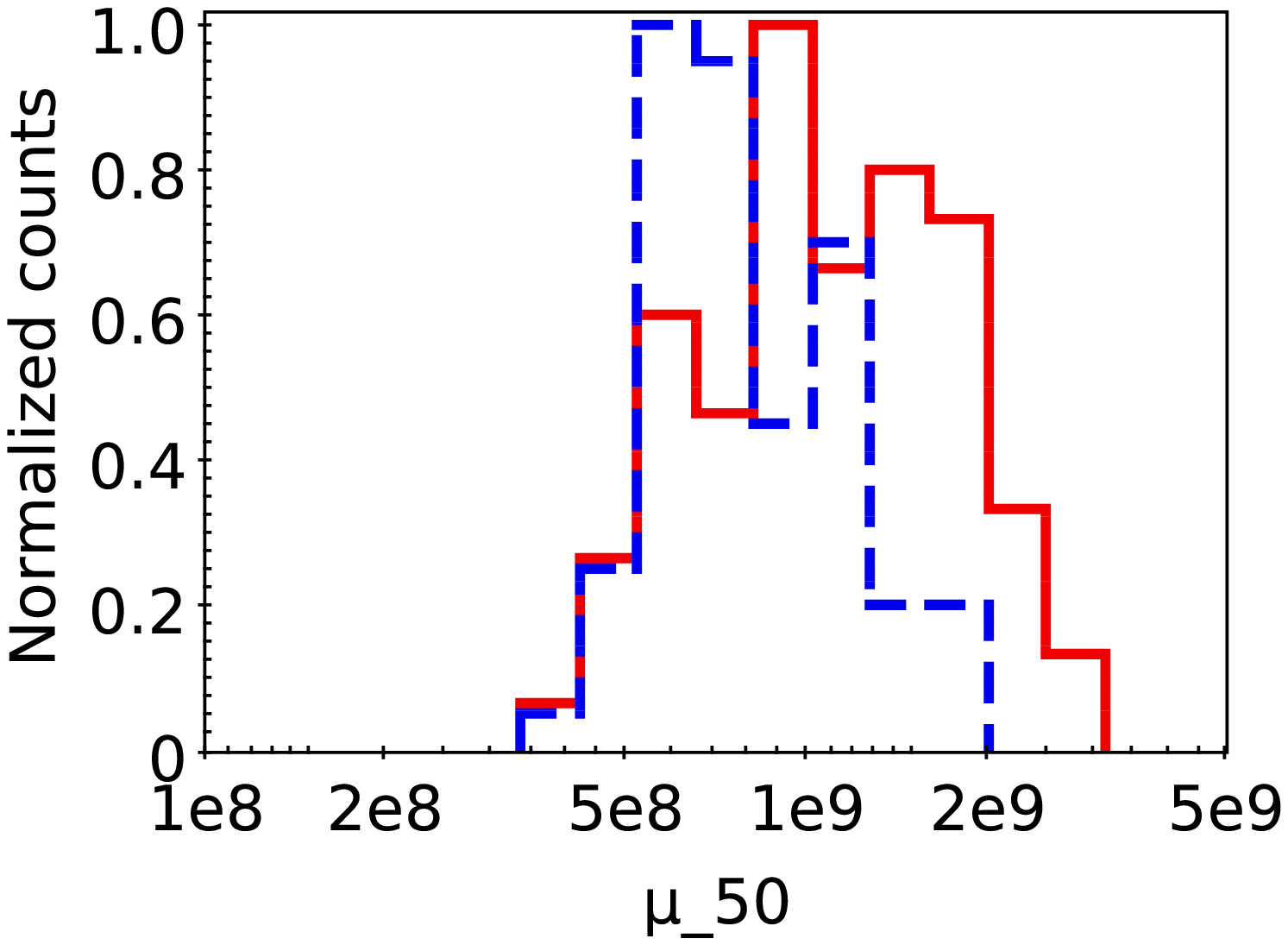,width=1.6in,height=1.3in}
\epsfig{file=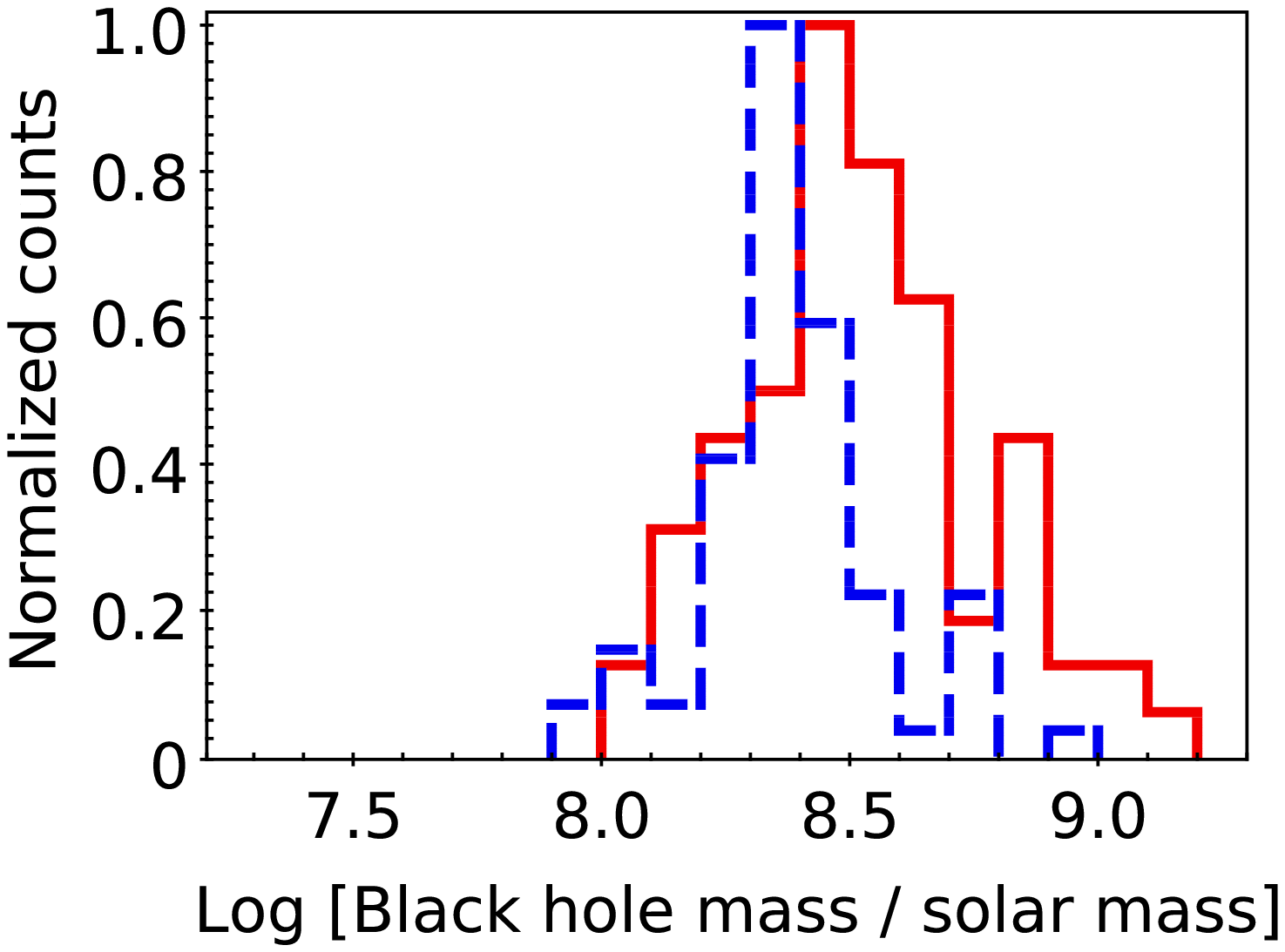,width=1.6in,height=1.3in}
\epsfig{file=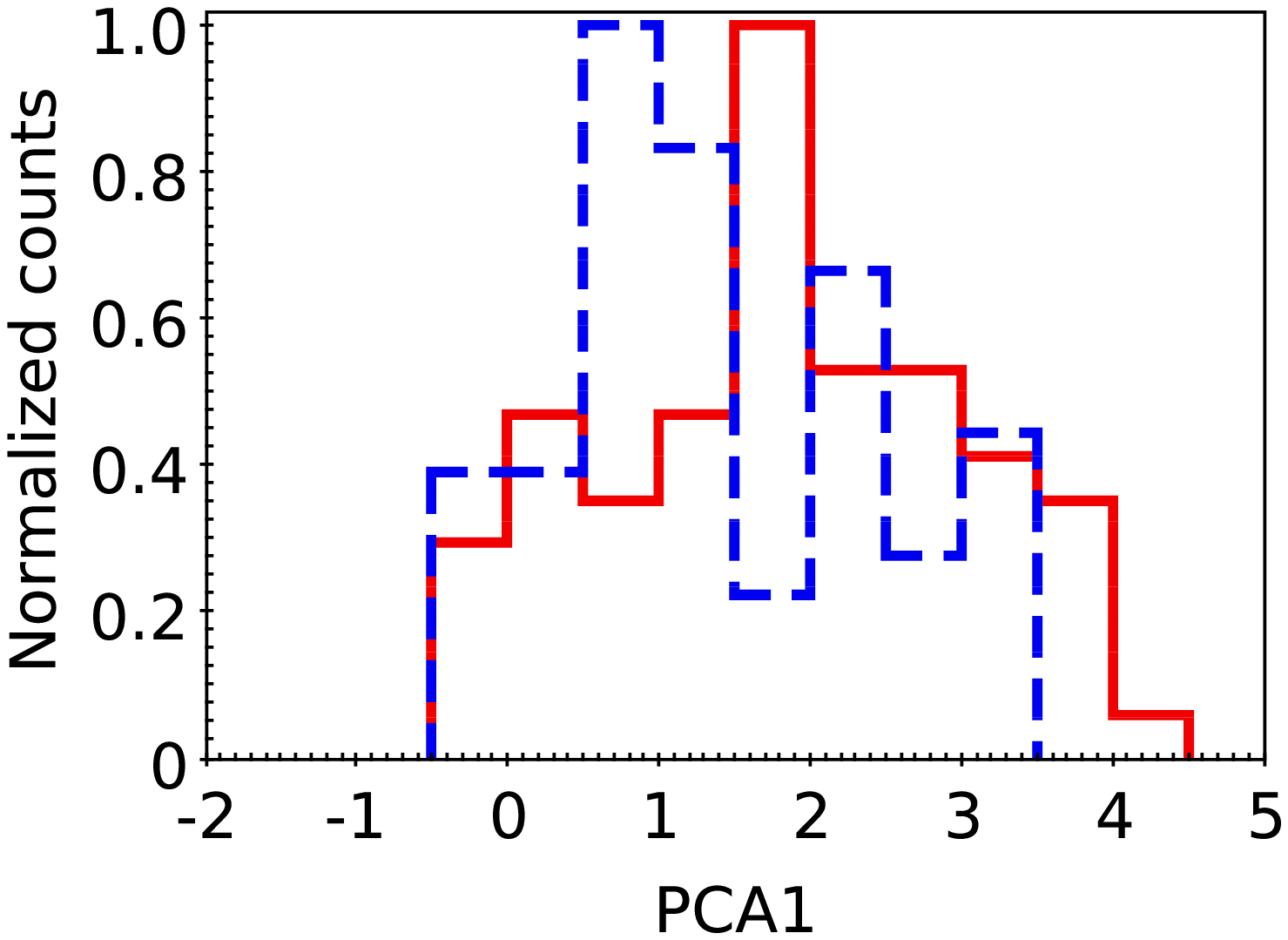,width=1.6in,height=1.3in}
\epsfig{file=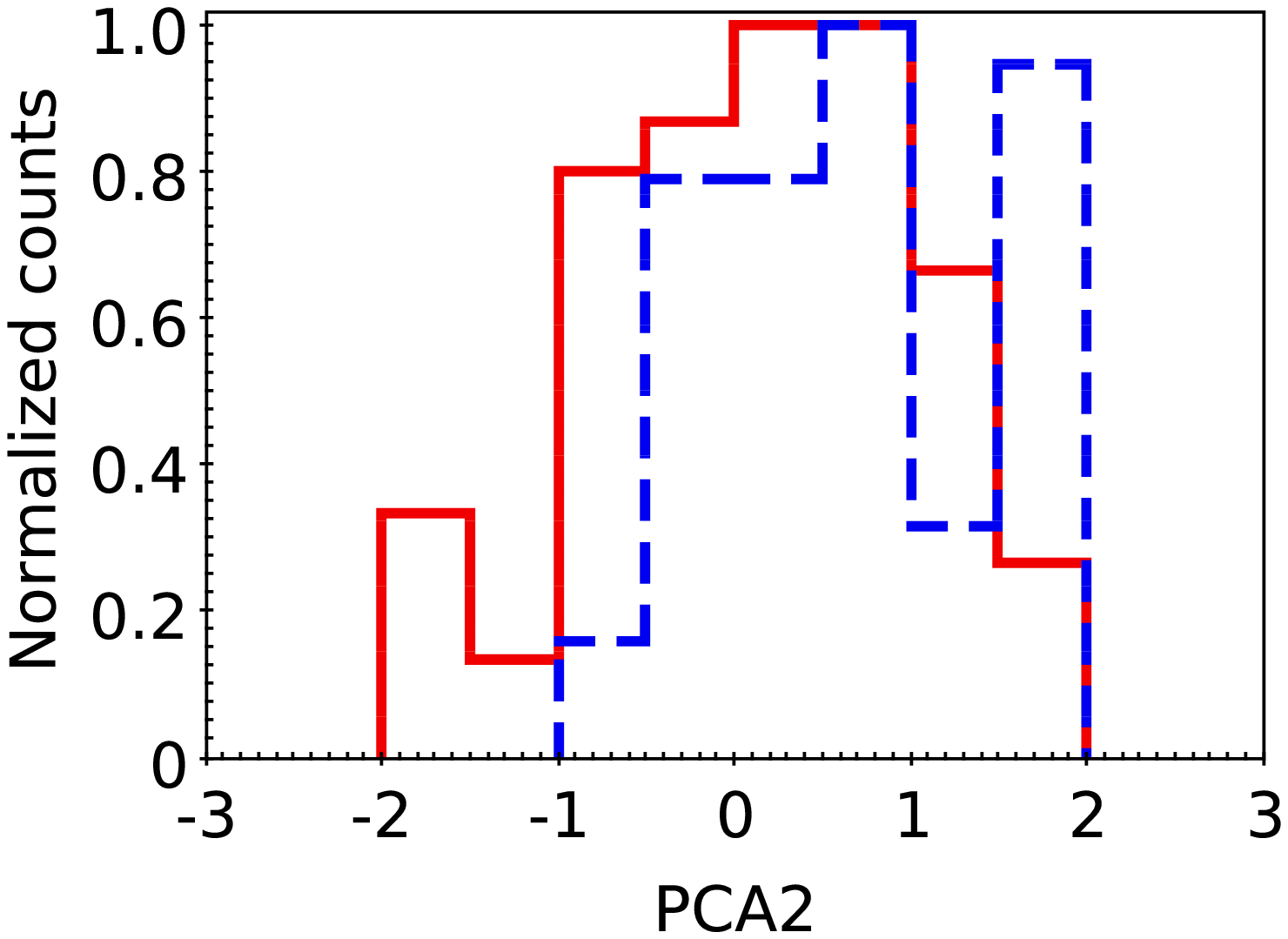,width=1.6in,height=1.3in}
\epsfig{file=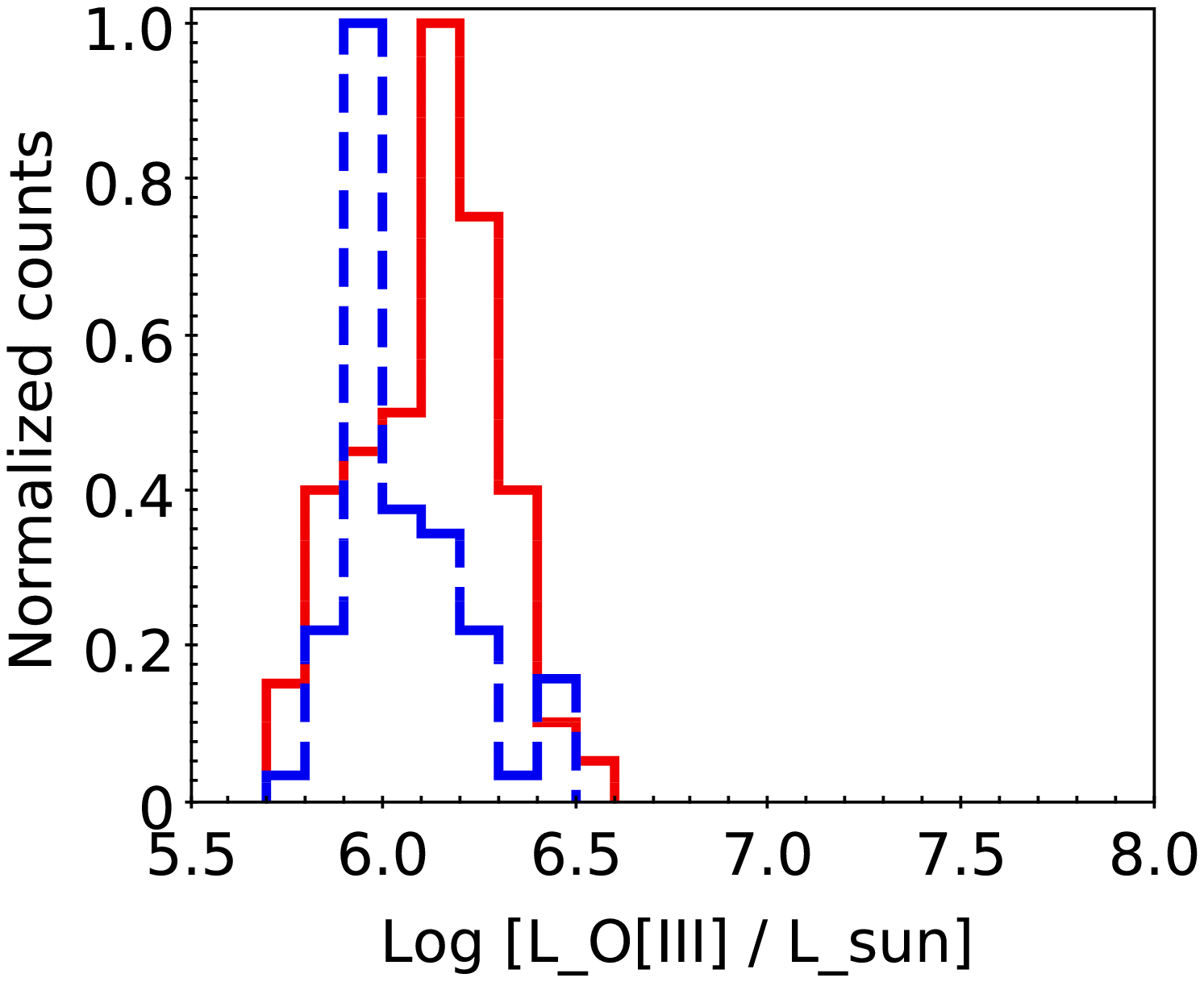,width=1.6in,height=1.3in}
\caption{Histograms of the host galaxy and environmental parameter
  distributions for FRI (red) and FRII (blue) radio galaxies matched
  in LERG classification and in the L$_{rad,t}$--M$_{\star}$ plane.  }
\label{histFR}
\end{figure*}

\subsection{FRI vs. FRII}

Here, we confine the analysis to FRI LERGs versus FRII LERGs. We want to
remove all the effects caused by HERG/LERG nature of FR radio galaxies,
and the LERG sample is numerous enough for both FR types to be well
represented, while the HERG sample size is small (especially the FRI
HERGs). As seen in the right-hand panel of  Fig.~$\ref{RadioOpt}$, in a plane of
L$_{rad,t}$--M$_{\star}$, there is a slight segregation between FRIs and
FRIIs, showing FRIIs have higher total radio luminosity on average 
 (the median values of log [L$_{rad,t}$] for the FRIIs and FRIs are 
24.44 and 24.29 W Hz$^{-1}$ respectively). The
matched sample has been constructed by finding all pairs within the error
in radio luminosity $\Delta$ log[L] = $\pm$ 0.2 and in mass $\Delta$
log[M] = $\pm$ 0.1, and then choosing randomly unique pairs of
FRI--FRII. The result of KS tests to investigate the significance of
differences in the distribution of host galaxy and environment parameters
between the two populations, averaged out of 1000 iterations, are
presented in Table~$\ref{table3}$. The histograms for each of the
parameters are presented in Fig.~$\ref{histFR}$.

There are a lot of differences with 99 percent significance presented in
Table~$\ref{table3}$. FRIs have higher core radio luminosity than FRIIs
with the same total radio luminosity, which emerges trivially from the
definition of FRIs as being core dominated and edge darkened. Concerning
the host galaxy properties, FRIs reside in smaller galaxies (lower
R$_{50}$) with higher concentration, higher mass surface density and
higher M$_{bh}$/M$_{\star}$ (higher M$_{bh}$ for the matched sample of
mass), all of which imply less disk-like structure for the host galaxy.
Concerning the environmental parameters, FRIs seem to lie in richer local
environment: the density, PCA1 and richness are all higher for them than
for FRIIs of the same mass, radio luminosity and excitation class, at high
significance. The difference in tidal interaction is not significant. PCA2
shows the opposite behaviour, being higher in FRIIs.

These results suggest that extrinsic parameters can be the main
driver of the morphological dichotomy. There are several indications for
that. The first one is that FRIs have more concentrated host galaxies
(higher R$_{90}$/R$_{50}$) with higher surface mass density ($\mu_{50}$),
indicating a greater density of material available to disrupt the radio
jets. The
second indication is that FRIs appear to reside in a denser galaxy
environment, since all the environmental parameters tracing this seem to
be higher for the FRIs compared to FRIIs. The only exception for that is
the PCA2 parameter which is higher for FRIIs; this might show that FRIIs
suffer a higher level of one-on-one interactions and are more likely to be
merger/interaction triggered than FRIs (Miraghaei et~al.\ 2014, 2015).

The picture that we can make from these results is that radio jets in
denser galaxies, and in denser environments like galaxy clusters and
groups, are much more susceptible to being disrupted and becoming FRI
(cf. Kaiser \& Best 2007). These are also the environment in which giant
elliptical galaxies have been formed, and so we observe less disk-like
structures in them. The FRI galaxies may consequently be expected to be
redder and less star forming, but these differences are not significant in
our datasets, and might need a bigger sample size to be discovered. These
results are consistent with the extrinsic scenario for the FR
dichotomy. The one surprising result is observing higher [OIII] luminosity
for FRI LERGs than matched FRII LERGs, with the high significance. This could
be explained in the context of the extrinsic scenario, if there is more cold gas surrounding
 the nucleus, which converts the radiated luminosity more efficiently into line radiation
  but which is also capable of disrupting the radio jets. However, this could alternatively 
  be due to higher levels of radiated luminosity from the core; this would not naturally fit
   into an extrinsic scenario, but could be interpreted as a selection effect caused by our 
   matching in total radio luminosity, and
FRIs being more core dominated sources, since the core radio luminosity
seems to be better correlated with the [OIII] line emission in FRIs than
total radio luminosity is (Baldi, Capetti \& Giovannini 2015; see also Section 4.3).

\subsection {HERG vs. LERG } 

\begin{figure}
\center
\epsfig{file=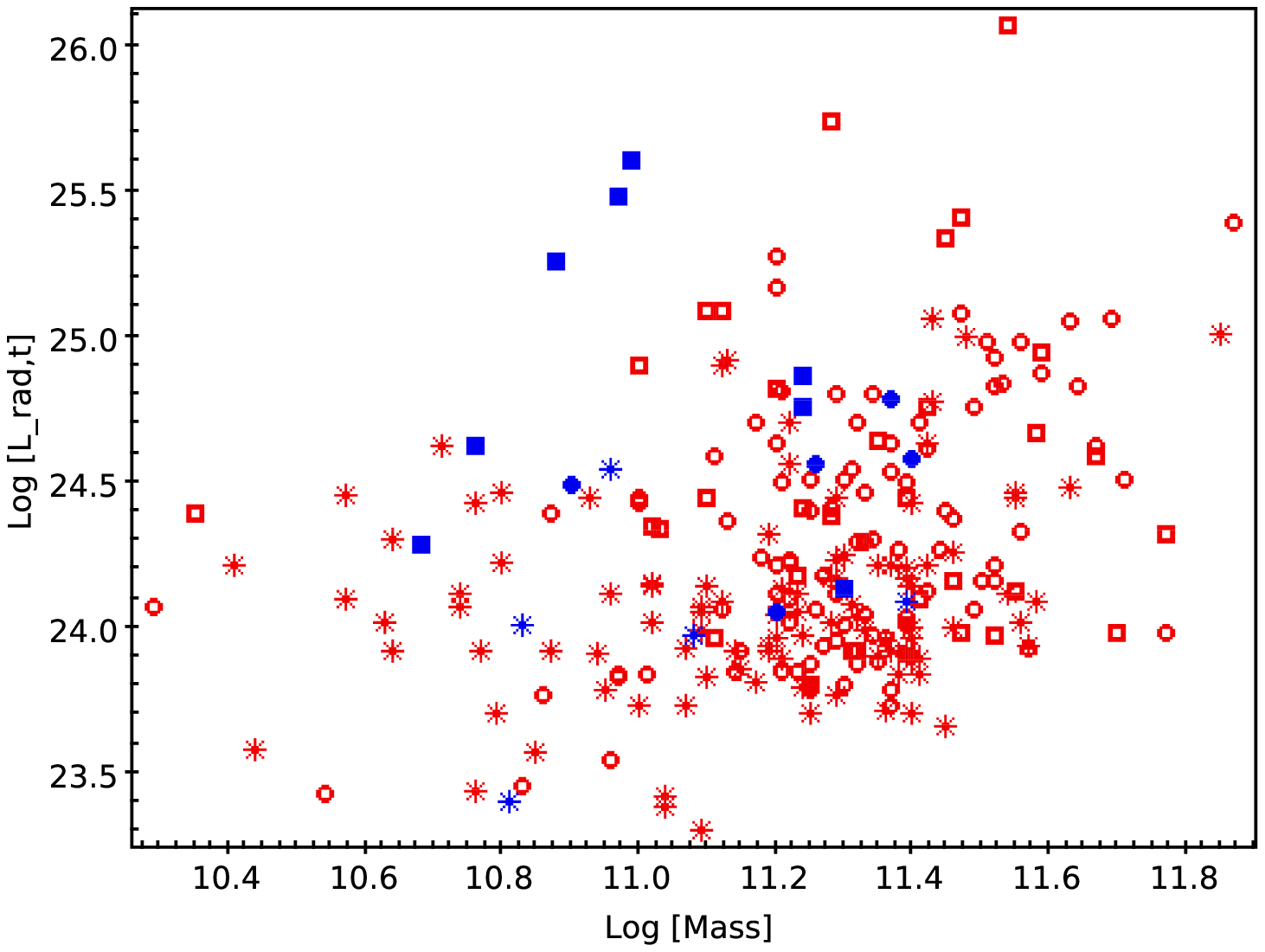,width=3.2in,height=2.6in}

\caption{Total radio luminosity versus stellar mass of FRIs (open and filled circles), 
FRIIs (open and filled squares) and compact radio AGN (stars). The colour red represents 
LERGs and blue represents HERGs.}
\label{HL}
\end{figure}

In order to compare HERGs and LERGs, we construct each sample by combining
HERG and LERG sources from three different classes of FRI, FRII and
compact radio AGN. We make the matched sample by cross-matching HERG
objects with LERG objects from the same class, to remove the morphological
effects caused by this method. Fig.~$\ref{HL}$ shows the radio luminosity versus stellar
mass distribution of all FRI, FRII and compacts separated into HERGs and
LERGs. There are relatively few HERG sources while LERGs are more
populated in each class of FRI/II and compact. A one-to-one matching
scheme thus result in a small sample of HERG/LERG and large uncertainties
for the comparison. Therefore, we cross-match each HERG with three
different LERGs, which is possible due to mismatch in HERG and LERG
numbers, and helps to improve the overall sample size and significance. We
also allow a wider matching tolerance for the differences in radio
luminosity ($\Delta$ log[L] = $\pm$ 0.5) and mass ($\Delta$ log[M] = $\pm$
0.2) that will help with the random selection of matches. Finally, as is
clear on Fig.~$\ref{HL}$, there are five low mass FRII HERGs with only a few FRII
LERGs around them, which are insufficient to match all 5 HERGs. Thus, in
each iteration we randomly choose two HERGs and cross-matched them with
the three FRII LERGs each, within a wider range of radio luminosity
($\Delta$ log[L] = $\pm $ 1.0) and mass ($\Delta$ log[M] = $\pm $ 0.4)
differences. By these methods, we have constructed significant-sized
samples of HERGs (15) and LERGs (45) with the same distribution of stellar
mass, total radio luminosity and morphology.

\begin{figure*}
\center
\epsfig{file=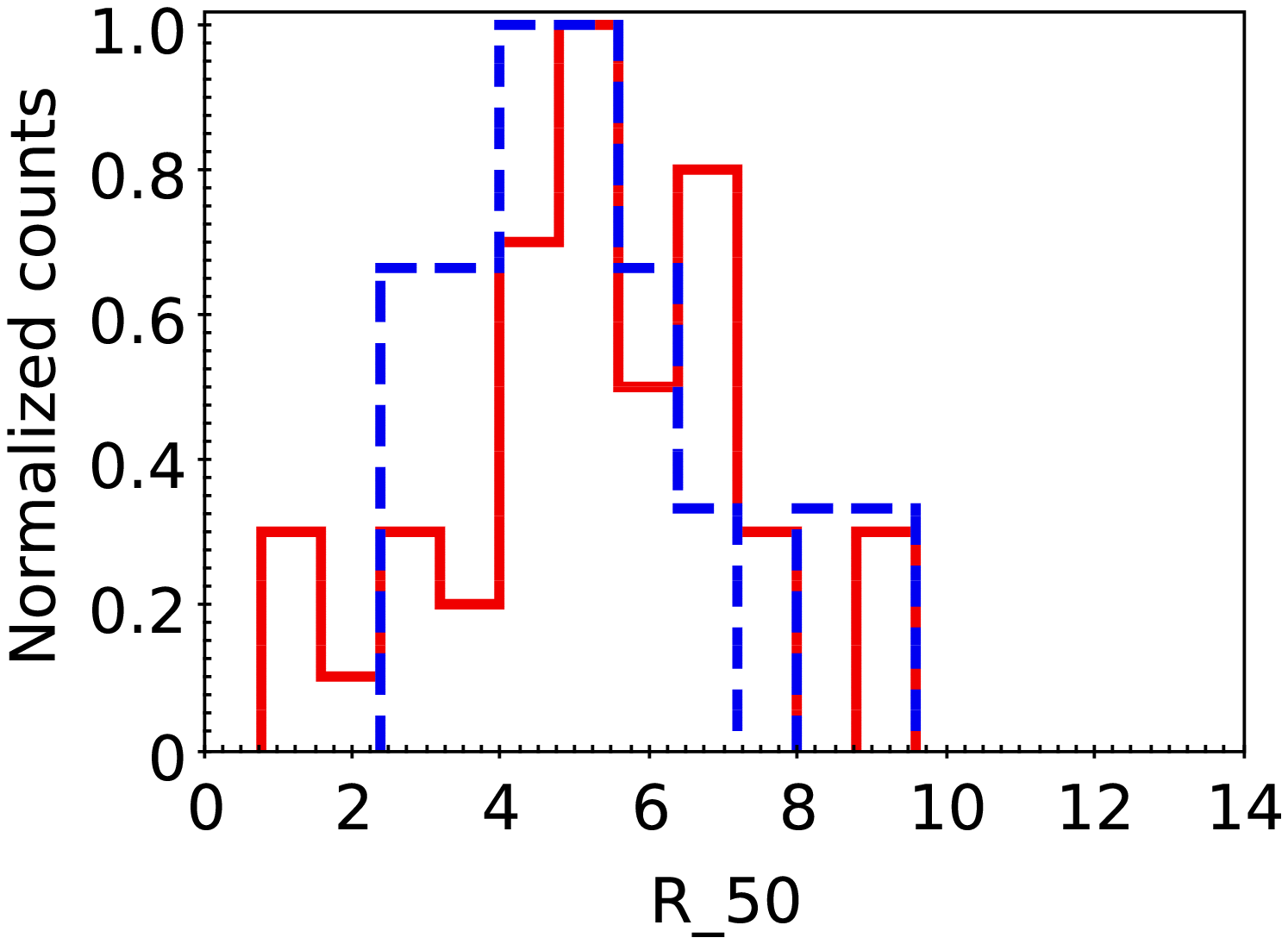,width=1.6in,height=1.3in}
\epsfig{file=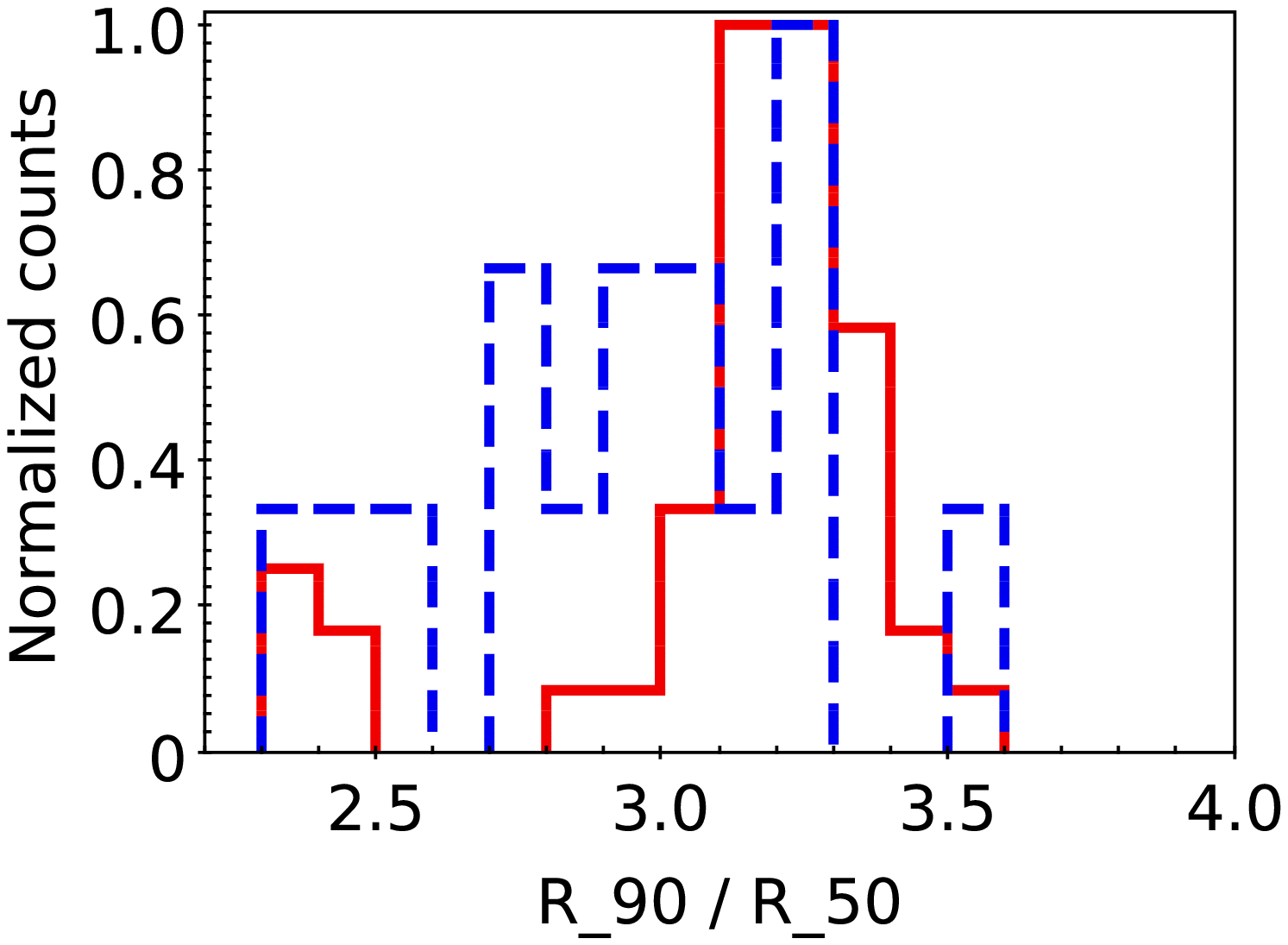,width=1.6in,height=1.3in}
\epsfig{file=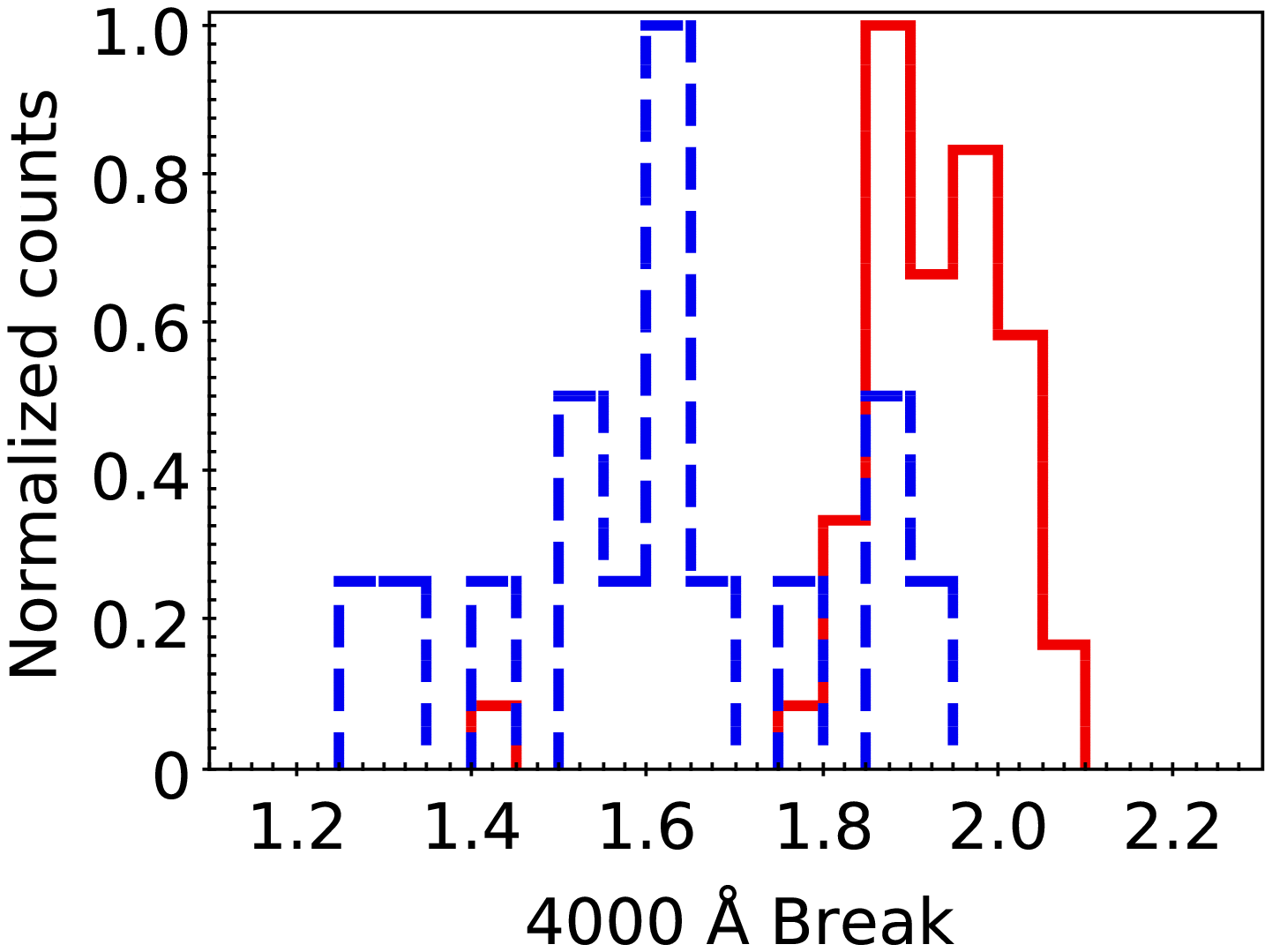,width=1.6in,height=1.3in}
\epsfig{file=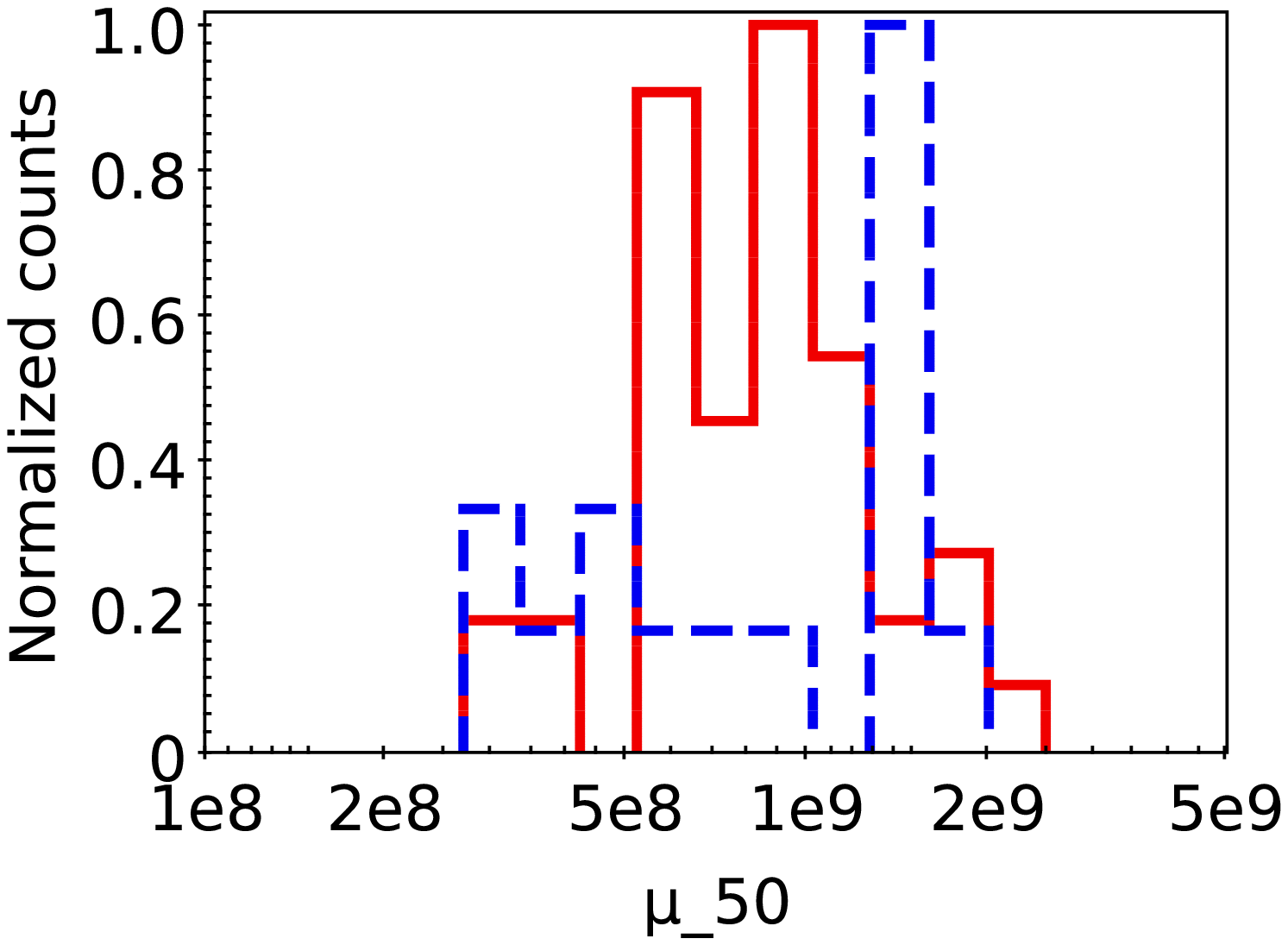,width=1.6in,height=1.3in}
\epsfig{file=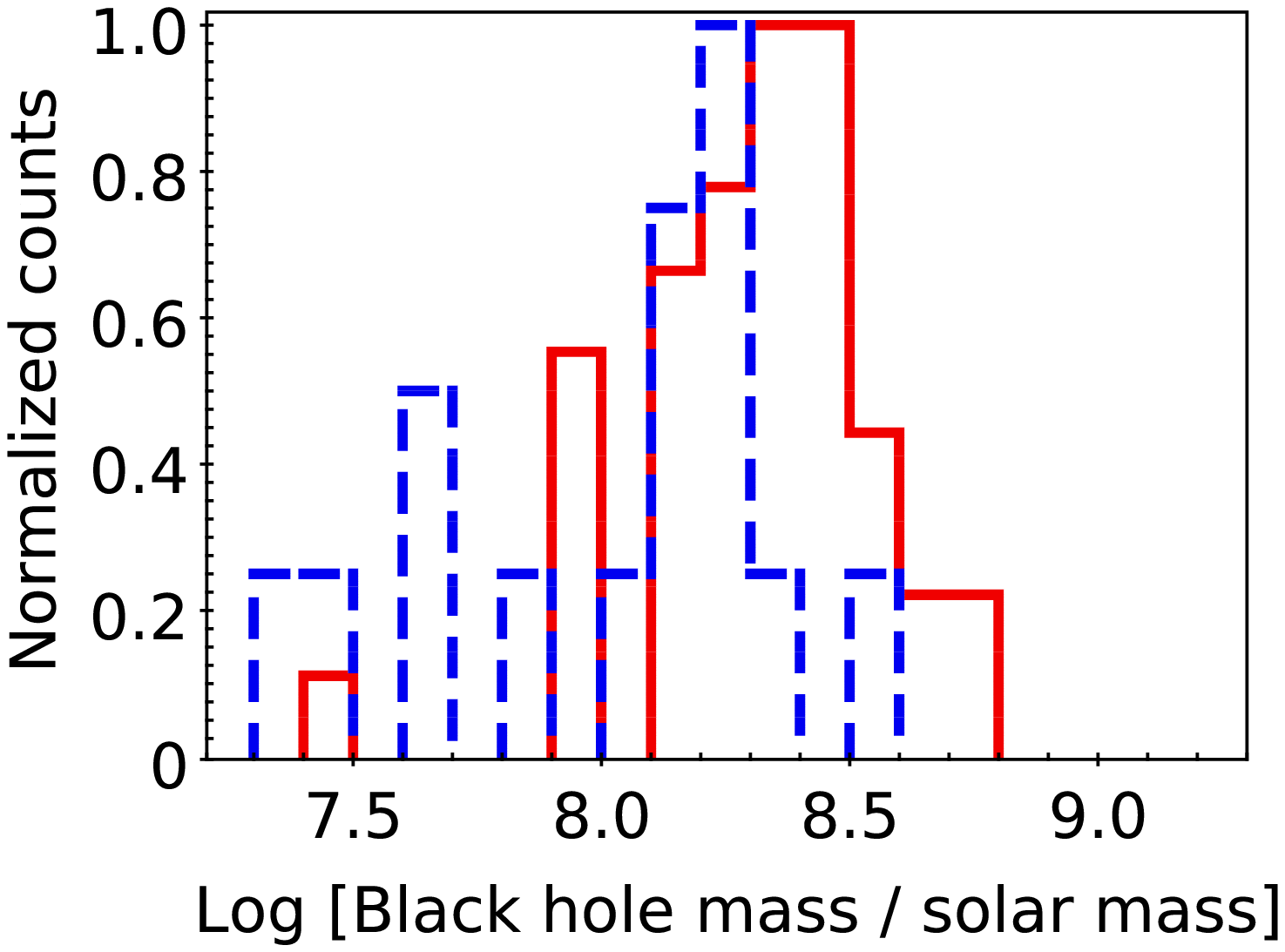,width=1.6in,height=1.3in}
\epsfig{file=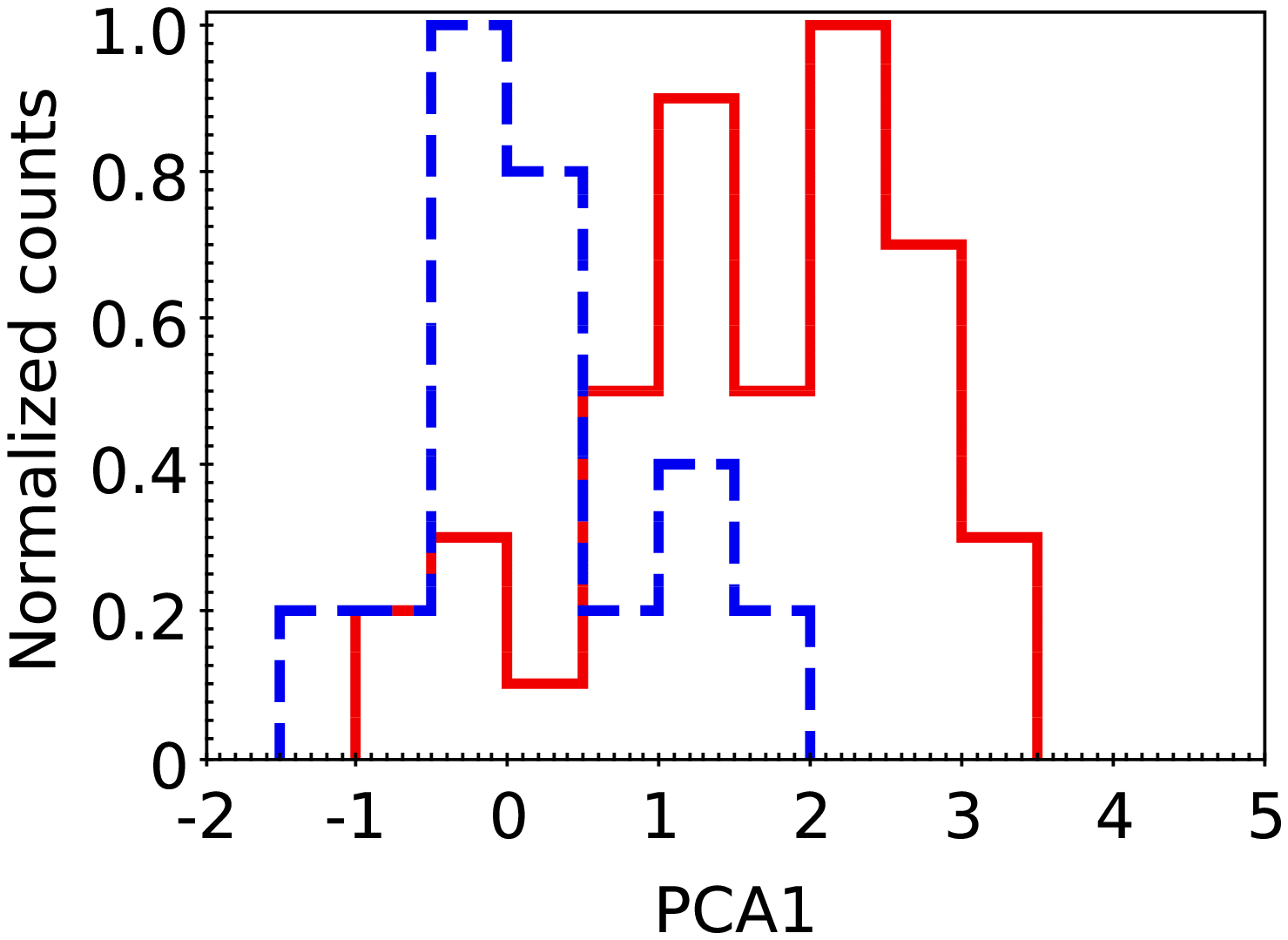,width=1.6in,height=1.3in}
\epsfig{file=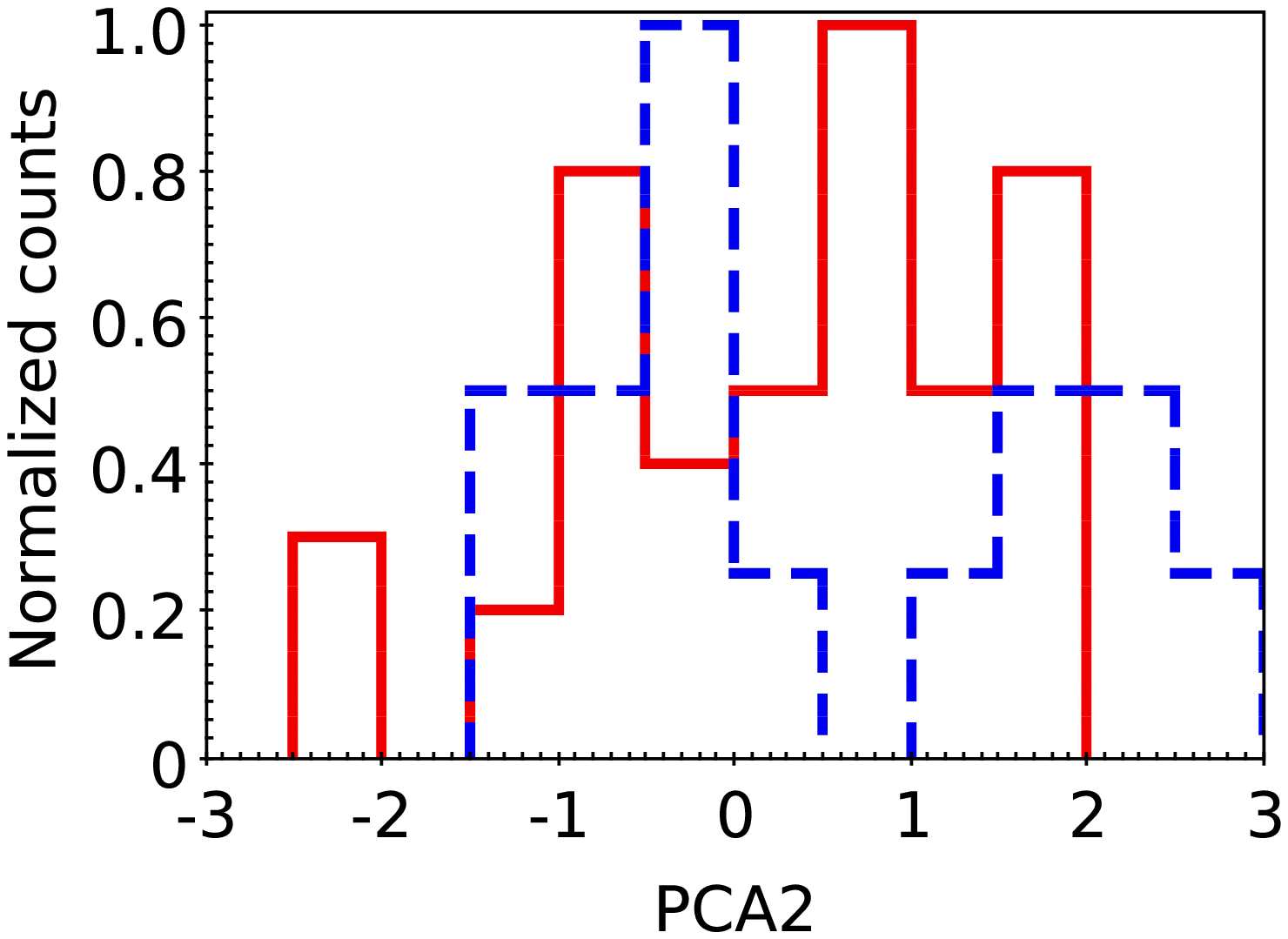,width=1.6in,height=1.3in}
\epsfig{file=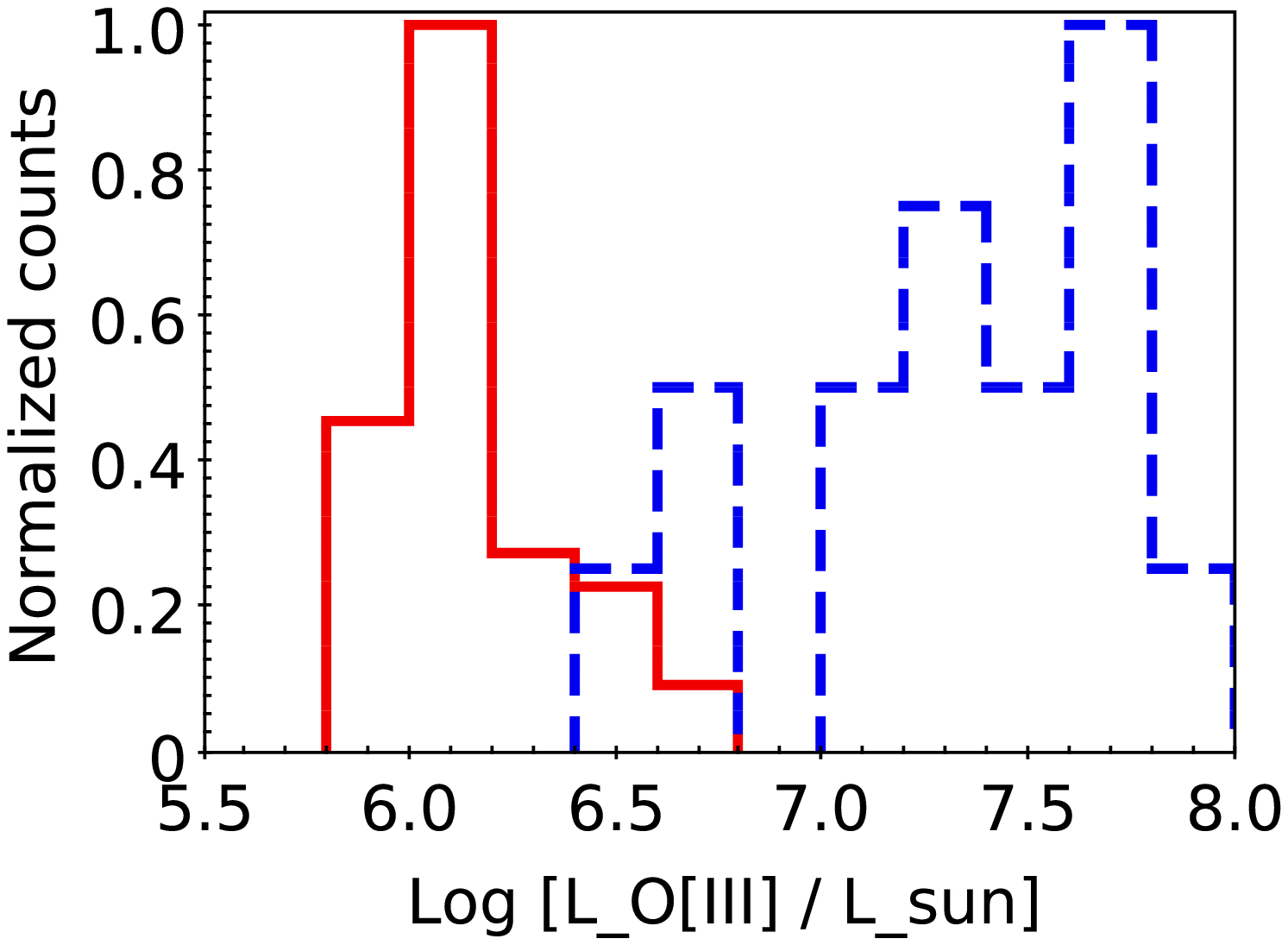,width=1.6in,height=1.3in}
\caption{Histogram of the host galaxy and environmental parameter
  distributions for LERG (red) and HERG (blue) radio sources, matching in
  radio morphology and in the L$_{rad,t}$--M$_{\star}$ plane.  }
\label{histHL}
\end{figure*}

The results of the comparison of host galaxy and environmental properties,
confirmed by KS test, are presented in Table~$\ref{table3}$ and the
histograms for each of the parameters are presented in
Fig.~$\ref{histHL}$. Differences with over 95 percent confidence have been detected
for both environmental and host galaxy parameters. In terms of host galaxy
properties, HERGs are younger with lower concentration and lower black
hole mass (thus, lower M$_{bh}$/M$_{\star}$) than LERGs, indicating that
they reside in more disky galaxies, as previously reported by Best \&
Heckman (2012). The significance in our study is lower for some
correlations than was found by Best \& Heckman, because of the smaller
sample size, but importantly we have eliminated any possible biases
associated with FRI/II classifications. Therefore, our results are robust.
HERGs also have higher [OIII] luminosity, as expected from their
definition as sources with a stronger ionising component. The environments
of HERGs appear to show lower density and tidal interactions than those of
LERGs. The significance of the PCA1 parameter analysis confirms the lower
density environment for HERGs, while the lack of any difference in PCA2
distributions shows that the apparently lower tidal interaction in HERGs
might be a projection effect. These environmental results are consistent
with those of Gendre et~al.\ (2013) who have reported low density
environments for HERGs independent of FR morphology.

It is worth mentioning that by comparing HERGs and LERGs using a sample of
FRII HERGs and FRII LERGs, without cross-matching for luminosity and mass,
we get the same result but with lower significance.

These results are consistent with the currently favoured description of
HERG and LERG origins, which associated the differences to
Eddington-scaled accretion rates on to the black hole (see discussions in
Heckman \& Best 2014). HERGs require high accretion rates fuelled from
extensive cold gas reservoirs; this gas-rich environment is more readily
available in later-type disky galaxies with lower concentration and higher
star formation, as seen in the data. Their low density environments are
also consistent with their fueling mechanisms, in a sense that in high
density environments galaxies tend to be gas-poor, due to a combination of
processes including stripping and strangulation (cf. Boselli \& Gavazzi
2006). In contrast, galaxy groups and clusters have giant elliptical
galaxies in their center with the high black hole masses, high
concentration, old stellar population and little cold gas remaining to
feed the central nuclei, but do have the cooling of hot gas which can
provide the low accretion rates necessary to fuel LERGs (e.g. Best \&
Heckman 2012): these are exactly the properties found for the LERG
sources.

These arguments can also help to explain the overlap of FRIIs with HERGs
and FRIs with LERGs; these have their origin in both radio luminosity and
environment. The higher accretion rates required to fuel HERGs also lead
to more powerful radio jets, which are more likely to be able to survive
the disrupting effects of their surrounding environments and become FRII
sources; only as minority form FRIs. In contrast, at the lower accretion
rates of LERGs, the lower power jets are more likely to be disrupted and
become FRIs, although there remains a significant population of LERG FRIIs
where the jets manage to survive. This connects to the host galaxy and
surrounding environment, responsible for disrupting the jets, which also
provides links between the FR classification and the excitation
state. FRIIs and HERGs both are favoured in lower density environments and
later-type galaxies, since these both offer a more plentiful supply of
cold gas to provide higher fuelling rates, and less potential to disrupt
the jets. FRIs and LERGs are developed in early type galaxies and higher
density environment, in both of which the gas supplies are likely to be
limited, and jets more easily disrupted.

\subsection {Extended vs. Compact }

\begin{figure*}
\center
\epsfig{file=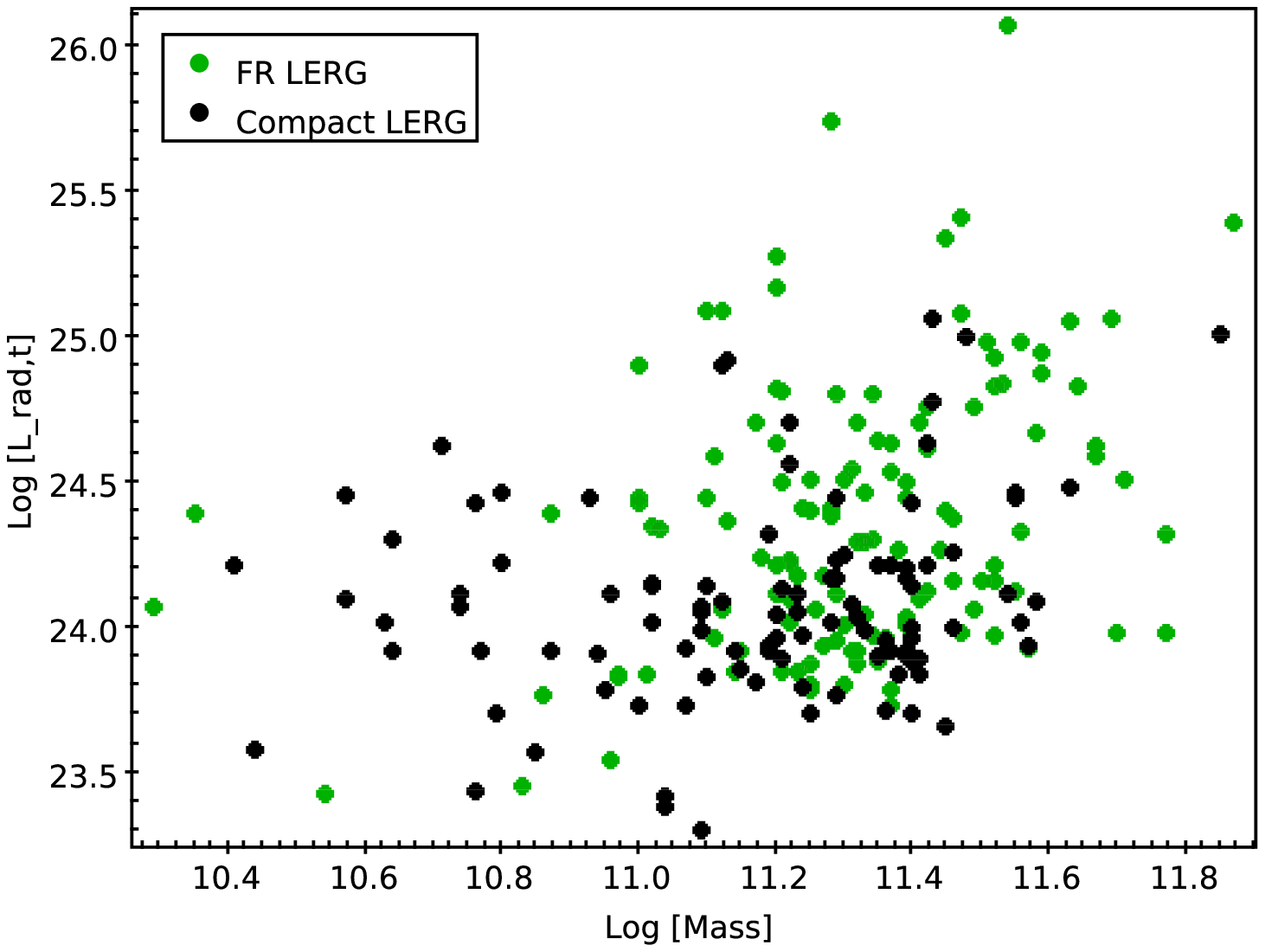,width=3.2in,height=2.6in}
\epsfig{file=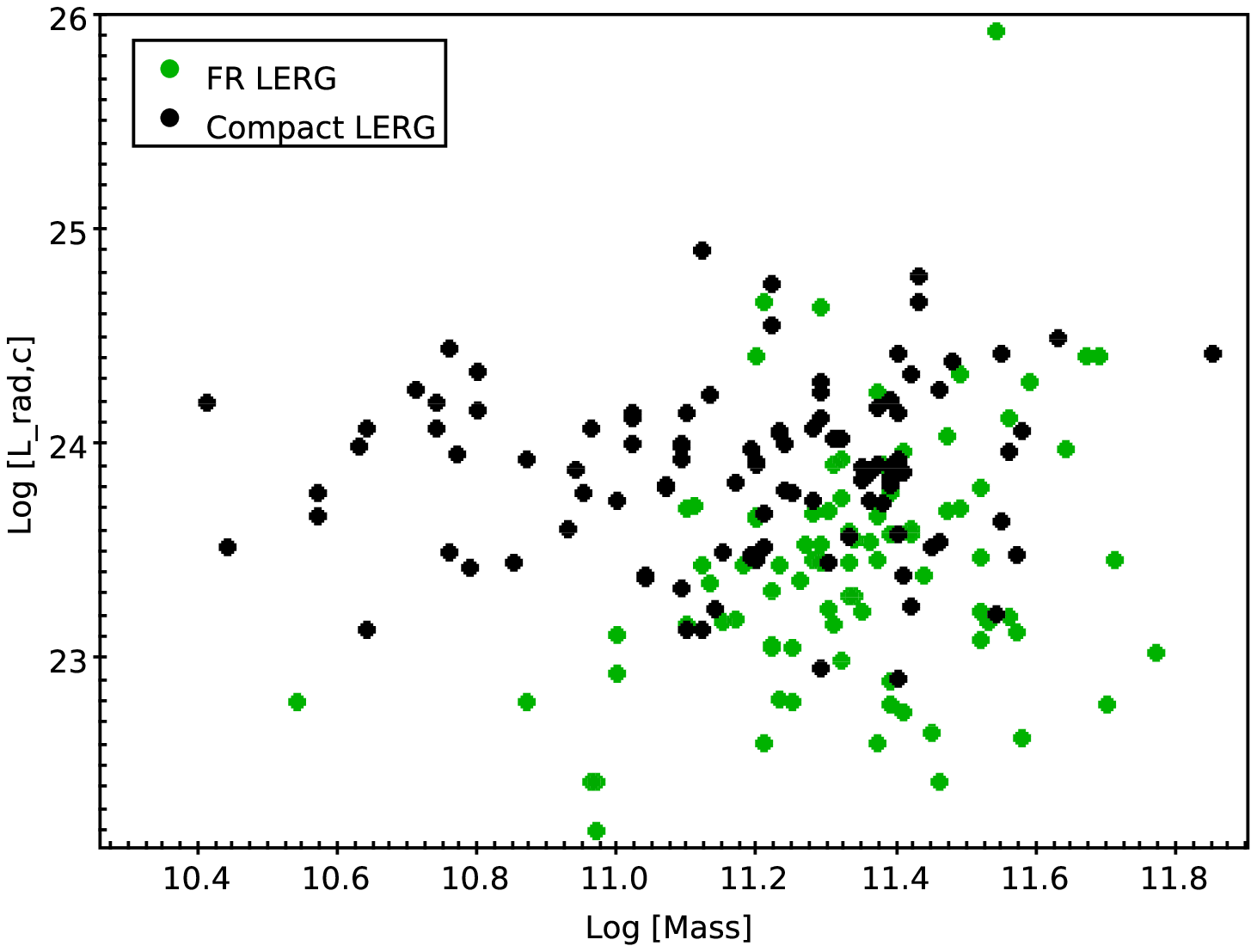,width=3.2in,height=2.6in}
\caption{Total radio luminosity (left panel) and core radio luminosity
  (right panel) versus the stellar mass of the extended radio galaxies
  (green) and compact radio galaxies (black).}
\label{compact}
\end{figure*}

\begin{figure*}
\center
\epsfig{file=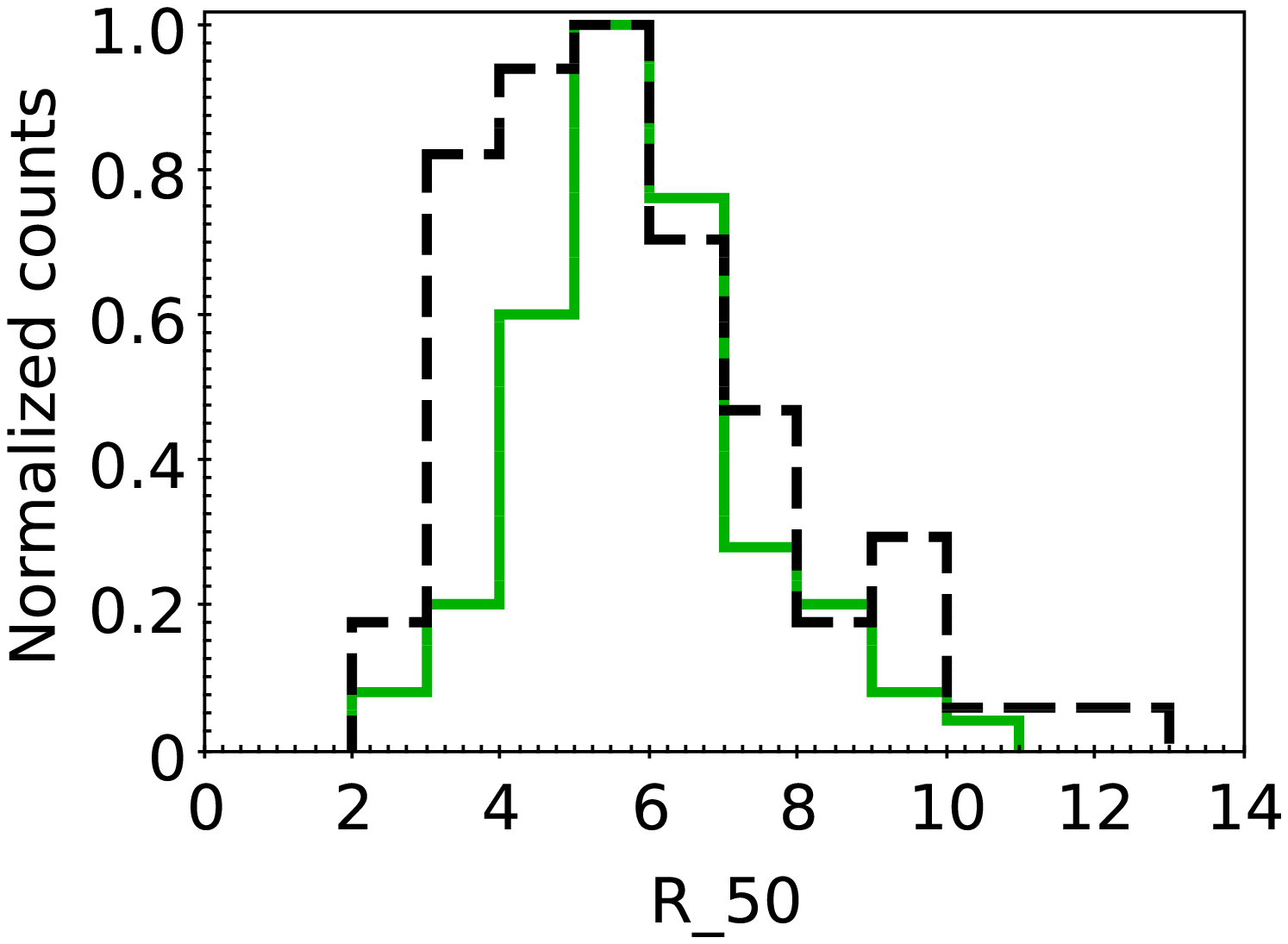,width=1.6in,height=1.3in}
\epsfig{file=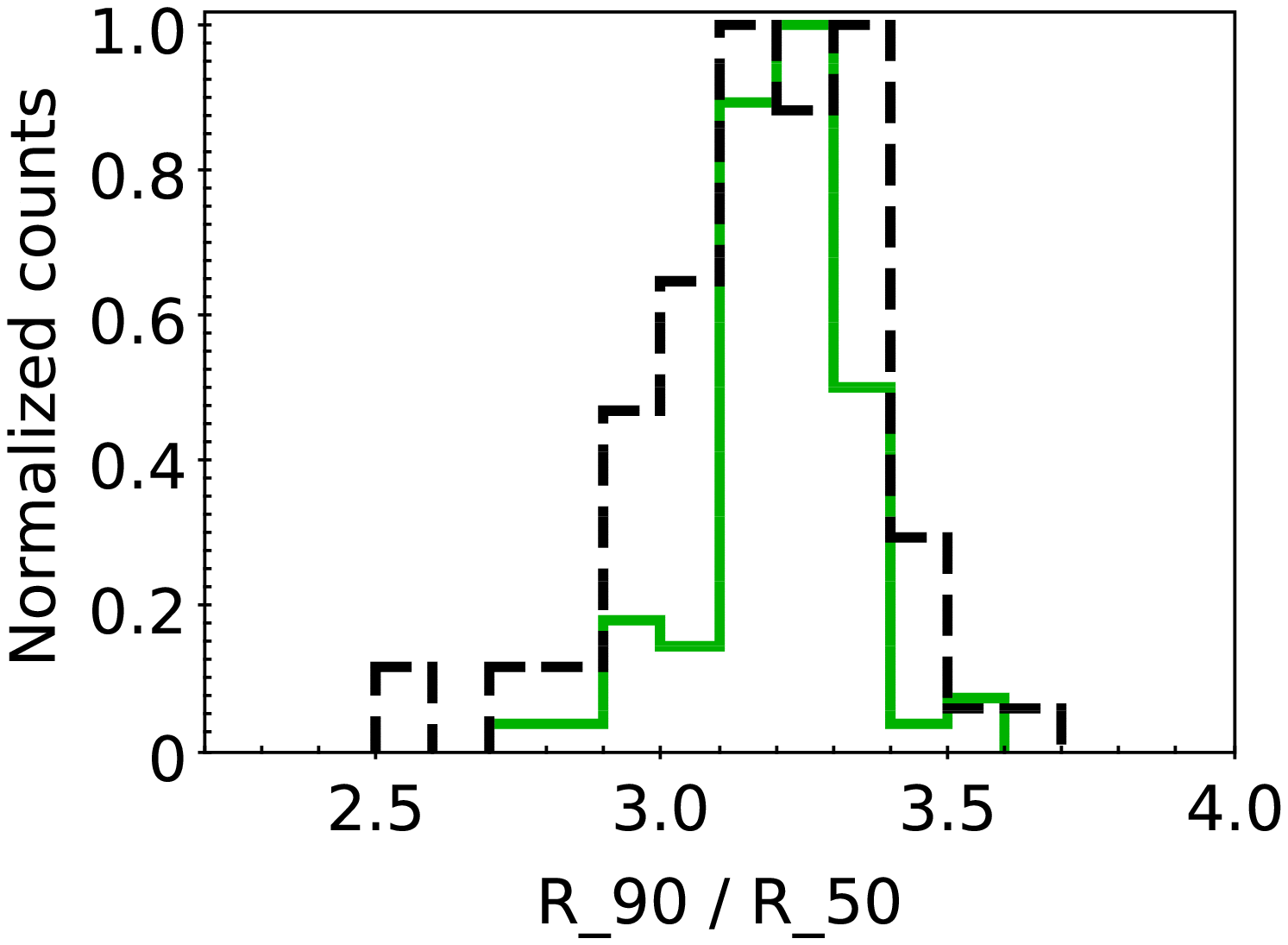,width=1.6in,height=1.3in}
\epsfig{file=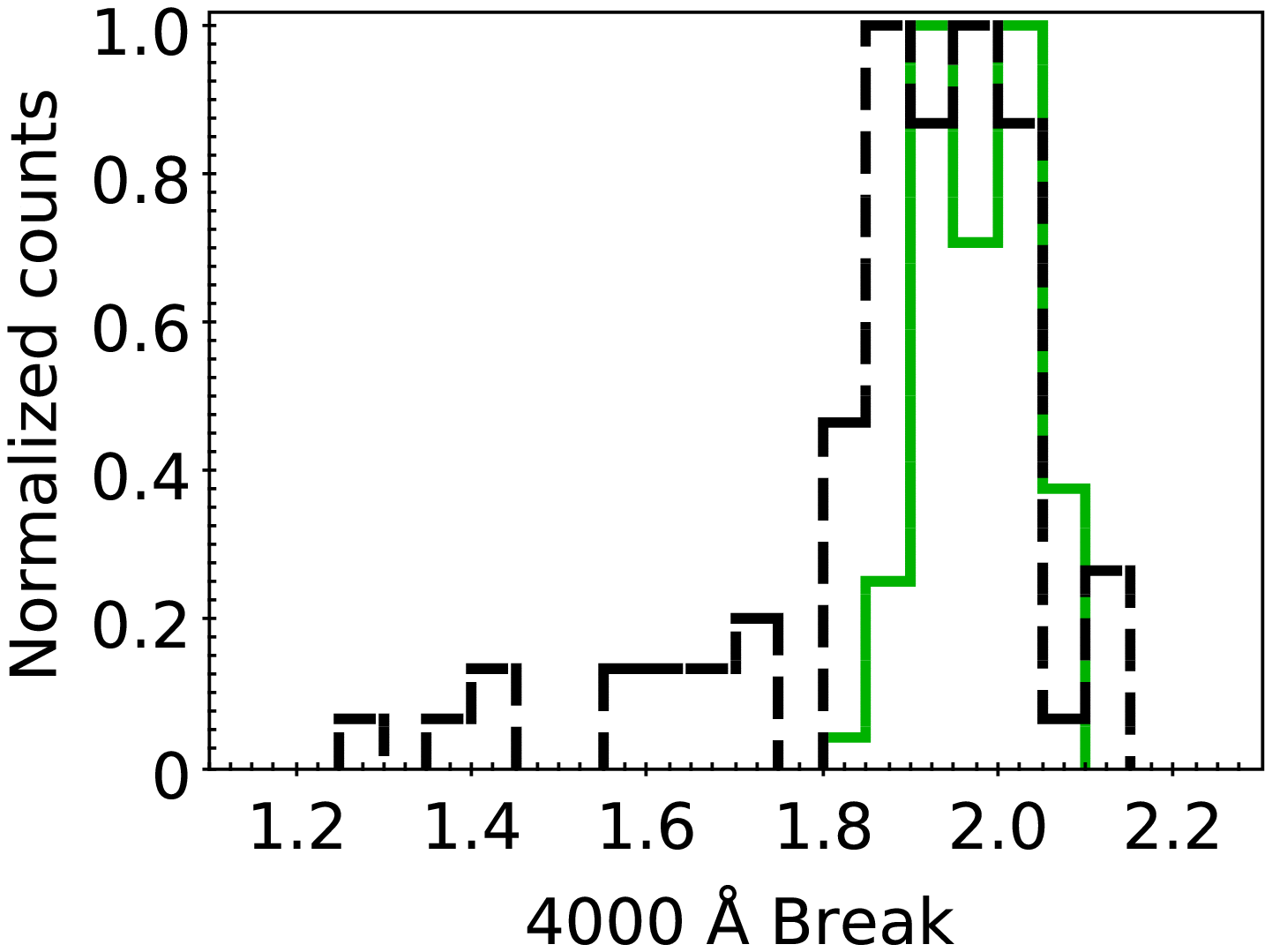,width=1.6in,height=1.3in}
\epsfig{file=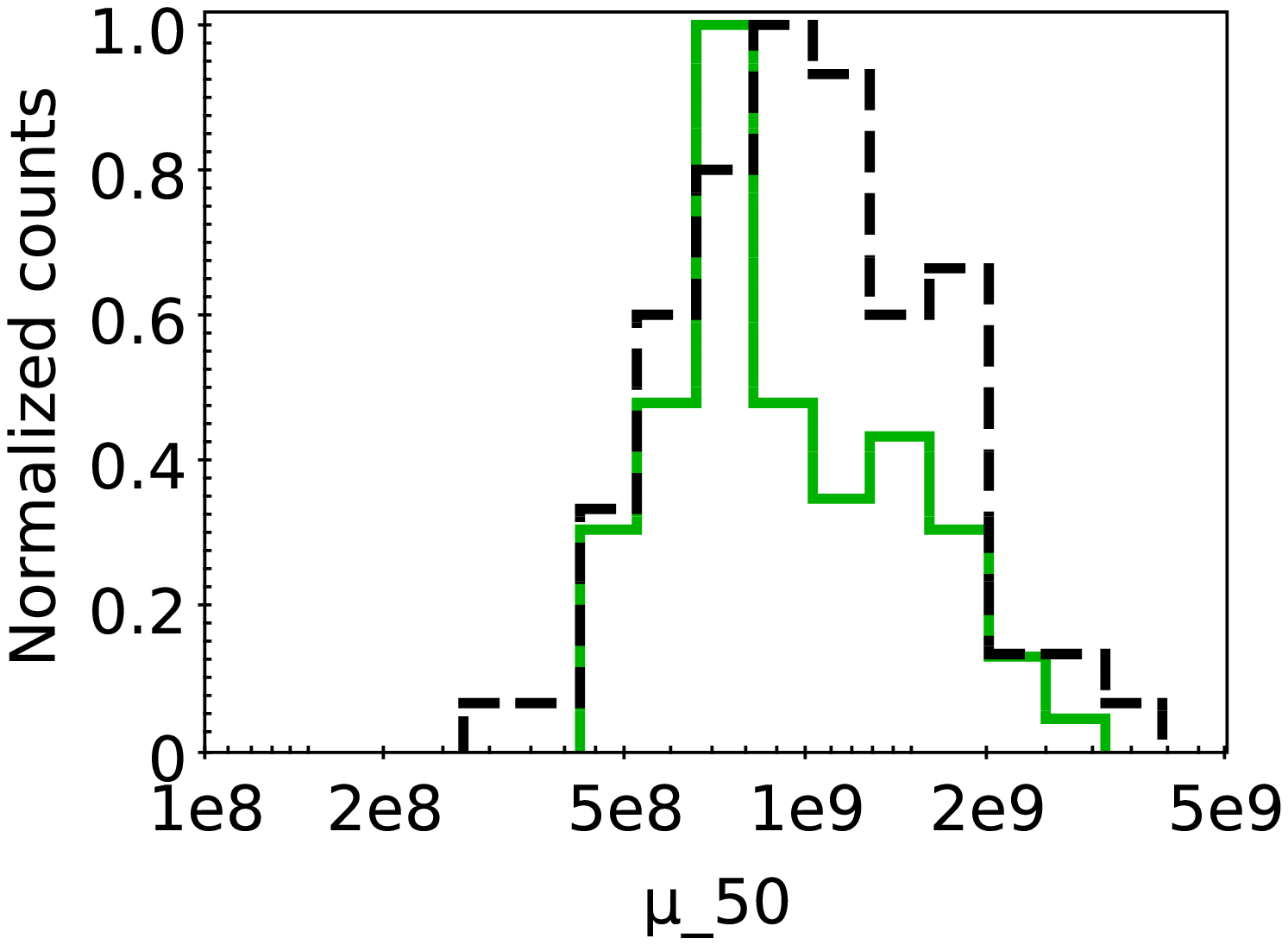,width=1.6in,height=1.3in}
\epsfig{file=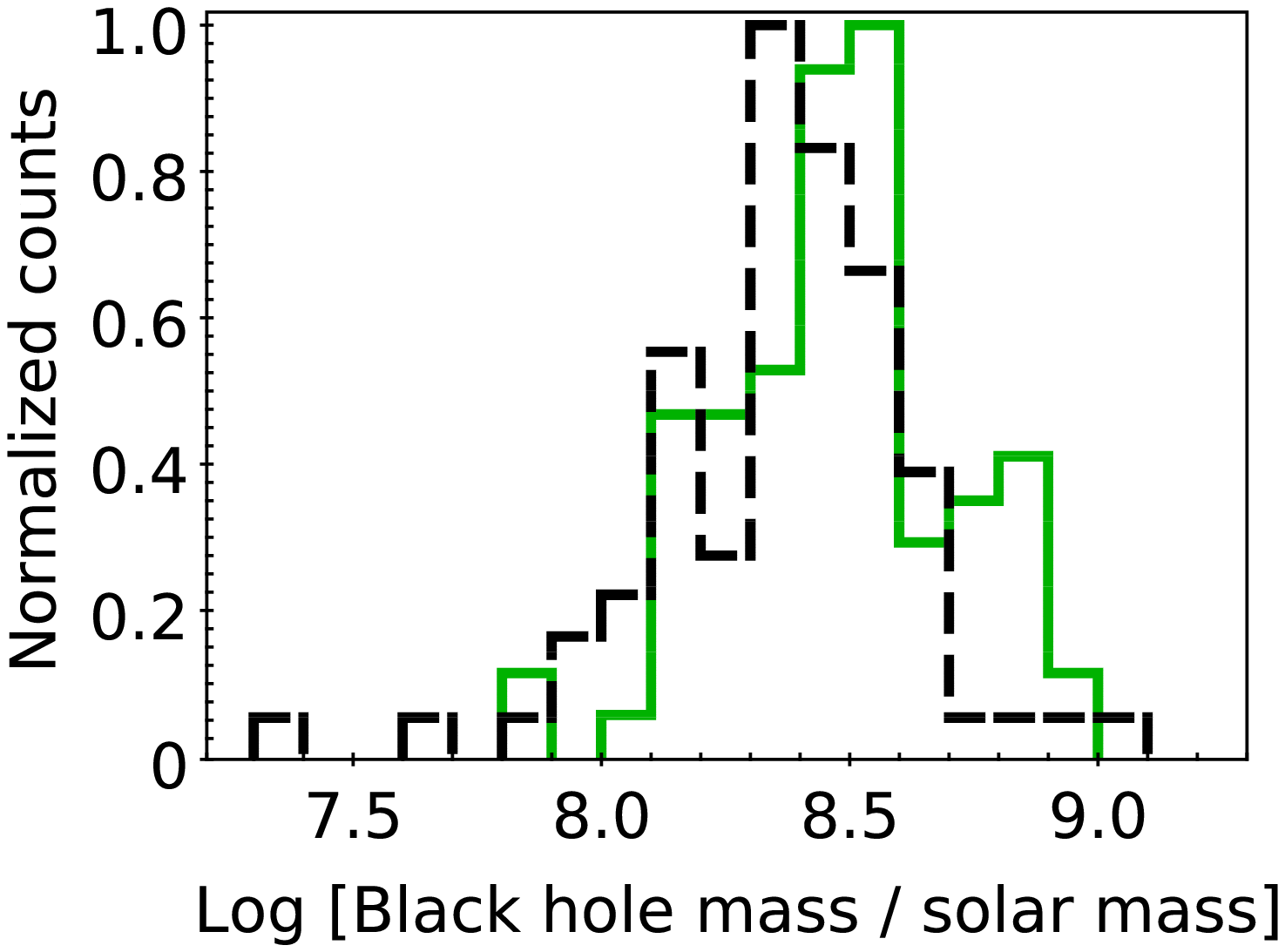,width=1.6in,height=1.3in}
\epsfig{file=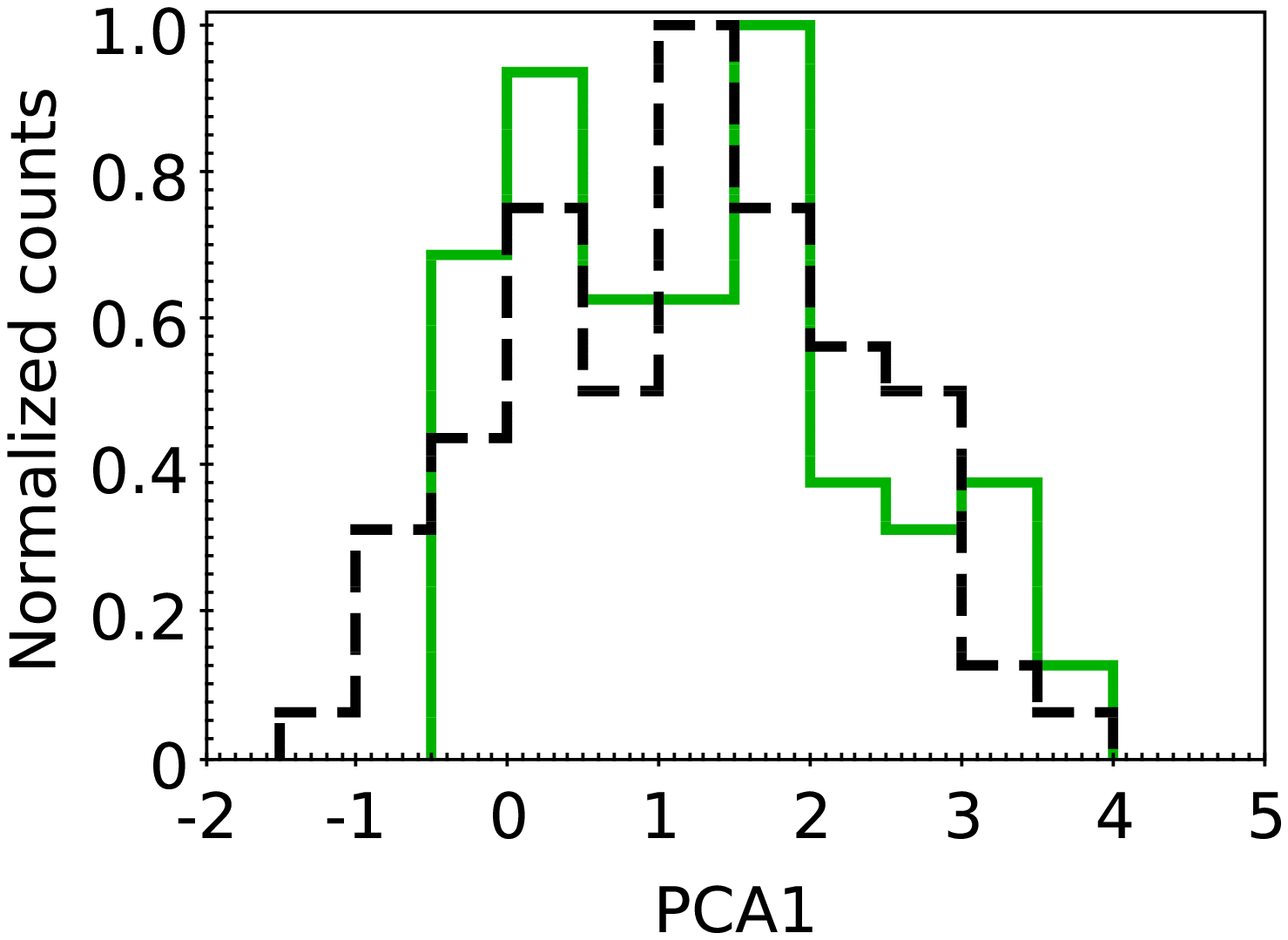,width=1.6in,height=1.3in}
\epsfig{file=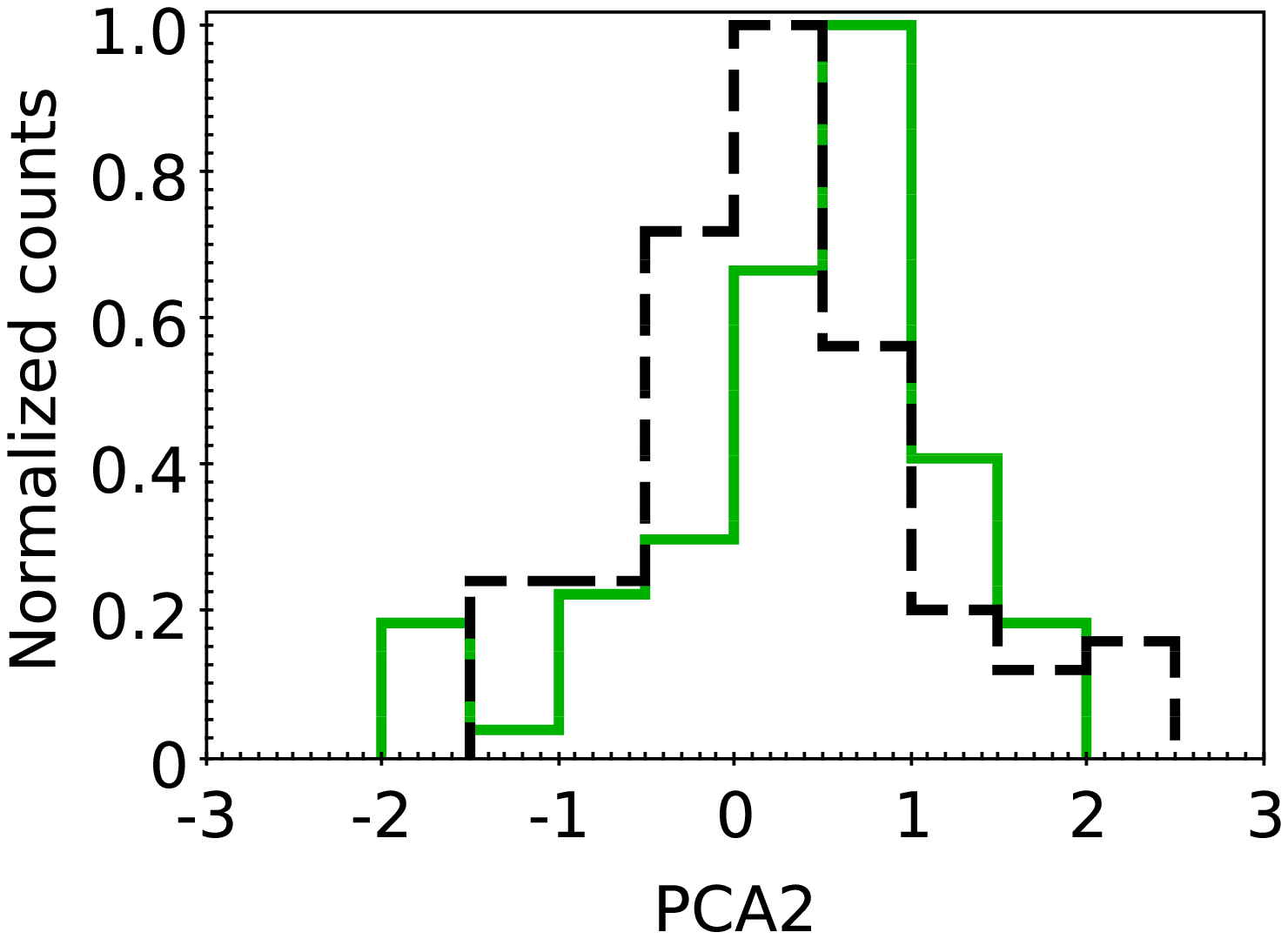,width=1.6in,height=1.3in}
\epsfig{file=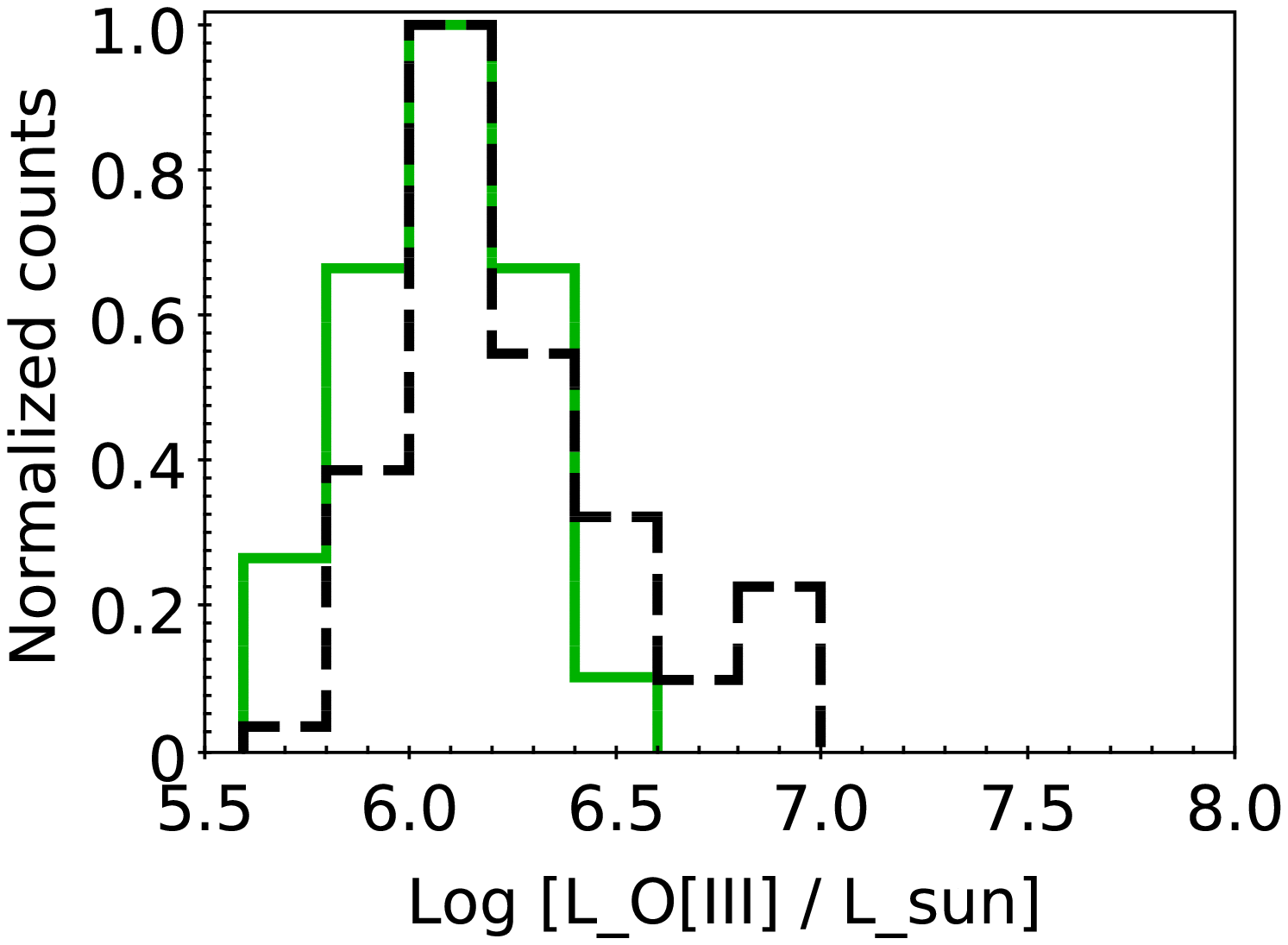,width=1.6in,height=1.3in}
\caption{Histogram of the host galaxy and environmental parameter
  distributions for compact (black) and extended (green) radio sources
  matched in the L$_{rad,t}$--M$_{\star}$ plane.  }
\label{histCom1}
\end{figure*}

\begin{figure*}
\center
\epsfig{file=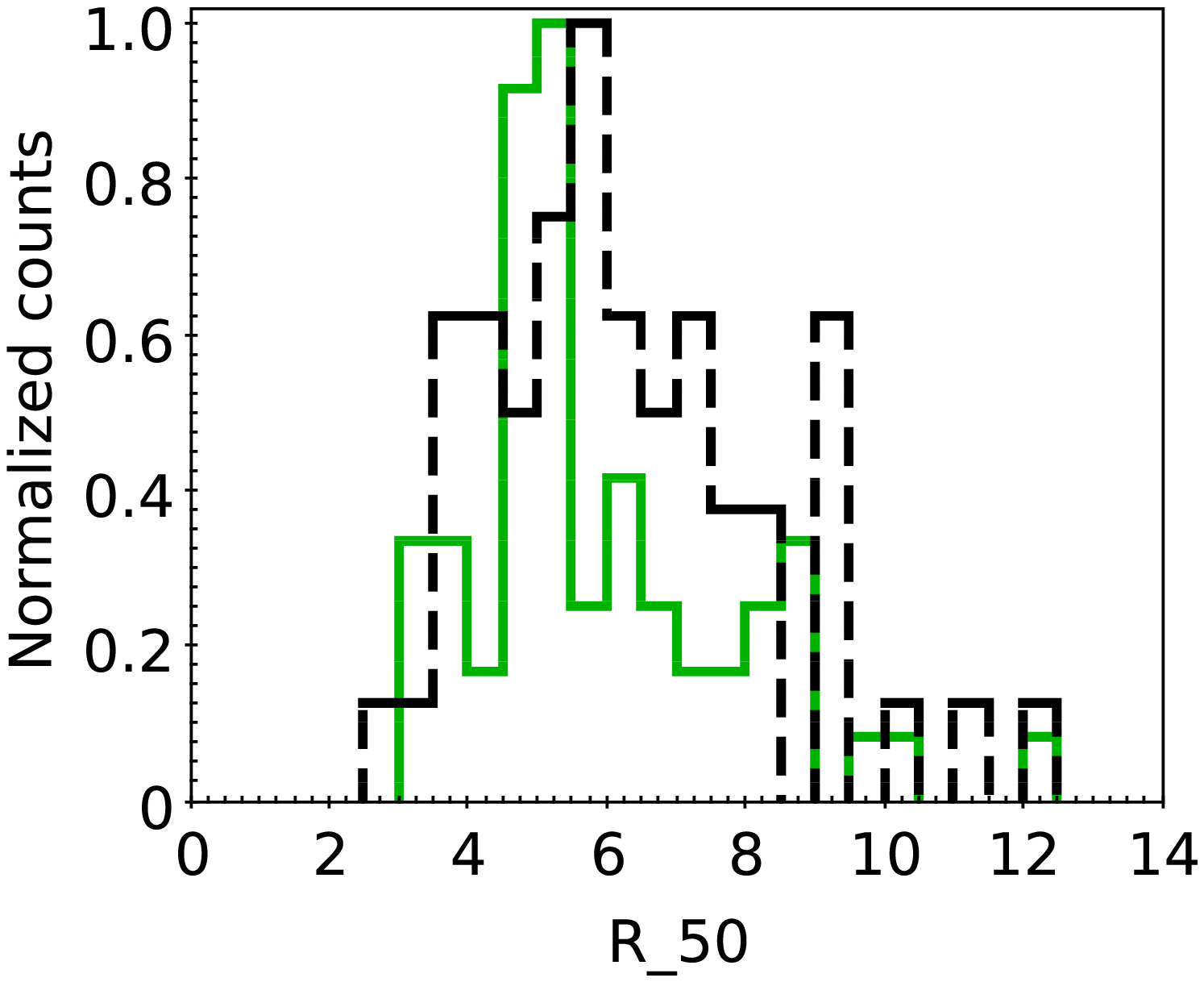,width=1.6in,height=1.3in}
\epsfig{file=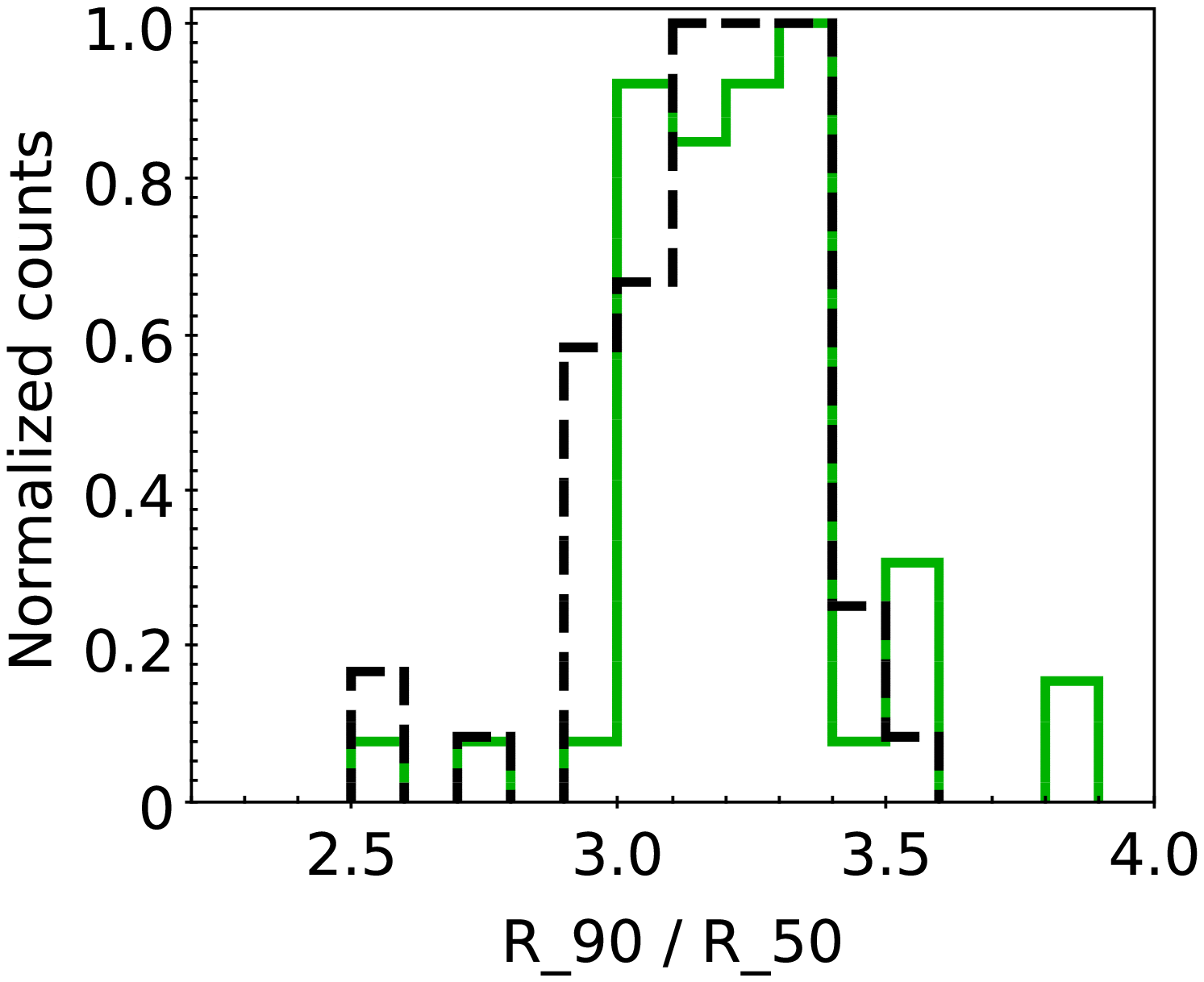,width=1.6in,height=1.3in}
\epsfig{file=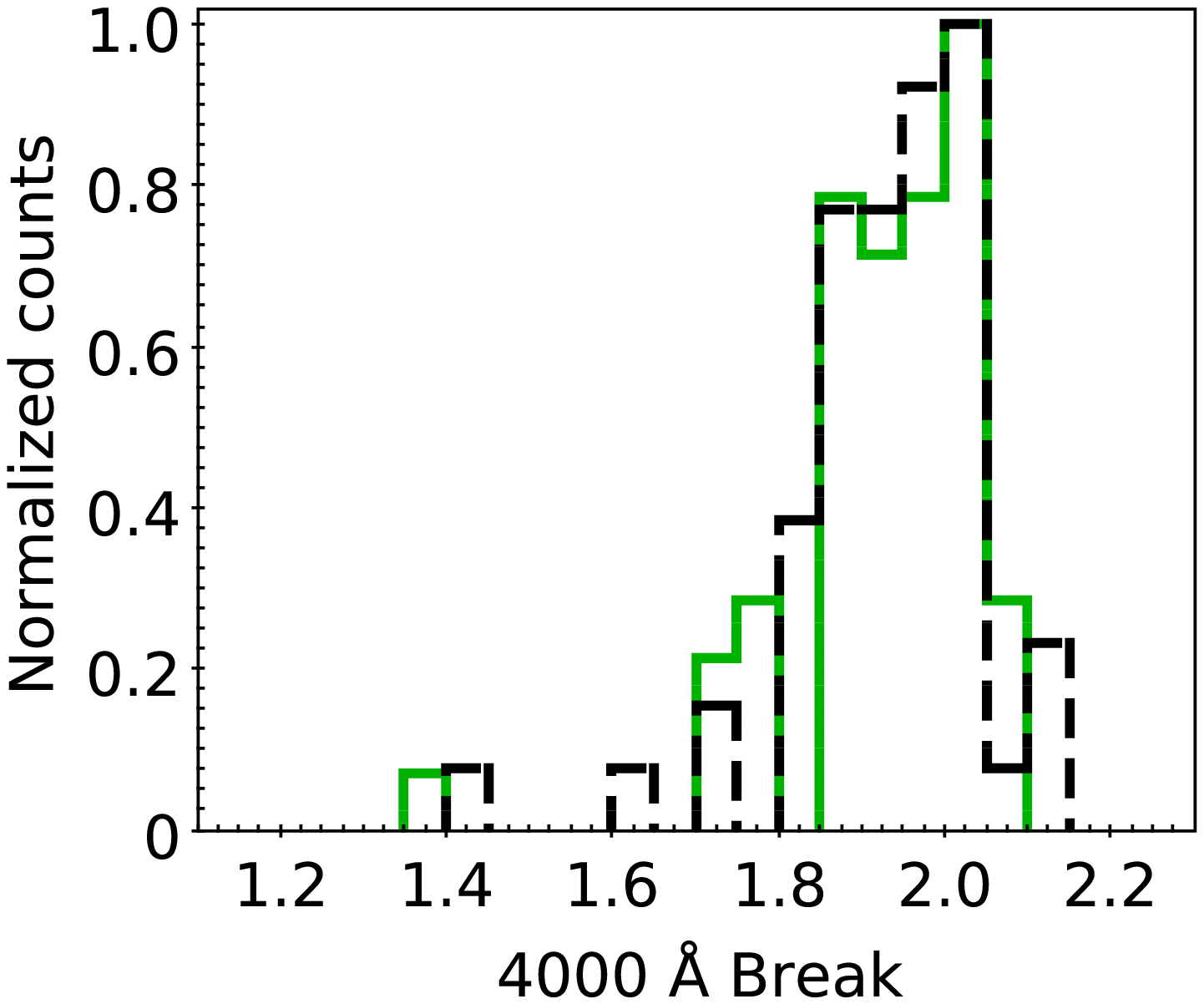,width=1.6in,height=1.3in}
\epsfig{file=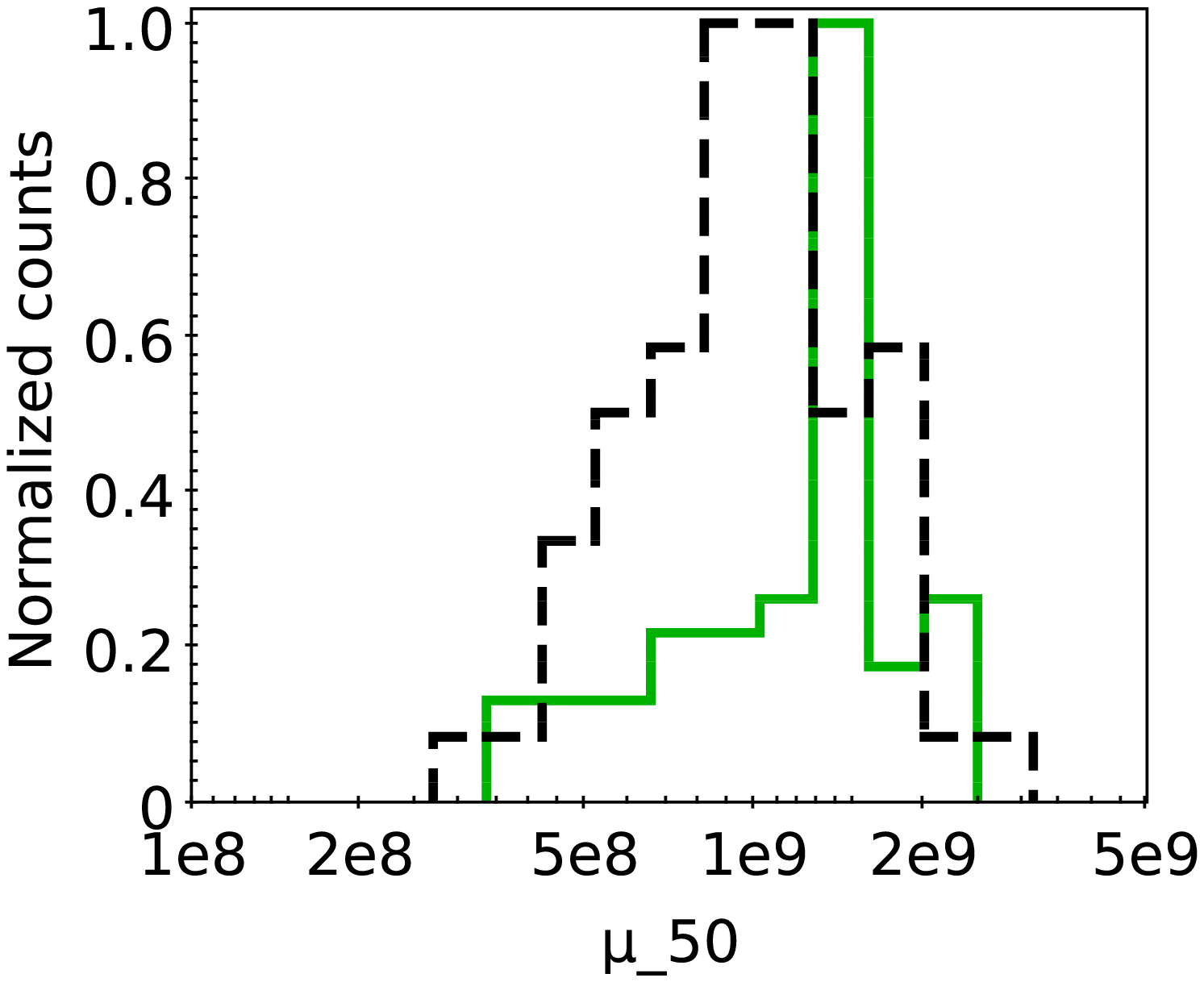,width=1.6in,height=1.3in}
\epsfig{file=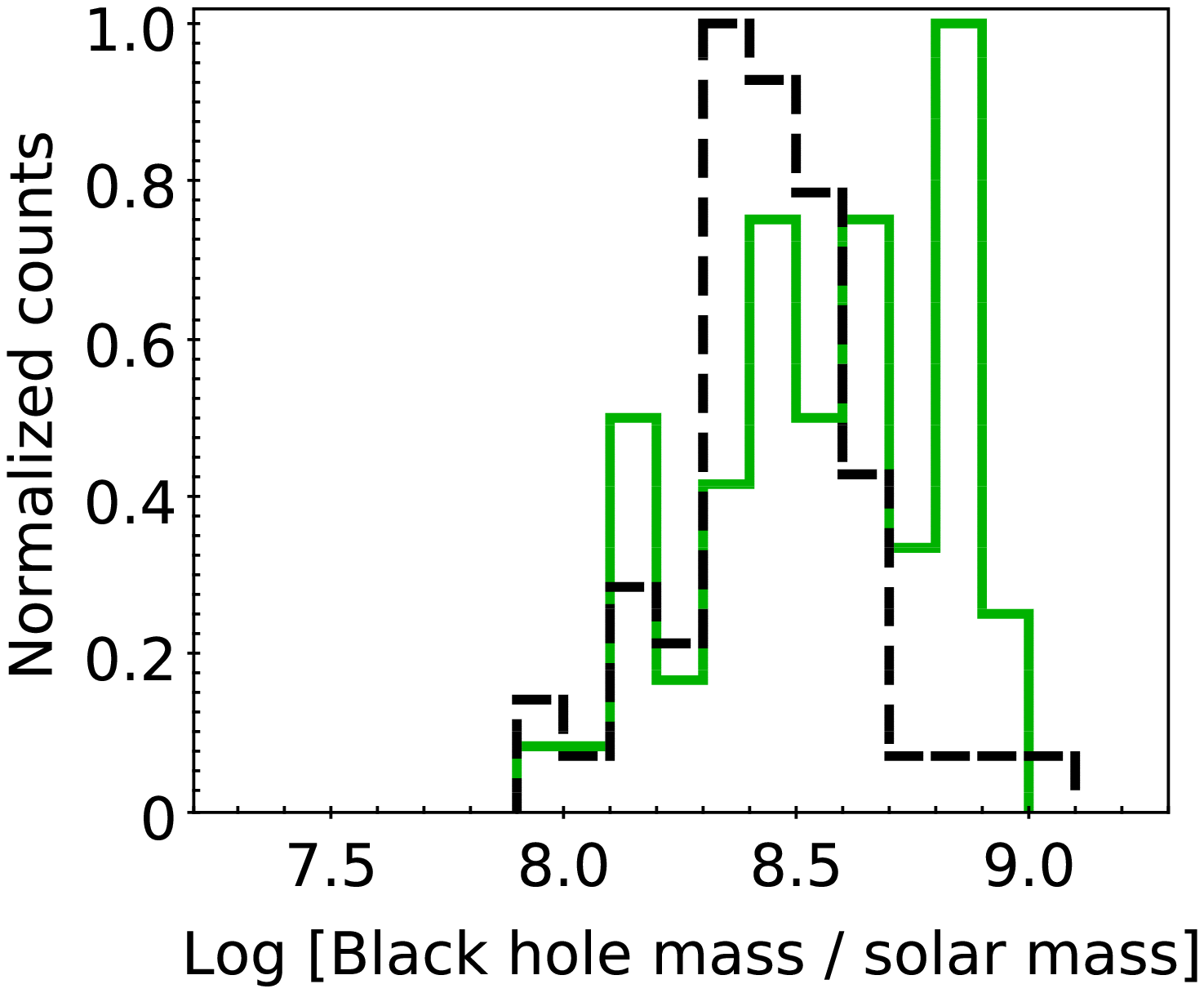,width=1.6in,height=1.3in}
\epsfig{file=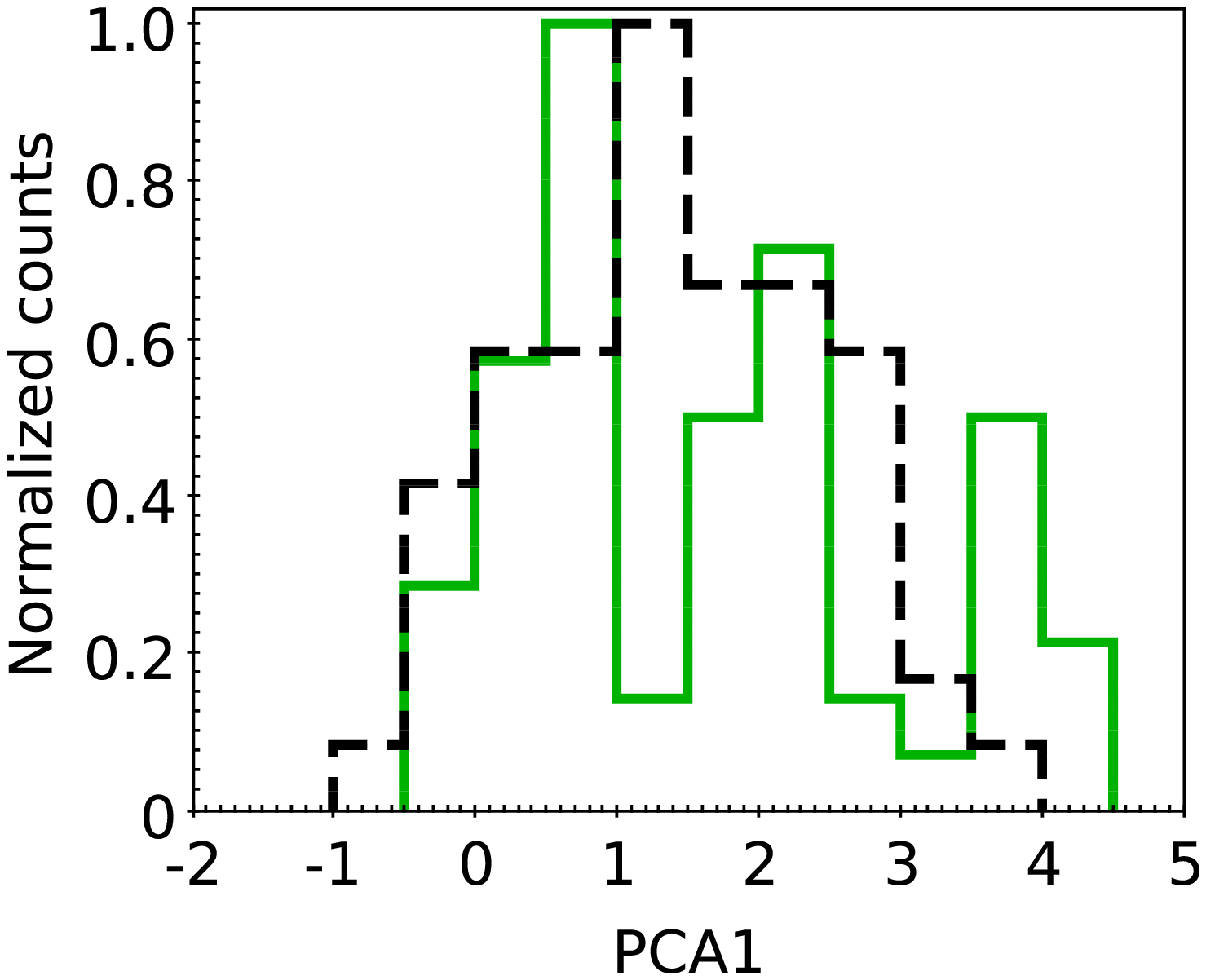,width=1.6in,height=1.3in}
\epsfig{file=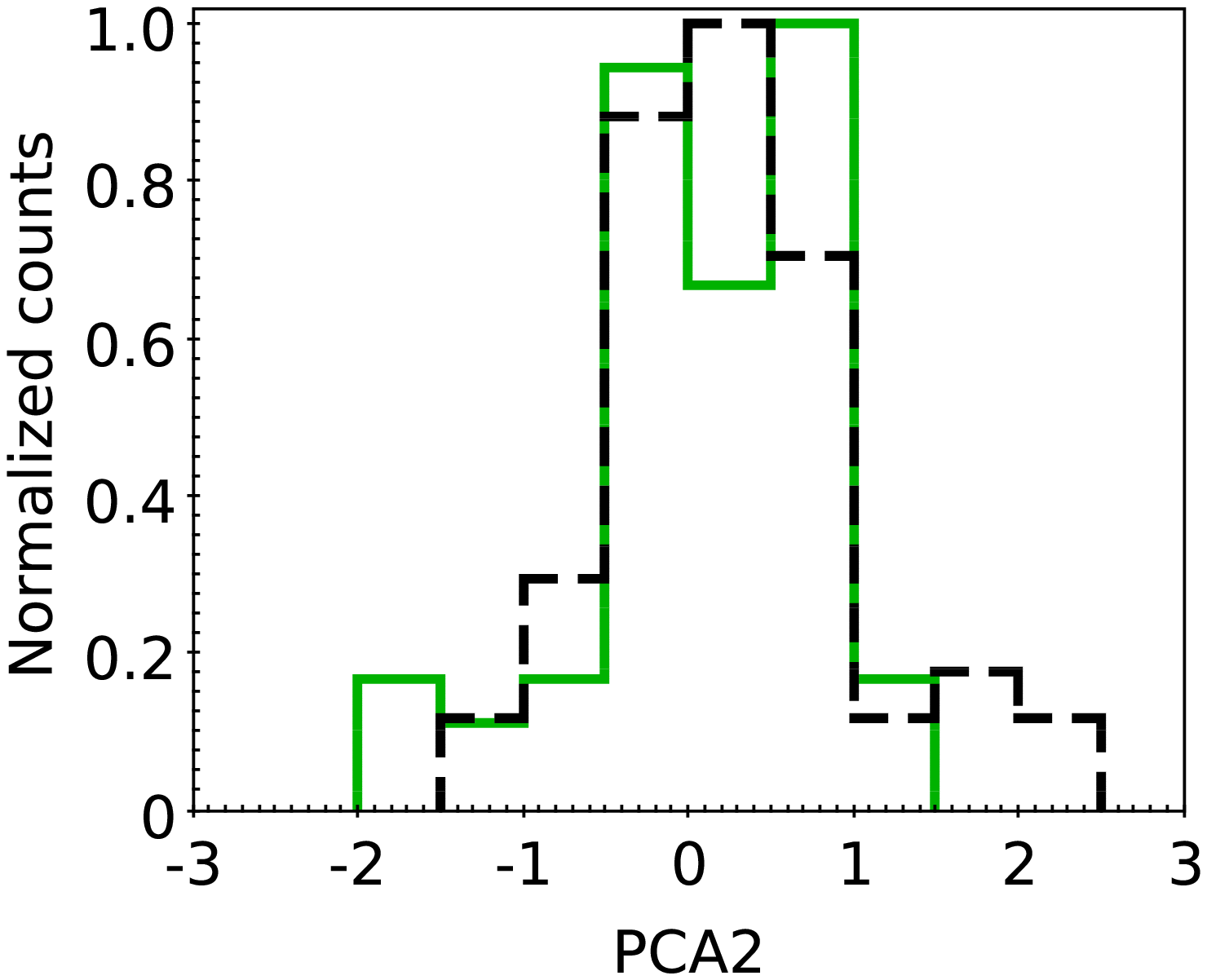,width=1.6in,height=1.3in}
\epsfig{file=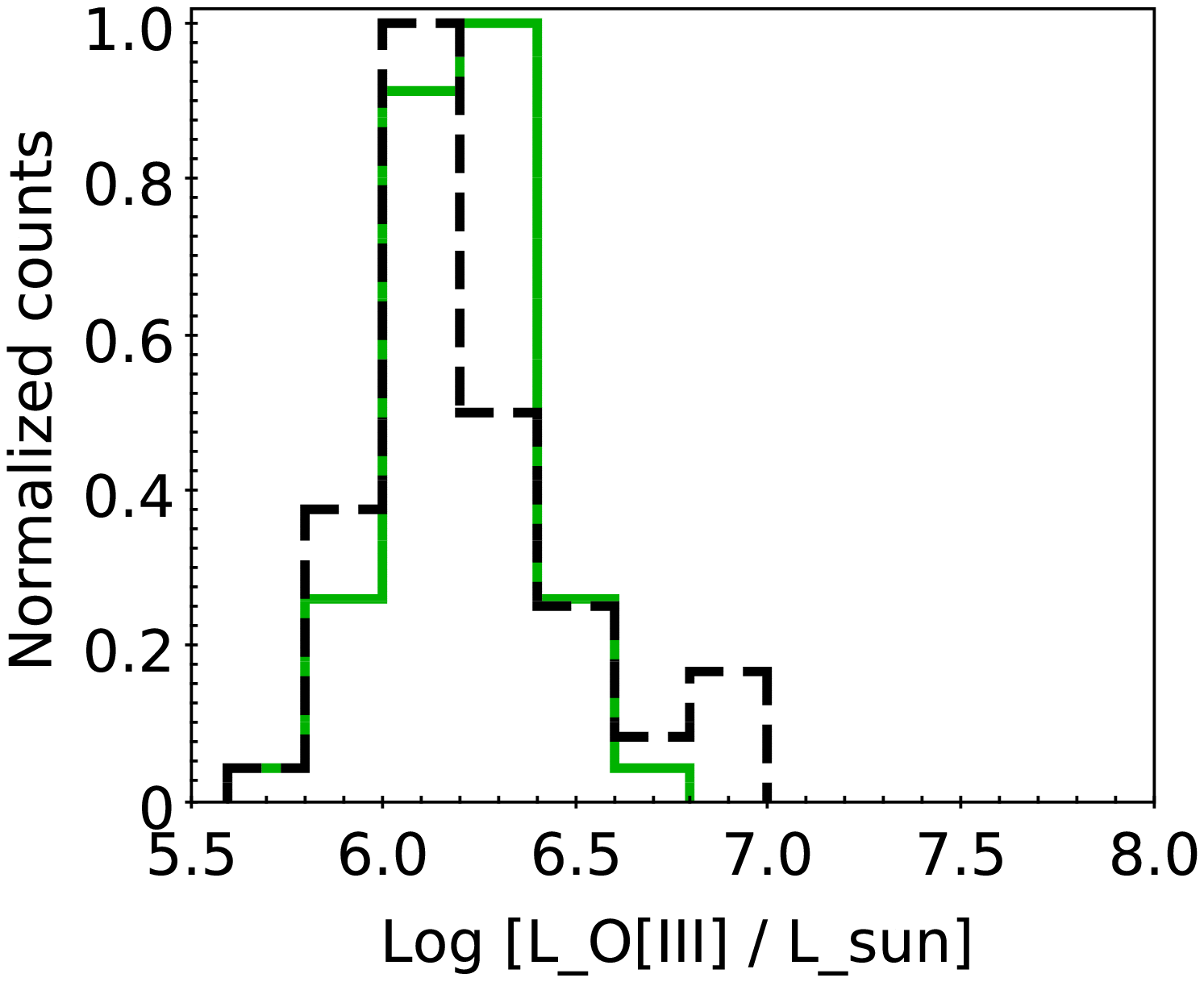,width=1.6in,height=1.3in}
\caption{Histogram of the host galaxy and environmental parameter
  distributions for compact (black) and extended (green) radio sources
  matched in the L$_{rad,c}$--M$_{\star}$ plane.  }
\label{histCom2}
\end{figure*}

In this section, we investigate compact radio AGN and compare them with
the extended FR radio sources. The main question that we address is which
scenarios for the origin of compact AGN fit best with the observations:
are the compact sources a fundamentally different class of objects, are
they FR radio galaxies at the early stage of their evolutions, or are they
short-lived sources which die before they extend to large distances?

The selection criteria for each class are presented in Section 2.  Fig.~$\ref{compact}$
shows the total and core radio luminosity of both the compact and extended
samples versus stellar mass. Compact sources have, on average, lower total
radio luminosity while extended radio sources have lower core radio
luminosity. Accordingly, we set up two different comparisons, creating a
sample matched in {\it total} radio luminosity and stellar mass, and a
second sample matched in {\it core} radio luminosity and stellar
mass. Both core radio luminosity and total radio luminosity have been used
to estimate the jet power (Kording, Jester \& Fender 2008). While there is
a tight correlation between these two parameters in low luminosity sources
(with little extended emission), the correlation shows a very large
scatter for the very luminous radio sources which are subject of this work
(see Fig.~$\ref{core}$, right panel). The core luminosity has been argued to be a
better gauge of jet power than total radio luminosity, as it is a measure
of instantaneous power, rather than something averaged over time and
influenced by environment; even at fixed jet power, the total luminosity
of a radio source evolves as the source grows, going first up then down
according to current models of radio source growth (e.g.\ Kaiser \&
Alexander 1999; Turner \& Shabala 2015).  On the other hand, core luminosity may sometimes be affected 
by relativistic beaming (see the discussion in, e.g., Marcha et~al.\ 2005; Sadler et~al.\ 2014). 
Total luminosity might be a good gauge if the compact sources were
simply small, caught early in their life and perhaps shorter lived than
FRI/II. Considering both core and total radio luminosity in this section
will help to identify whether the results obtained by the two methods
match or differ, thus giving an idea which is the better comparison.
 
In order to make matched samples we adopted the same method that we used
in Section 4.1. Here, we make the matched samples in both the
L$_{rad,t}$--M$_{\star}$ and L$_{rad,c}$--M$_{\star}$ planes with
matching-tolerance limits of $\Delta$ log[L] = $\pm$ 0.2 and $\Delta$
log[M] = $\pm$ 0.1.  The KS-test results for the comparison of host galaxy
and environment parameters are listed in Table~$\ref{table3}$, and
histograms for each of the parameters are presented in Figs.~$\ref{histCom1}$ and $\ref{histCom2}$.

If we assume that total radio luminosity is a good measure for the average
jet power, then this match in total radio luminosity will select a sample
of compact and extended objects matched in jet power, and then we can
investigate which characteristics drive the compact-extended
dichotomy. Compact and extended objects with the same distribution of
total radio luminosity show $>$95\% significant differences in only
4000$\AA$ break and [OIII] luminosity: compact objects have younger
stellar populations and higher line luminosity, both of which imply there
is more cold gas available either for star formation or AGN fueling in
these objects. There is also a consistent trend of differences (but below
95 percent significance in each individual case) in concentration, colour
and size of the host galaxy, all of which point towards compact radio
sources being found in galaxies with stronger disk-like components,
consistent with the higher star-formation rates. No significant
environmental differences have been detected. Prestage \& Peacock (1988)
previously argued that compact radio sources lie in regions of lower
galactic density than extended sources; our results show a weak trend in
that direction, but at below 95$\%$ statistical significance. It is
possible that their result was partially driven by stellar mass and/or
radio luminosity differences between their samples. Finally, the results
show that the matched samples of compact and extended sources have similar
distribution of black hole mass, which supports the correlation of black
hole mass and average jet power.

Interesting results come out when we matched the radio luminosity of the
core for both samples. In this case, the only significant difference that
is observed between the two samples is in the black hole mass. There is no
difference in [OIII] luminosity. This result shows that core radio
luminosity is correlated to the [OIII] luminosity, as previously discussed
by Baldi, Capetti \& Giovannini (2015). They showed that compact LERGs lie
in the same region of the [OIII] versus core luminosity plane that FRI LERGs
do. Therefore, when we matched in core radio luminosity we consequently
matched in [OIII] luminosity. The second important result is that again no
environmental differences have been detected. This completely rules out a
scenario in which jet disruption by the dense galactic and intergalactic
environment causes the radio morphological differences between compact and
extended objects. It also rules out any model in which equivalent jets are
launched in the two cases, but that in low density environments the lack
of a strong working surface causes the jet to escape without producing
luminous extended radio emission (giving the impression of a compact
source). 

The most important result is that, by having the same environmental and
host galaxy properties observed in core-matched compact and extended
sources, the morphological differences appear to have their origin in the
black hole mass. The lower black hole mass in compact sources seems to be
less efficient at launching stable large-scale radio jets, or is able to
support these jets for much shorter periods of time. This result is
consistent with these of Baldi, Capetti \& Giovannini (2016) who claim compact
sources (or type FR0s, as they named compact LERGs) have smaller jet
Lorentz factor compared to FRIs. On a broader scale, a black-hole mass {\it vs} 
jet launching efficiency correlation would also explain the very strong correlation seen
between black hole mass and the fraction of galaxies that host radio-loud AGN (e.g. Best et al. 2005b). 

Generally, the robust conclusions out of these two comparisons are that
compact objects can not simply be FR radio galaxies at the early stage of
their evolutions, or viewed at small angle to their axis, as these models 
could not account for the observed differences in the host galaxy parameters of the compact
and extended sources in our samples. (More specifically, some of the
compact objects may well be caused by one of these effects, but the full
population cannot be -- there must also be other effects at work).
Furthermore, the differences are not driven in any way by different
environments of the sources. Rather, there must be a fundamental
difference between the objects, with the compact objects either being
short-lived radio sources disrupted before they expand to large scale, or
objects which do not efficiently launch large-scale radio jets, perhaps
due to their lower mass black holes.

\begin{figure*}
\center
\epsfig{file=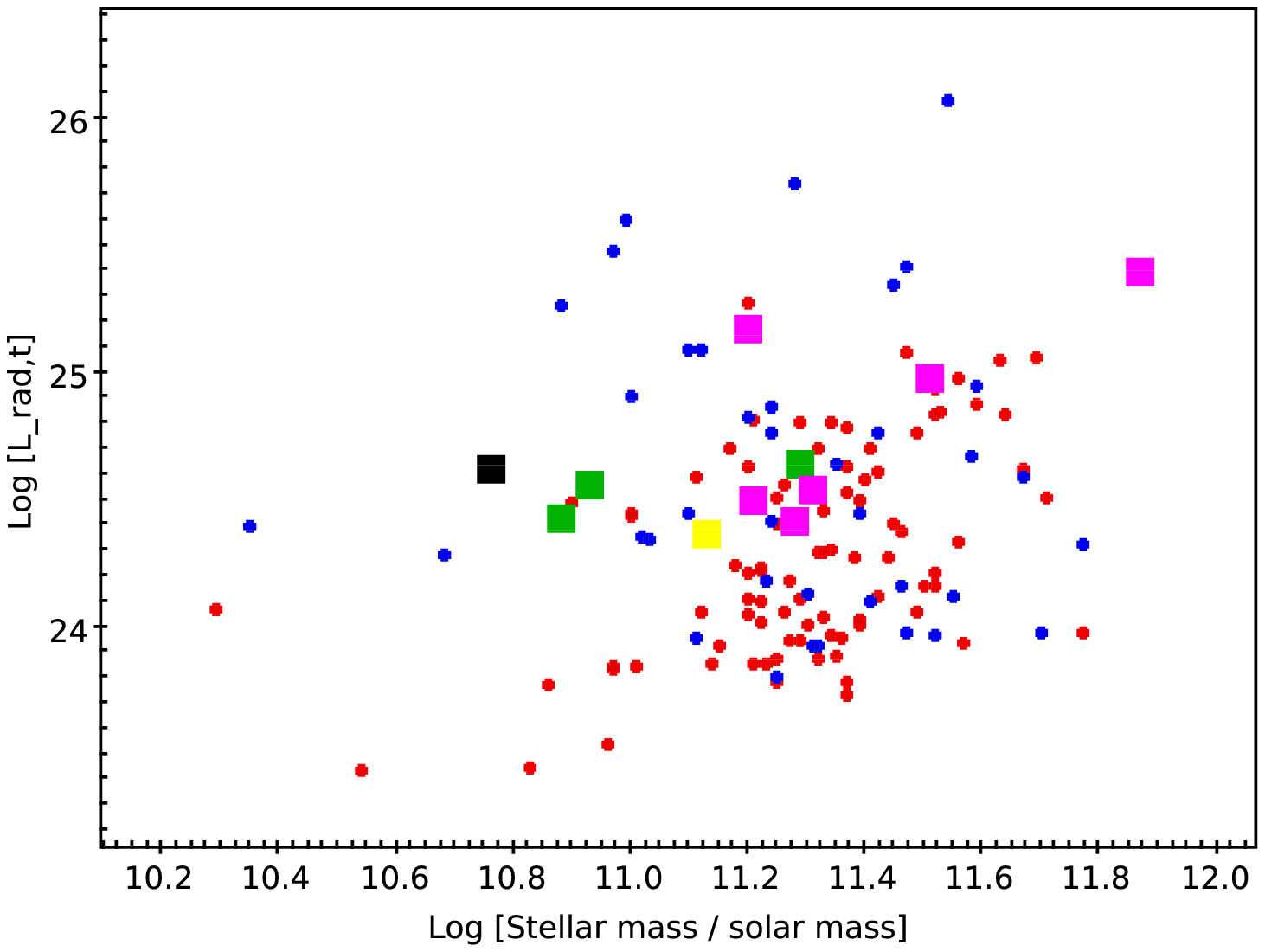,width=3.2in,height=2.6in}
\epsfig{file=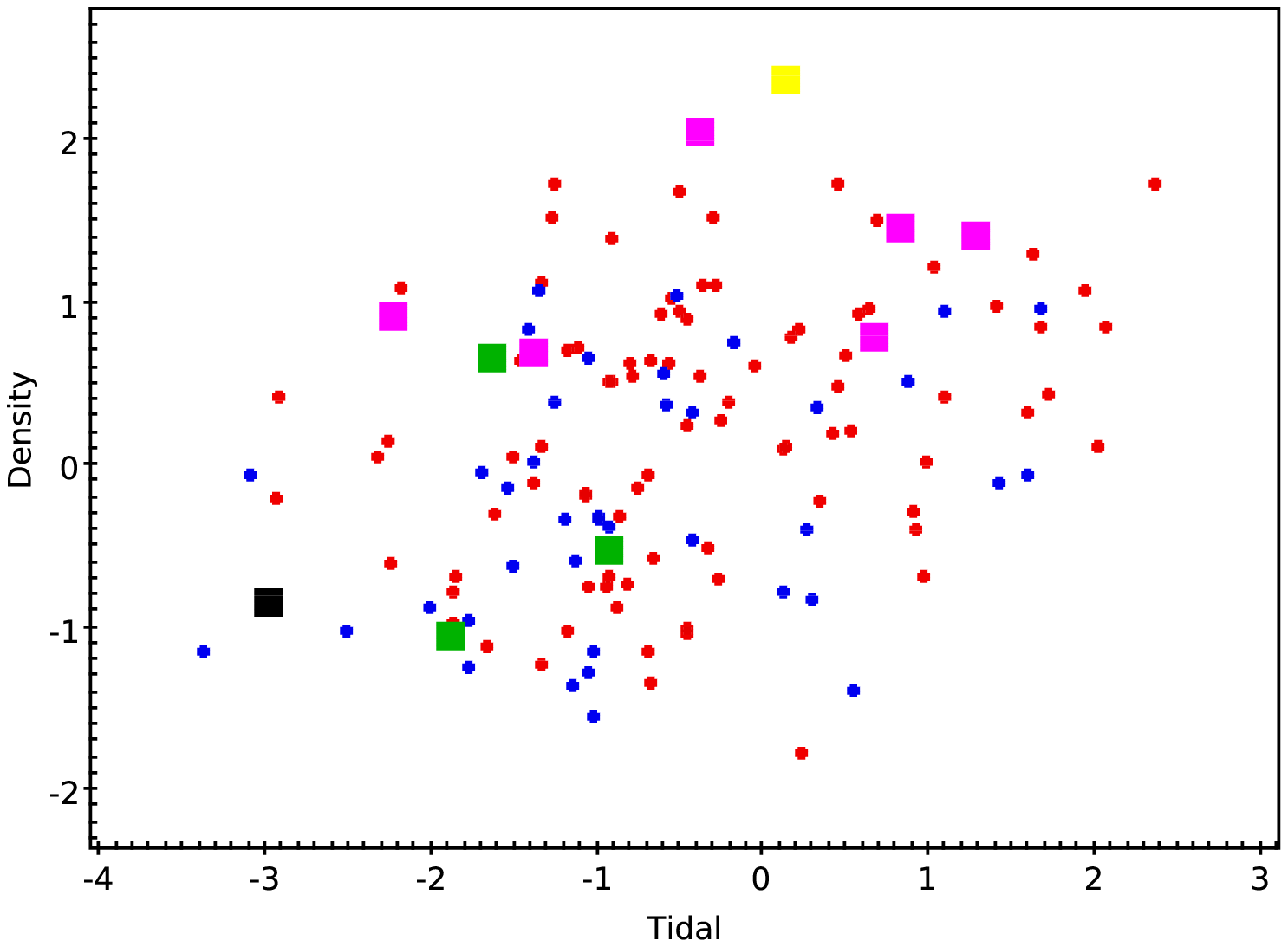,width=3.2in,height=2.6in}
\epsfig{file=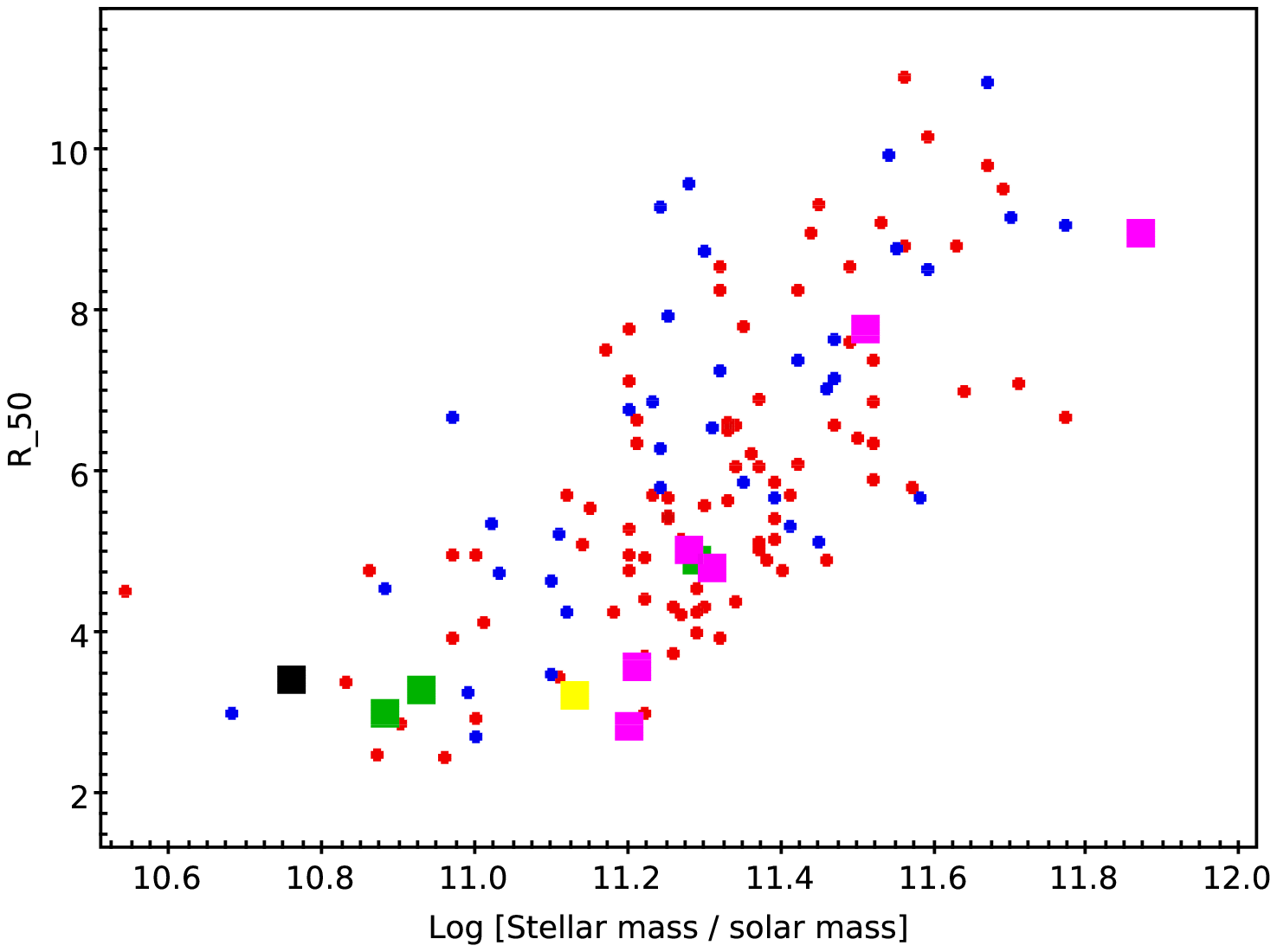,width=3.2in,height=2.6in}
\epsfig{file=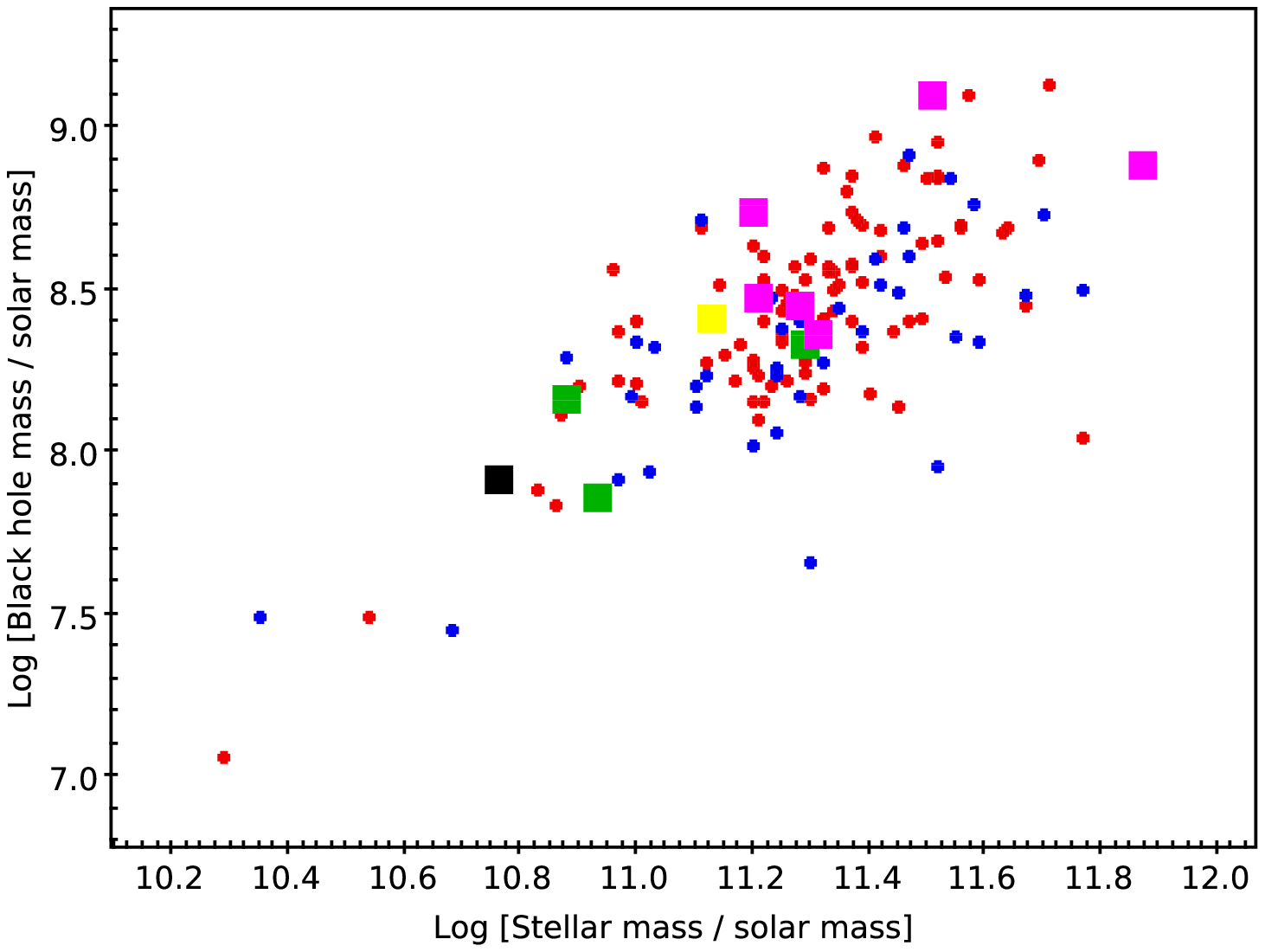,width=3.2in,height=2.6in}

\caption{The host galaxy and environment properties of FRI (red circle),
  FRII (blue circle), HT (yellow square), WAT (pink square), hybrid (green
  square) and D-D (black square) radio galaxies. }
\label{WAT}
\end{figure*}

\section{Beyond the normal FR radio galaxies}

As we have described in Secton 2.1, there are several extended sources
with different and more complex morphologies compared to the normal class
of FRI/II radio galaxies; these are flagged as double-double (D-D) sources
(Schoenmakers et~al.\ 2000), head-tail (HT) sources (Rudnick \& Owen
1976), wide-angle tailed (WAT) sources (Owen \& Rudnick 1976) and FR
hybrid (FRH) sources (Gopal-Krishna \& Wiita 2000). In this section, we
focus on these types of radio sources and explore the host galaxy and the
environment properties of them to see what light this may shed on what
causes such complex morphologies. For this purpose, we applied the
redshift cuts described in Section 2, which greatly decreases the number
of sources but provides an unbaised framework to look at these sources
among normal FRI/II radio galaxies. Fig.~$\ref{WAT}$ shows the results. Some
examples of these sources in our sample are shown in Fig.~$\ref{14}$.

Given the small sample sizes, only qualitative conclusions can be drawn.
The WATs and HT lie systematically towards the lower part of the
R$_{50}$--M$_{\star}$ distribution, suggesting that they have smaller host
galaxies compared to FRIs and FRIIs. They also seem to have higher total
radio luminosity. The clearest result, however, comes from the environment
properties of these sources. HT and WATs are found to reside in the
densest environments, which is exactly as expected since these are
understood to be shaped when the radio jet emission is bent by the
relative movement of the galaxies through the intra-cluster medium.
Therefore, WAT and HT can be efficiently used to identify overdensities
(Blanton et~al.\ 2000, 2001; Dehghan et~al.\ 2014; O'Brien et~al.\ 2016)
especially in the distant universe where the current resolution and
sensitivity of X-ray observations do not allow deep exploration. This
study provides a rich sample of WAT (53) and HT (9) radio sources (see
Table~\ref{table1}), distributed over the redshift range of 0.03-0.4, and
deeper radio surveys will soon allow these to be selected to higher
redshifts.

Likewise, although there is only one D-D source, it is interesting that
this lies exactly at the lowest density part of the density--tidal plane.
Again this is what would be expected, since D-D sources are usually giant
radio galaxies, and low density environments allow these to be achieved by
the jet expanding freely. The FRHs are not found in any special region of
parameter space, although we can't make any strong statements on the basis
of just a few sources. More information about these FRHs could be gleaned
by deriving and examining host galaxy and environment parameters for the
wider sample of 35 sources presented over the full redshift range in
Table~\ref{table1}, but this is beyond the scope of the current paper.

\section{Summary and conclusions}
\label{sec:summery}

We have studied powerful radio galaxies with a wide range of radio
structures, from compact to very extended double-lobe radio sources, and
with a very different optical spectrum, in order to understand the origin
of the observed differences. The radio sources and their corresponding
host galaxies were obtained from Best \& Heckman (2012) who have
cross-matched DR7 of the SDSS with the NVSS and FIRST catalogues. The
radio galaxy sample has been divided into compact and extended radio
sources according to their radio morphologies. The extended radio galaxies
known as Fanaroff--Riley (FR) radio galaxies have been visually divided
into type I (FRI) and type II (FRII), with a few additional sources
classified as hybrid, wide-angle tailed, head-tail and
double-double radio galaxies. The resultant catalogue, which is
presented here, provides a precious sample of over a thousand
FR-classified radio galaxies brighter than $S_{\rm 1.4 GHz} = 40$\,mJy,
out to $z \approx 0.4$.

The subset of radio sources with $0.03 < z < 0.1$ have also been divided
into high excitation (HERG) and low excitation (LERG) radio galaxies
according to their optical spectrum. HERG and LERG sources are understood
to correspond to the sources with high rate of cold accretion flow and low
rate of hot accretion flow respectively. The purpose of this paper was
to investigate the differences in the host galaxies and environment of the
FRI/II and compact sources with HERG and LERG nature separately, in order
to disentangle which effects cause each of the FRI/II, compact/extended
and HERG/LERG dichotomies.

We investigated the FRI/FRII dichotomy using a sample of FRI LERGs and
FRII LERGs with the same stellar mass and total radio luminosity
distribution, to remove any biases caused by HERG/LERG nature, mass and
radio luminosity. We show that FRIs are hosted by smaller galaxies with
higher concentration, higher mass surface density, and higher black hole
to stellar mass ratio than the FRIIs, consistent with the galaxies
possessing less disky structure. The environment of the FRI radio galaxies
show higher density and richness. All the results are consistent with the
models that employ extrinsic parameters (i.e. jet disruption by the 
interstellar and intergalactic media) to explain the FRI/FRII
dichotomy. Previous studies that focussed on intrinsic differences were
all biased by the HERG/LERG classification in the sense that most of them
compare FRI LERGs with the FRII HERGs.
   
We investigated the environment and the host galaxy properties of HERGs
and LERGs using a sample of combined FRI/II and compact HERGs with FRI/II
and compact LERGs, matched in classification, mass and total radio
luminosity. We confirm that HERGs are hosted by galaxies with smaller
4000$\AA$ break, higher [OIII] luminosity and lower black holes mass with
bluer colour and lower concentration compared to the LERGs -- independent
of FR classification. These all indicate that HERGs are found in more
star-forming and disky galaxies. The environments of LERGs display higher
density compared to the HERGs. These results support the hypothesis that the AGN fueling
source is the main origin of HERG/LERG dichotomy. In dense environments
and massive elliptical galaxies the AGN fueling source is believed to be
primarily the hot intergalactic gas which cools and accretes on to the
central black hole at a low accretion rate, giving rise to LERGs. In low
density environments (without hot haloes), depending on the availability
of cold gas, HERG radio sources may form. Therefore, HERG sources are
found in more star-forming and disky galaxies, more typically in lower
density environments, with a higher prevalance of cold gas.

We also investigated the compact/extended dichotomy by comparing a
combined sample of FRI and II LERGs with a sample of compact LERGs,
matched in stellar mass and either core or total radio luminosity.  In
neither case did we find any difference in the AGN environment, indicating
that this is not a cause of the dichotomy. We confirm that the [OIII]
luminosity distributions are the same when matched in core radio
luminosity but not in total radio luminosity, suggesting that the core
radio luminosity is the better measure of the current accretion power. In
the core luminosity matched samples, the only parameter which showed a
significant difference between compact and extended radio sources is the
black hole mass: compact objects harbour lower mass black holes. This
result implies that lower mass black holes are either less efficient at
launching stable large-scale radio jets (consistent with the
interpretation of Baldi et~al. 2016), or able to do so for a shorter time
such that these sources are short-lived.

Finally, we explored the host galaxy and environment properties of radio
galaxies with more complex and interesting morphologies such as wide-angle tailed, 
head-tail, double-double, and FR hybrid. Although the samples are too small to draw quantitative
conclusions, we confirm that HT and WAT reside in very dense regions
compared to the whole population, offering the prospect to identify
over-dense regions such as galaxy clusters and groups through radio
observation alone. This will be a powerful tool in next-generation radio
surveys.

\section*{Acknowledgments}
H. M. would like to thank Institute for Astronomy Royal Observatory
Edinburgh for partial financial support. PNB is grateful for support from
the UK STFC via grant ST/M001229/1.

\label{lastpage}
\end{document}